\documentclass[aps,onecolumn,nofootinbib,tightenlines,unsortedaddress]{revtex4}

\usepackage{amssymb,amstext,amsmath,amsthm,slashed}
\usepackage[dvips]{graphicx}
\usepackage{latexsym}
\usepackage{psfrag}
\usepackage{amsfonts}
\usepackage{bbm} 
\usepackage{color}
\usepackage{verbatim}
\usepackage{dcolumn}
\usepackage{tabularx}
\usepackage{leftidx}
\usepackage{tensor}
\usepackage{boxedminipage}
\usepackage{fancyvrb}
\usepackage{float}
\usepackage{subfigure}
\usepackage{wrapfig}
\usepackage{booktabs}
\usepackage{dsfont} 
\usepackage{mathrsfs}
\usepackage{varwidth}
\usepackage[hidelinks]{hyperref}

\begin{document}

\title{Flaring of Blazars from an Analytical, Time-dependent Model for \\ Combined Synchrotron and Synchrotron Self-Compton Radiative \\ Losses of Multiple Ultrarelativistic Electron Populations}

\vskip.5in

\author{Christian R\"oken\footnote{e-mail: christian.roeken@mathematik.uni-regensburg.de}}
\affiliation{\vspace{-0.4cm} Universit\"at Regensburg, Fakult\"at f\"ur Mathematik, 93040 Regensburg, Germany}

\author{Florian Schuppan}
\affiliation{\vspace{-0.4cm} Ruhr-Universit\"at Bochum, Institut f\"ur Theoretische Physik, Lehrstuhl IV: Weltraum- und Astrophysik, 44780 Bochum, Germany}

\author{Katharina Proksch}
\affiliation{\vspace{-0.4cm} Georg-August-Universit\"at G\"ottingen, Institut f\"ur Mathematische Stochastik, 37077 G\"ottingen, Germany}

\author{Sebastian Sch\"oneberg}
\affiliation{\vspace{-0.4cm} Ruhr-Universit\"at Bochum, Institut f\"ur Theoretische Physik, Lehrstuhl IV: Weltraum- und Astrophysik, 44780 Bochum, Germany}

\date{September 2017}

\begin{abstract}
	
\vspace{0.2cm}

\noindent \textbf{\footnotesize ABSTRACT.} \, We present a fully analytical, time-dependent leptonic one-zone model that describes a simplified radiation process of multiple interacting ultrarelativistic electron populations, accounting for the flaring of GeV blazars. In this model, several mono-energetic, ultrarelativistic electron populations are successively and instantaneously injected into the emission region, i.e., a magnetized plasmoid propagating along the blazar jet, and subjected to linear, time-independent synchrotron radiative losses, which are caused by a constant magnetic field, and nonlinear, time-dependent synchrotron self-Compton radiative losses in the Thomson limit. Considering a general (time-dependent) multiple-injection scenario is, from a physical point of view, more realistic than the usual (time-independent) single-injection scenario invoked in common blazar models, as blazar jets may extend over tens of kiloparsecs and, thus, most likely pick up several particle populations from intermediate clouds. We analytically compute the electron number density by solving a kinetic equation using Laplace transformations and the method of matched asymptotic expansions. Moreover, we explicitly calculate the optically thin synchrotron intensity, the synchrotron self-Compton intensity in the Thomson limit, as well as the associated total fluences. In order to mimic injections of finite duration times and radiative transport, we model flares by sequences of these instantaneous injections, suitably distributed over the entire emission region. Finally, we present a parameter study for the total synchrotron and synchrotron self-Compton fluence spectral energy distributions for a generic three-injection scenario, varying the magnetic field strength, the Doppler factor, and the initial electron energy of the first injection in realistic parameter domains, demonstrating that our model can reproduce the typical broad-band behavior seen in observational data.

\end{abstract}

\maketitle

\tableofcontents

\section{Introduction}

\noindent Blazars are among the most energetic phenomena in nature, representing the most extreme type in the class of active galactic nuclei \cite{UP}. They feature relativistic jets that extend over tens of kiloparsecs and are directed toward the general direction of Earth. Observations of their radiation emission show very high luminosities, rapid variabilities, and high polarizations. Moreover, apparent superluminal characteristics can be detected along the first few parsecs of the jets.
The main components of blazar jets are magnetized plasmoids, which are assumed to arise in the Blandford-Znajek and the Blandford-Payne process \cite{BlZn, BP}, constituting the major radiation zones. These plasmoids pick up -- and interact with -- particles of interstellar and intergalactic clouds along their trajectories \cite{PoSch}, giving rise to the emission of a series of strong flares.

A blazar spectral energy distribution (SED) consists of two broad non-thermal radiation components in different domains. The low-energy spectral component, ranging from radio to optical or X-ray energies, is usually attributed to synchrotron radiation of relativistic electrons subjected to ambient magnetic fields. The origin of the high-energy spectral component, covering the X-ray to $\gamma$-ray regime, is still under debate. It can, for instance, be modeled by inverse Compton radiation coming from low-energy photon fields that interact with the relativistic electrons \cite{B07, DSS}. This process can be described either by a synchrotron self-Compton (SSC) model (see, e.g.,  \cite{SchlRö, RS09a} and references therein), where the electrons scatter their self-generated synchrotron photons, or by so-called external Compton models like \cite{DS, sbr94, bea00}, where the seed photons are generated in the accretion disk, the broad-line region, or the dust torus of the central black hole (some models also consider ambient fields of IR radiation from diffuse, hot dust, see, e.g., \cite{SBMM}). Aside from these leptonic scenarios, the high-energy component can also be modeled via proton-synchrotron radiation or the emission of $\gamma$-rays arising from the decay of neutral pions formed in interactions of protons with ambient matter (see, e.g., \cite{Bea13, ws15, Cer2} and references therein). Mixed models including both leptonic as well as hadronic processes are also considered in the literature (e.g., \cite{Cer1, DaLi}).  

A major task is to properly understand and to account for the distinct variability patterns of the non-thermal blazar emission at all frequencies with different time scales ranging from years down to a few minutes, where the shortest variability time scales are usually observed for the highest energies of the spectral components, as in PKS 2155-304 \cite{AeaH07, AeaH09} and Mrk 501 \cite{aeaM07} in the TeV range, or Mrk 421 \cite{C04} in the X-ray domain. So far, only multi-zone models, which feature an internal structure of the emission region with various radiation zones caused by collisions of moving and stationary shock waves, have been proposed to explain the extreme short-time variability of blazars (see, e.g., \cite{smm, mar, ggpk, gtbc09, gub09, BG12}). In the framework of one-zone models, however, extreme short-time variability can, a priori, not be realized as the duration of the injection into a plasmoid of finite size (with characteristic radius $\mathcal{R}_0$) and the light crossing (escape) time in this region are naturally of the order $\mathcal{O}(2 \, \mathcal{R}_0/(c \, \mathcal{D}))$, $c$ being the speed of light in vacuum and $\mathcal{D}$ the bulk Doppler factor \cite{ESR}. This leads to a minimum time scale for the observed flare duration, which may exceed the short-time variability scale in the minute range by several magnitudes. In particular, using the typical parameters $\mathcal{R}_0 = 10^{15} \, \textnormal{cm}$ and $\mathcal{D} = 10$, we find $2 \, \mathcal{R}_0/(c \, \mathcal{D}) \approx 1.9 \, \textnormal{h}$ in the observer frame. Thus, this type of model can only be used to account for variability on larger time scales, that is, from years down to hours. 

Both one-zone and multi-zone models have been studied extensively over the last decades in order to explain the variability but also the SEDs of blazars, incorporating leptonic and hadronic interactions. In these studies, analytical as well as numerical approaches were used, containing, among others, various radiative loss and acceleration processes, details of radiative transport, diverse injection patterns, different cross sections for particle interactions, as well as particle decay and pair production/annihilation (see the review article \cite{B10} and the references therein). Thus, the literature on this matter is quite comprehensive. However, models featuring pick-up processes of any kind usually assume only a single injection of particles into the emission region as the cause of the flaring. This may be unrealistic as blazar jets, which may extend over tens of kiloparsecs, most likely intersect with several clouds, leading to multiple injections. In such an intricate situation, it is of particular interest to have a self-consistent description of the particle's radiative cooling, the radiative transport, et cetera. Therefore, we propose a simple, but fully analytical, time-dependent leptonic one-zone model featuring multiple uniform injections of nonlinearly interacting ultrarelativistic electron populations, which undergo combined synchrotron and SSC radiative losses (for previous works, see \cite{RökenSchlickeiser, RS09a}). More precisely, we assume that the blazar radiation emission originates in spherically shaped and fully ionized plasmoids, which feature intrinsic randomly-oriented, but constant, large-scale magnetic fields and propagate ultrarelativistically along the general direction of the jet axis. These plasmoids pass through -- and interact with -- clouds of the interstellar and intergalactic media, successively and instantaneously picking up multiple mono-energetic, spatially isotropically distributed electron populations, which are subjected to linear, time-independent synchrotron radiative losses via interactions with the ambient magnetic fields and to nonlinear, time-dependent SSC radiative losses in the Thomson limit. This is the first time an analytical model that describes combined synchrotron and SSC radiative losses of several subsequently injected, interacting injections is presented (for a small-scale, purely numerical study on multiple injections see \cite{lk00}). We point out that because the SSC cooling is a collective effect, that is, the cooling of a single electron depends on the entire ensemble within the emission region, injections of further particle populations into an already cooling system give rise to alterations of the overall cooling behavior \cite{RökenSchlickeiser, zs13}. Moreover, as we do employ Dirac distributions for the time profile of the source function of our kinetic equation and, further, do not consider any details of radiative transport, we mimic injections of finite duration times and radiative transport by partitioning each flare into a sequence of instantaneous injections, which are appropriately distributed over the entire emission region, using the quantities computed here. 

The article is organized as follows. In Section \ref{SEC2}, an approximate analytical solution of the time-dependent, relativistic kinetic equation of the volume-averaged differential electron number density is derived. Based on this solution, the optically thin synchrotron intensity, the SSC intensity in the Thomson limit, as well as the corresponding total fluences are calculated in Sections \ref{SEC3} and \ref{SEC4}. We explain how to mimic finite injection durations and radiative transport by multiple use of our results in Section \ref{MSTVBL}, and show a parameter study for the total synchrotron and SSC fluence SEDs. Section \ref{S&O} concludes with a summary and an outlook. Supplementary material, which is required for the computations of the electron number density and the synchrotron and SSC intensities, is given in Appendices \ref{app:LM}-\ref{LTOF}. Moreover, in Appendix \ref{numcode}, we briefly describe the plotting algorithm employed for the creation of the fluence SEDs.

\section{The Relativistic Kinetic Equation} \label{SEC2}

\noindent The kinetic equation for the volume-averaged differential electron number density $n = n(\gamma, t)$ (where $t$ is the time, $\gamma := E_{\textnormal{e}}/(m_{\textnormal{e}} \, c^2)$ the normalized electron energy, and [n] = $\textnormal{cm}^{-3}$) of $m$ ultrarelativistic, mono-energetic, instantaneously injected and spatially isotropically distributed electron populations in the rest frame of a non-thermal radiation source with dominant magnetic field self-generation and radiative loss rate $L = L(\gamma, t)$ (with $[L] = \textnormal{s}^{-1}$) reads \cite{Kard} 
\begin{equation} \label{KE}
\frac{\partial n}{\partial t} - \frac{\partial}{\partial \gamma} \bigl(L \, n\bigr) =  \sum_{i = 1}^{m} \, q_i \, \delta(\gamma - \gamma_i) \, \delta(t - t_i) \, ,
\end{equation}
where $\delta(\cdot)$ is the Dirac distribution, $q_i$ (for which $[q_i] = \textnormal{cm}^{-3}$) the $i$th injection strength, $\gamma_i := E_{\textnormal{e}, i}/(m_{\textnormal{e}} \, c^2) \gg 1$ the $i$th normalized initial electron energy, and $t_i$ the $i$th injection time for $i: \, 1 \leq i \leq m$. In this work, radiative losses in form of both a linear, time-independent synchrotron cooling process (with a constant magnetic field $\boldsymbol{B}$) and a nonlinear, time-dependent SSC cooling process in the Thomson limit
\begin{equation} \label{Cool}
L = L_\textnormal{syn.} + L_\textnormal{SSC} = \gamma^2 \left(D_0 + A_0 \int_0^{\infty} \gamma^2 \, n \, \textnormal{d}\gamma\right) 
\end{equation}
are considered. The respective cooling rate prefactors yield $D_0 = 1.3 \times 10^{- 9} \, b^2 \, \textnormal{s}^{- 1}$ and $A_0 = 1.2 \times 10^{- 18} \, b^2 \, \textnormal{cm}^3 \, \textnormal{s}^{- 1}$, which depend on the nondimensional magnetic field strength $b := \|\boldsymbol{B}\| / \textnormal{Gauss} = \textnormal{const.}$ \cite{BG, S09}. In the present context, the term \textit{linear} refers to the fact that the synchrotron loss term does not depend on the electron number density $n$, thus, resulting in a linear contribution to the partial differential equation (PDE) (\ref{KE}), whereas the SSC loss term depends on an energy integral containing $n$, yielding a nonlinear contribution. The synchrotron loss term can, in principle, be modeled as nonlinear and time-dependent, too. This can be achieved by making an equipartition assumption between the magnetic and particle energy densities \cite{sl07}. We point out that, except for TeV blazars, where Klein-Nishina effects drastically reduce the SSC cooling strength above a certain energy threshold, the dominant contribution of the SSC energy loss rate always originates in the Thomson regime \cite{S09}, justifying our initial restriction. But as a consequence, this limits the applicability of our model to at most GeV blazars and bounds the normalized initial electron energies from above by $\gamma_i < 1.9 \times 10^4 \, b^{-1/3}$. We also note that the only accessible energy in the synchrotron and SSC cooling processes is the kinetic energy of the electrons. Thus, $\gamma$ denotes the kinetic component of the normalized total energy $E_\textnormal{tot.}/(m_{\textnormal{e}} c^2)$, i.e., 
\begin{equation*}
\gamma = \frac{E_\textnormal{tot.} - m_{\textnormal{e}} \, c^2}{m_{\textnormal{e}} \, c^2} = \gamma_\textnormal{tot.} - 1 \, .
\end{equation*}
Then, $\gamma_\textnormal{tot.} \in [1, \infty)$ implies $\gamma \in [0, \infty)$. For this reason, the lower integration limit in (\ref{Cool}) is zero. Furthermore, the Dirac distributions $\delta(\gamma - \gamma_i)$ and $\delta(t - t_i)$ in Eq.\ (\ref{KE}) determine a sequence of energy and time points $(\gamma_i, t_i)_{i \in \{1, ..., m\}}$. They are to be understood in the distributional sense, that is, each is a linear functional on the space of smooth test functions $\varphi$ on $\mathbb{R}$ with compact support
\begin{equation*}
C_0^{\infty}(\mathbb{R}) := \bigl\{\varphi \, | \, \varphi \in C^{\infty}(\mathbb{R}) \, , \,\,\, \textnormal{supp} \, \varphi \,\,\, \textnormal{compact}\bigr\} \, .
\end{equation*}
More precisely, for $k \in \mathbb{R}$, one can rigorously define them as the mapping
\begin{equation*}
\delta: \, C_0^{\infty}(\mathbb{R}) \rightarrow \mathbb{R}\, , \,\,\, \varphi \mapsto \varphi(0)
\end{equation*}
with the integral of the Dirac distribution against a test function given by
\begin{equation*}
\int_{- \infty}^{\infty} \varphi(k) \, \delta(k) \, \textnormal{d}k =  \varphi(0) \,\,\,\,\,\,\,\, \textnormal{for all} \,\,\,\,\,\,\,\, \varphi \in C_0^{\infty}(\mathbb{R}) \, .    
\end{equation*}

\subsection{Formal Solution of the Relativistic Kinetic Equation} \label{IIa}

\noindent In terms of the function $R(\gamma, t) := \gamma^2 \, n(\gamma, t)$ and the variable $x := 1/\gamma$, we can rewrite Eq.\ (\ref{KE}) in the form
\begin{equation}\label{TER} 
\frac{\partial R}{\partial t} + J \, \frac{\partial R}{\partial x} = \sum_{i = 1}^{m} \, q_i \, \delta(x - x_i) \, \delta(t - t_i) \, ,
\end{equation}
where
\begin{equation}\label{F}
J = J(t) := D_0 + A_0 \int_{0}^{\infty} \frac{R(x, t)}{x^2} \, \textnormal{d}x \, . 
\end{equation}
Defining the strictly increasing, continuous function $G = G(t)$ via 
\begin{equation}\label{GF}
0 < \frac{\textnormal{d}G}{\textnormal{d}t} := J \, ,
\end{equation}
Eq.\ (\ref{TER}) becomes
\begin{equation}\label{dglr}
\frac{\partial R}{\partial G} + \frac{\partial R}{\partial x} = \sum_{i = 1}^{m} \, q_i \, \delta(x - x_i) \, \delta(G - G_i) 
\end{equation}
with $G_i := G(t_i)$. Albeit the nonlinear kinetic equation (\ref{KE}) is now transformed into a linear PDE, its solution can obviously serve as a Green's function only for the single-injection scenario with $m = 1$. Consequently, in order to solve the generalized kinetic equation
\begin{equation*}
\frac{\partial n}{\partial t} - \frac{\partial}{\partial \gamma} \bigl(L \, n\bigr) =  Q 
\end{equation*}
for $m > 1$, where $Q = Q(\gamma, t)$ is a more realistic source function, one cannot simply use Green's method. Applying successive Laplace transformations with respect to $x$ and $G$ to Eq.\ (\ref{dglr}) yields the solution
\begin{equation}\label{solR}
R(x, G) = \sum_{i = 1}^{m} \, q_i \, H(G - G_i) \, \delta(x - x_i - G + G_i) \, ,
\end{equation}
where $H(\cdot)$ is the Heaviside step function 
\begin{equation*}
H(k - k_0) := \begin{cases} 0 & \, \textnormal{for} \,\,\,\,\, k < k_0 \\  
1 & \, \textnormal{for} \,\,\,\,\, k > k_0 \end{cases}
\end{equation*} 
with $k, k_0 \in \mathbb{R}$, which has a jump discontinuity at $k = k_0$. A detailed derivation of this solution can be found in Appendix \ref{app:LM}. In order to determine the function $G$, we substitute solution (\ref{solR}) into (\ref{F}) and, by using (\ref{GF}), obtain the ordinary differential equation (ODE) 
\begin{equation} \label{GE}
\frac{\textnormal{d}G}{\textnormal{d}t} = D_0 + A_0 \, \sum_{i = 1}^{m} \, \frac{q_i \, H(G - G_i)}{\bigl(G - G_i + x_i\bigr)^2} \, . 
\end{equation}
This equation can be regarded as a compact notation for the set of $m$ piecewise-defined ODEs
\begin{equation} \label{EFGE}
\begin{split}
\begin{cases} 
\displaystyle \frac{\textnormal{d}G}{\textnormal{d}t} = D_0 + A_0 \, \frac{q_1}{\bigl(G - G_1 + x_1\bigr)^2} & \,\,\,\,\,\,\, \textnormal{for} \,\,\,\,\,\,\, G_1 \leq G < G_2 \\ \\ 
\displaystyle \frac{\textnormal{d}G}{\textnormal{d}t} = D_0 + A_0 \, \Biggl(\frac{q_1}{\bigl(G - G_1 + x_1\bigr)^2} + \frac{q_2}{\bigl(G - G_2 + x_2\bigr)^2}\Biggr) & \,\,\,\,\,\,\, \textnormal{for} \,\,\,\,\,\,\, G_2 \leq G < G_3 \\ \\
\hspace{1.5cm} \vdots \hspace{2.5cm} \vdots \hspace{2.5cm} \vdots & \hspace{2cm} \vdots\\ \\
\displaystyle \frac{\textnormal{d}G}{\textnormal{d}t} = D_0 + A_0 \, \Biggl(\frac{q_1}{\bigl(G - G_1 + x_1\bigr)^2} + \frac{q_2}{\bigl(G - G_2 + x_2\bigr)^2} + \ldots + \frac{q_m}{\bigl(G - G_m + x_m\bigr)^2}\Biggr) & \,\,\,\,\,\,\, \textnormal{for} \,\,\,\,\,\,\, G_m \leq G < \infty \, .
\end{cases}
\end{split}
\end{equation}
In Section \ref{SecIIb}, we present an approximate analytical solution for the general case of $j$ injections, i.e., for the interval $G_j \leq G < G_{j + 1}$, where $j \in \{1, ..., m\}$ and $G_1 = 0$, $G_{m + 1} = \infty$, employing the method of matched asymptotic expansions. Having a separate analytical solution for each ODE of (\ref{EFGE}), in Section \ref{C}, these are connected successively requiring continuity at the transition points. Finally, by substituting (\ref{solR}), the electron number density results in 
\begin{equation}\label{edd}
\begin{split}
n(\gamma, t) & = \gamma^{- 2} \, R(\gamma, t) = \gamma^{- 2} \, \sum_{i = 1}^m \, q_i \, H\bigl(G(t) - G_i\bigr) \, \delta\left(\frac{1}{\gamma} - \frac{1}{\gamma_i} - G(t) + G_i\right) \\ \\
& = \sum_{i = 1}^m \, q_i \, H\left(t - t_i\right) \, \delta\left(\gamma - \frac{\gamma_i}{1 + \gamma_i \, \bigl(G(t) - G_i\bigr)}\right) \, .
\end{split}
\end{equation}

\subsection{Solution of the Relativistic Kinetic Equation for $\bigl\{t \in \mathbb{R}_{\geq 0} \, | \, t_j \leq t < t_{j + 1} \,\,\, \textnormal{with} \,\,\, j \in \{1, ..., m\}\bigr\}$} \label{SecIIb}

\noindent In the interval $G_j \leq G < G_{j + 1}$, Eq.\ (\ref{GE}) gives
\begin{equation} \label{GEm}
\frac{\textnormal{d}G}{\textnormal{d}t} = D_0 + A_0 \, \sum_{i = 1}^{j} \, \frac{q_i}{\bigl(G - G_i + x_i \bigr)^2} \, . 
\end{equation}
In order to solve this equation, we introduce the time-dependent sets $S_1$, $S_2$, and $S_3$ that contain indices $i : 1 \leq i \leq j$ belonging to either injections in the near-injection domain (NID) $0 \leq G - G_i \ll x_i$, the intermediate-injection domain (IID) $G - G_i \approx x_i$, or the far-injection domain (FID) $G - G_i \gg x_i$, respectively. We may now rewrite Eq.\ (\ref{GEm}) in terms of these sets by continuing their domains of validity to $G_i \leq G < \textnormal{min}\bigl(G_{j + 1}, G_{\textnormal{T}}^{(\textnormal{N} \rightarrow \textnormal{I})}(i)\bigr)$ for the NID, $G_{\textnormal{T}}^{(\textnormal{N} \rightarrow \textnormal{I})}(i) \leq G < \textnormal{min}\bigl(G_{j + 1}, G_{\textnormal{T}}^{(\textnormal{I} \rightarrow \textnormal{F})}(i)\bigr)$ for the IID, and $G_{\textnormal{T}}^{(\textnormal{I} \rightarrow \textnormal{F})}(i) \leq G < G_{j + 1}$ for the FID with the NID-IID and IID-FID transition values $G_{\textnormal{T}}^{(\textnormal{N} \rightarrow \textnormal{I})}(i) := G_i + x_i/\xi_i^{(\textnormal{N} \rightarrow \textnormal{I})}$ and $G_{\textnormal{T}}^{(\textnormal{I} \rightarrow \textnormal{F})}(i) := G_i + \xi_i^{(\textnormal{I} \rightarrow \textnormal{F})} x_i$, where $\xi_i^{(\textnormal{N} \rightarrow \textnormal{I})}, \xi_i^{(\textnormal{I} \rightarrow \textnormal{F})} > 1$ are positive constants, taking into account that the $(j + 1)$th injection can occur in each domain. However, for the computations below, we still use the former \textit{much less than}, \textit{approximately equal}, and \textit{much greater than} relations when we carry out approximations. Then, Eq.\ (\ref{GEm}) can be represented formally by 
\begin{equation} \label{GEmform}
\frac{\textnormal{d}G}{\textnormal{d}t} = D_0 + A_0 \Biggl(\sum_{u \in S_1} \, \frac{q_u}{\bigl(G - G_u + x_u\bigr)^2} + \sum_{v \in S_2} \, \frac{q_v}{\bigl(G - G_v + x_v\bigr)^2} + \sum_{w \in S_3} \, \frac{q_w}{\bigl(G - G_w + x_w\bigr)^2}\Biggr) \, . 
\end{equation}
Next, we express the NID and the FID summands by the Taylor series of the function $(1 + z)^{- 2}$ evaluated at the point $z = 0$ (generalized geometric series) 
\begin{equation*}
\frac{1}{(1 + z)^2} = \sum_{n = 0}^{\infty} \, (- 1)^n \, (n + 1) \, z^n \,\,\,\,\,\, \textnormal{for} \,\,\,\,\,\, |z| < 1 \, ,
\end{equation*}
and the IID summands by the Taylor series of the same function evaluated at the point $z = 1$
\begin{equation*}
\frac{1}{(1 + z)^2} = \sum_{n = 0}^{\infty} \, (- 1)^n \, \frac{n + 1}{2^{n + 2}} \, (z - 1)^n \,\,\,\,\,\, \textnormal{for} \,\,\,\,\,\, |z - 1| < 2 \, , 
\end{equation*}
and suitably approximate these series by considering only the leading- and next-to-leading-order terms. This results in  
\begin{align} 
\frac{1}{(G - G_u + x_u)^2} & = \frac{1}{x_u^2} \, \sum_{n = 0}^{\infty} \, (n + 1) \, \biggl(- \frac{G - G_u}{x_u}\biggr)^n \approx \frac{1}{x_u^2} \, \biggl(1 - \frac{2 \, (G - G_u)}{x_u}\biggr) \label{APPROX1} \\ \nonumber\\
\frac{1}{(G - G_v + x_v)^2} & = \frac{1}{x_v^2} \, \sum_{n = 0}^{\infty} \, \frac{n + 1}{2^{n + 2}} \, \biggl(1 - \frac{G - G_v}{x_v}\biggr)^n \approx \frac{1}{4 \, x_v^2} \, \biggl(2 - \frac{G - G_v}{x_v}\biggr) \label{APPROX2} \\ \nonumber \\
\frac{1}{(G - G_w + x_w)^2} & = \frac{1}{(G - G_w)^2} \, \sum_{n = 0}^{\infty} \, (n + 1) \, \biggl(- \frac{x_w}{G - G_w}\biggr)^n \approx \frac{1}{(G - G_w)^2} \, \biggl(1 - \frac{2 \, x_w}{G - G_w}\biggr) \label{APPROX3}
\end{align}
for all $u \in S_1$, $v \in S_2$, and $w \in S_3$. To find an approximate analytical solution to Eq.\ (\ref{GEmform}), we require the function $G$ to be extricable from the respective sums. This is already achieved with the approximations (\ref{APPROX1}) and (\ref{APPROX2}). The latter approximation (\ref{APPROX3}), however, needs further modifications as $G$ still couples to $G_w$ in the denominator. To this end, we employ both geometric and generalized geometric series and again consider only the leading- and next-to-leading-order terms
\begin{equation} \label{APPROX4}
\frac{1}{(G - G_w + x_w)^2} \approx \frac{1}{G^2} \, \, \sum_{k = 0}^{\infty} \, (k + 1) \, \biggl(\frac{G_w}{G}\biggr)^k \, \Biggl[1 - \frac{2 \, x_w}{G} \, \sum_{l = 0}^{\infty} \, \biggl(\frac{G_w}{G}\biggr)^l \, \Biggr] \approx \frac{1}{G^2} \, \biggl(1 + \frac{2 \, (G_w - x_w)}{G}\biggr) \, .
\end{equation}
The approximation errors and optimal values for the interval parameters $\xi_i^{(\textnormal{N} \rightarrow \textnormal{I})}$ and $\xi_i^{(\textnormal{I} \rightarrow \textnormal{F})}$ are given in Appendix \ref{AEaOD}. Substituting (\ref{APPROX1}), (\ref{APPROX2}), and (\ref{APPROX4}) into the ODE (\ref{GEmform}), we obtain
\begin{equation} \label{FA}
\frac{\textnormal{d}G}{\textnormal{d}t} \approx \mathscr{C}_1(j; S_1, S_2) + \mathscr{C}_2(j; S_1, S_2) \, G + \frac{\mathscr{C}_3(j; S_3)}{G^2} + \frac{\mathscr{C}_4(j; S_3)}{G^3} 
\end{equation}
with the ``constants'' 
\begin{equation} \label{constants}
\begin{split}
& \mathscr{C}_1(j; S_1, S_2) := D_0 + A_0 \, \Biggl(\sum_{u \in S_1} \, \frac{q_u}{x_u^2} \, \biggl[1 + \frac{2 \, G_u}{x_u}\biggr] + \frac{1}{2} \, \sum_{v \in S_2} \, \frac{q_v}{x_v^2} \, \biggl[1 + \frac{G_v}{2 \, x_v}\biggr]\Biggr)\, , \\ \\ 
& \mathscr{C}_2(j; S_1, S_2) := - A_0 \, \Biggl(2 \, \sum_{u \in S_1} \, \frac{q_u}{x_u^3} + \frac{1}{4} \, \sum_{v \in S_2} \, \frac{q_v}{x_v^3}\Biggr) \, , \\ \\
& \mathscr{C}_3(j; S_3) := A_0 \, \sum_{w \in S_3} \, q_w \, , \,\,\, \textnormal{and} \,\,\,\,\, \mathscr{C}_4(j; S_3) := 2 \, A_0 \, \sum_{w \in S_3} \, q_w \, (G_w - x_w) \, ,
\end{split}
\end{equation}
which depend on the actual numbers of elements of $S_1$, $S_2$, and $S_3$. Note that the explicit dependences of these constants on the total number of injections as well as on (the numbers of elements of) the sets $S_1$, $S_2$, and $S_3$ are suppressed in the subsequent calculations for simplicity if possible and given if necessary. We derive an approximate analytical solution of Eq.\ (\ref{FA}) by computing separate solutions for the NID, IID, and FID of the $j$th injection, which, in case $G_{\textnormal{T}}^{(\textnormal{N} \rightarrow \textnormal{I})}(j) < G_{\textnormal{T}}^{(\textnormal{I} \rightarrow \textnormal{F})}(j) < G_{j + 1}$, are glued together continuously at the transition points $G_{\textnormal{T}}^{(\textnormal{N} \rightarrow \textnormal{I})}(j)$ and $G_{\textnormal{T}}^{(\textnormal{I} \rightarrow \textnormal{F})}(j)$ (that define the transitions of the $j$th injection between $S_1$ and $S_2$ as well as between $S_2$ and $S_3$) corresponding to the transition times $t_{\textnormal{T}}^{(\textnormal{N} \rightarrow \textnormal{I})}(j)$ and $t_{\textnormal{T}}^{(\textnormal{I} \rightarrow \textnormal{F})}(j)$ specified in Appendix \ref{NIRFIRTT}. For $G_{\textnormal{T}}^{(\textnormal{N} \rightarrow \textnormal{I})}(j) < G_{j + 1} \leq G_{\textnormal{T}}^{(\textnormal{I} \rightarrow \textnormal{F})}(j)$, we regard only the NID and IID solutions with the proper continuous gluing at $G_{\textnormal{T}}^{(\textnormal{N} \rightarrow \textnormal{I})}(j)$, and for $ G_{j + 1} \leq G_{\textnormal{T}}^{(\textnormal{N} \rightarrow \textnormal{I})}(j) < G_{\textnormal{T}}^{(\textnormal{I} \rightarrow \textnormal{F})}(j)$, we consider solely the NID solution. Moreover, we have to update the above constants (\ref{constants}) each time an injection $i : 1 \leq i \leq j$ crosses over from its NID to its IID at $t = t_{\textnormal{T}}^{(\textnormal{N} \rightarrow \textnormal{I})}(i)$ or from its IID to its FID at $t = t_{\textnormal{T}}^{(\textnormal{I} \rightarrow \textnormal{F})}(i)$, as these transitions cause changes in either the numbers of elements of the sets $S_1$ and $S_2$ or of $S_2$ and $S_3$. In the following, due to the particular structure of Eq.\ (\ref{FA}) and for reasons of analytical solvability, the general strategy is to use both the leading- and next-to-leading-order contributions only if $S_1$, or $S_2$, or $S_1$ and $S_2$, or $S_3$ has to be taken into account, whereas only the leading-order terms are considered if $S_1$ and $S_3$, or $S_2$ and $S_3$, or $S_1$ and $S_2$ and $S_3$ have to be employed.

\subsubsection{Near-injection Domain Solution} \label{niasec}

\noindent In the NID $G_j \leq G < \textnormal{min}\bigl(G_{j + 1}, G_{\textnormal{T}}^{(\textnormal{N} \rightarrow \textnormal{I})}(j)\bigr)$, at least the $j$th injection is an element of $S_1$. Further, one may -- but does not necessarily need to -- have injections in $S_2$ and $S_3$. Therefore, we distinguish the following four cases 
\begin{align} 
\frac{\textnormal{d}G}{\textnormal{d}t} & = \mathscr{C}_1(j; S_1) + \mathscr{C}_2(j; S_1) \, G & & \hspace{-3.3cm} \textnormal{for} \,\,\,\,\,\,\, S_2 = \emptyset = S_3 \label{NIDODE1} \\ \nonumber \\
\frac{\textnormal{d}G}{\textnormal{d}t} & = \mathscr{C}_1(j; S_1, S_2) + \mathscr{C}_2(j; S_1, S_2) \, G & & \hspace{-3.3cm} \textnormal{for} \,\,\,\,\,\,\, S_2 \not= \emptyset, S_3 = \emptyset \label{NIDODE2} \\ \nonumber \\
\frac{\textnormal{d}G}{\textnormal{d}t} & = \mathscr{C}_1(j; S_1) + \frac{\mathscr{C}_3(j; S_3)}{G^2} & & \hspace{-3.3cm} \textnormal{for} \,\,\,\,\,\,\, S_2 = \emptyset, S_3 \not= \emptyset \label{NIDODE3} \\ \nonumber \\
\frac{\textnormal{d}G}{\textnormal{d}t} & = \mathscr{C}_1(j; S_1, S_2) + \frac{\mathscr{C}_3(j; S_3)}{G^2} & & \hspace{-3.3cm} \textnormal{for} \,\,\,\,\,\,\, S_2 \not= \emptyset \not= S_3 \, . \label{NIDODE4}
\end{align}
Since Eqs.\ (\ref{NIDODE1}) and (\ref{NIDODE2}) as well as Eqs.\ (\ref{NIDODE3}) and (\ref{NIDODE4}) are identical except for the present constants, only two different kinds of ODEs have to be solved. Starting with Eqs.\ (\ref{NIDODE1}) and (\ref{NIDODE2}), we use separation of variables in order to obtain
\begin{equation*} 
\int \frac{\textnormal{d}G}{\mathscr{C}_1 + \mathscr{C}_2 \, G} = \frac{1}{\mathscr{C}_2} \, \ln{(\mathscr{C}_1 + \mathscr{C}_2 \, G)} = t + c_1 \, , \,\,\,\,\, c_1 = \textnormal{const.} \in \mathbb{R} \, ,
\end{equation*}
which is equivalent to 
\begin{equation*} 
G(t) = \frac{1}{|\mathscr{C}_2(j)|} \Bigl[\mathscr{C}_1(j) - \exp{\bigl(- |\mathscr{C}_2(j)| \, (t + c_1)\bigr)}\Bigr] \,\,\,\,\,\,\, \textnormal{for} \,\,\,\,\,\,\, t_j \leq t < \textnormal{min}\bigl(t_{j + 1}, t_{\textnormal{T}}^{(\textnormal{N} \rightarrow \textnormal{I})}(j)\bigr) \, .
\end{equation*}
As this solution converges strictly against the value $\mathscr{C}_1(j)/|\mathscr{C}_2(j)|$, which can be smaller than the NID transition value $G_{\textnormal{T}}^{(\textnormal{N} \rightarrow \textnormal{I})}(j)$, it is not suitable for covering this domain. Including a second-order term in the ODE under consideration, that is, working with the equation  
\begin{equation} \label{MODGODE}
\frac{\textnormal{d}G}{\textnormal{d}t} = \mathscr{C}_1 + \mathscr{C}_2 \, G + \mathscr{C} \, G^2 \, ,
\end{equation}
where $\mathscr{C} = \textnormal{const.} \in \mathbb{R}$, leads to a solution in form of a tangent function, which may have several poles in the NID, rendering it useless, too. Adding further contributions to (\ref{MODGODE}) yields analytically non-solvable ODEs. Hence, we have to employ the simpler ODEs 
\begin{align} 
\frac{\textnormal{d}G}{\textnormal{d}t} & = \mathscr{C}_1(j; S_1) & & \hspace{-4.3cm} \textnormal{for} \,\,\,\,\,\,\, S_2 = \emptyset = S_3 \label{NIDODE1NEW} \\ \nonumber \\
\frac{\textnormal{d}G}{\textnormal{d}t} & = \mathscr{C}_1(j; S_1, S_2) & & \hspace{-4.3cm} \textnormal{for} \,\,\,\,\,\,\, S_2 \not= \emptyset, S_3 = \emptyset \label{NIDODE2NEW} 
\end{align}
with the linear solution 
\begin{equation} \label{NIRSOL1NEW}
G(t) = \mathscr{C}_1(j) \, t + c_2(j) \,\,\,\,\,\,\, \textnormal{for} \,\,\,\,\,\,\, t_j \leq t < \textnormal{min}\bigl(t_{j + 1}, t_{\textnormal{T}}^{(\textnormal{N} \rightarrow \textnormal{I})}(j)\bigr)
\end{equation}
and $c_2 = \textnormal{const.} \in \mathbb{R}$. For Eqs.\ (\ref{NIDODE3}) and (\ref{NIDODE4}), we also apply separation of variables and extend the integrand with the multiplicative identity $\mathscr{C}_1/\mathscr{C}_1$ and the neutral element $\mathscr{C}_3 - \mathscr{C}_3$, leading to
\begin{equation} \label{FIE}
\int \frac{G^2 \, \textnormal{d}G}{\mathscr{C}_3 + \mathscr{C}_1 \, G^2} = \frac{1}{\mathscr{C}_1} \, \biggl(G - \mathscr{C}_3 \int \frac{\textnormal{d}G}{\mathscr{C}_3 + \mathscr{C}_1 \,  G^2}\biggr) = \frac{1}{\mathscr{C}_1} \, \Biggl[G - \sqrt{\frac{\mathscr{C}_3}{\mathscr{C}_1}} \, \arctan{\Biggl(\sqrt{\frac{\mathscr{C}_1}{\mathscr{C}_3}} \, G\Biggr)}\Biggr] = t + c_3 \, ,
\end{equation}
where $c_3 = \textnormal{const.} \in \mathbb{R}$. One way of deriving an approximate analytical solution of this transcendental equation is to determine $G$ asymptotically for small and large arguments of the $\textnormal{arctan}$ function (in case both asymptotic ends exist for $G : G_j \leq G < \textnormal{min}(G_{j + 1}, G_{\textnormal{T}}^{(\textnormal{N} \rightarrow \textnormal{I})}(j))$) and extrapolate these solutions up to an intermediate transition point, where they are connected requiring continuity, such that the entire domain of definition is covered. For small arguments $\sqrt{\mathscr{C}_1/\mathscr{C}_3} \, G \ll 1$, we use the third-order approximation $\arctan{(z)}_{| z \ll 1} \approx z - z^3/3$, as the linear term in Eq.\ (\ref{FIE}) and the linear contribution in the approximation of the $\textnormal{arctan}$ function cancel each other out. This leads to the asymptotic solution   
\begin{equation} \label{SAS}
G(t) \simeq (3 \, \mathscr{C}_3 \, t + \mathfrak{c}_0)^{1/3} \, , \,\,\,\,\, \mathfrak{c}_0 = \textnormal{const.} \in \mathbb{R} \, .
\end{equation}
For large arguments $\sqrt{\mathscr{C}_1/\mathscr{C}_3} \, G \gg 1$, the $\textnormal{arctan}$ function can be well approximated by $\pi/2$. We directly obtain an asymptotic solution of the form 
\begin{equation} \label{LAS}
G(t) \simeq \mathscr{C}_1 \, t + \mathfrak{c}'_0 \, , \,\,\,\,\, \mathfrak{c}'_0 = \textnormal{const.} \in \mathbb{R} \, .
\end{equation}
By means of the condition $\sqrt{\mathscr{C}_1/\mathscr{C}_3} \, G = 1$, we derive the NID transition value
\begin{equation} \label{TV1}
G_{\textnormal{T}}^{(\textnormal{N})}(j) = \sqrt{\frac{\mathscr{C}_3(j)}{\mathscr{C}_1(j)}} \, .
\end{equation}
This value specifies the upper bound of the domain of validity of (\ref{SAS}) and the lower bound of the domain of validity of (\ref{LAS}). The corresponding transition time can be directly computed by substituting (\ref{TV1}) into (\ref{FIE}), in which the constant $c_3$ had to be fixed via the initial condition $G(t = t_j) = G_j$ resulting in
\begin{equation*}
c_3 = \frac{1}{\mathscr{C}_1} \, \Biggl[G_j - \sqrt{\frac{\mathscr{C}_3}{\mathscr{C}_1}} \, \arctan{\Biggl(\sqrt{\frac{\mathscr{C}_1}{\mathscr{C}_3}} \, G_j\Biggr)}\Biggr] - t_j \, .
\end{equation*}
This yields 
\begin{equation*} 
t_{\textnormal{T}}^{(\textnormal{N})}(j) = t_j + \frac{1}{\mathscr{C}_1(j)} \, \Biggl(\sqrt{\frac{\mathscr{C}_3(j)}{\mathscr{C}_1(j)}} \, \Biggl[1 - \frac{\pi}{4} + \arctan{\Biggl(\sqrt{\frac{\mathscr{C}_1(j)}{\mathscr{C}_3(j)}} \, G_j\Biggr)}\Biggr] - G_j\Biggr) \, .
\end{equation*}
We point out that in general, one does not have the ordered sequence $G_j < G_{\textnormal{T}}^{(\textnormal{N})}(j) < G_{\textnormal{T}}^{(\textnormal{N} \rightarrow \textnormal{I})}(j)$ because $G_{\textnormal{T}}^{(\textnormal{N})}(j)$ can in principle be larger than or equal to $G_{\textnormal{T}}^{(\textnormal{N} \rightarrow \textnormal{I})}(j)$ as well as smaller than or equal to $G_j$. These cases arise when only one asymptotic end of the arctan function exists. Hence, for $G_j < G_{\textnormal{T}}^{(\textnormal{N})}(j) < G_{\textnormal{T}}^{(\textnormal{N} \rightarrow \textnormal{I})}(j)$, the solution of Eq.\ (\ref{FIE}) is approximately given by  
\begin{equation} \label{S1a}
G(t) = \begin{cases} \bigl(3 \, \mathscr{C}_3(j) \, t + c_4(j)\bigr)^{1/3} & \, \textnormal{for} \,\,\,\,\, t_j \leq t < \textnormal{min}\bigl(t_{j + 1}, t_{\textnormal{T}}^{(\textnormal{N})}(j)\bigr) \\  
\mathscr{C}_1(j) \, t + c_5(j) & \, \textnormal{for} \,\,\,\,\, \textnormal{min}\bigl(t_{j + 1}, t_{\textnormal{T}}^{(\textnormal{N})}(j)\bigr) \leq t < \textnormal{min}\bigl(t_{j + 1}, t_{\textnormal{T}}^{(\textnormal{N} \rightarrow \textnormal{I})}(j)\bigr) \, , \,\,\,\,\, c_4, c_5 = \textnormal{const.} \in \mathbb{R} \, ,\end{cases}
\end{equation}
for $G_j < G_{\textnormal{T}}^{(\textnormal{N} \rightarrow \textnormal{I})}(j) \leq G_{\textnormal{T}}^{(\textnormal{N})}(j)$ by
\begin{equation} \label{S1c}
G(t) = \bigl(3 \, \mathscr{C}_3(j) \, t + c_6(j)\bigr)^{1/3} \,\,\,\,\,\,\, \textnormal{for} \,\,\,\,\, t_j \leq t < \textnormal{min}\bigl(t_{j + 1}, t_{\textnormal{T}}^{(\textnormal{N} \rightarrow \textnormal{I})}(j)\bigr) \, , \,\,\,\,\, c_6 = \textnormal{const.} \in \mathbb{R} \, ,
\end{equation}
and for $G_{\textnormal{T}}^{(\textnormal{N})}(j) \leq G_j$ by
\begin{equation} \label{S1b}
G(t) = \mathscr{C}_1(j) \, t + c_7(j) \,\,\,\,\,\,\, \textnormal{for} \,\,\,\,\, t_j \leq t < \textnormal{min}\bigl(t_{j + 1}, t_{\textnormal{T}}^{(\textnormal{N} \rightarrow \textnormal{I})}(j)\bigr) \, , \,\,\,\,\, c_7 = \textnormal{const.} \in \mathbb{R} \, .
\end{equation}
The integration constants $c_2$ and $c_4, ..., c_7$ are determined via the proper initial and transition conditions in Appendix \ref{app:-Constants:Gluing-Updating}. In order to compute the constants $\mathscr{C}_1(j; S_1)$, $\mathscr{C}_1(j; S_1, S_2)$, $\mathscr{C}_2(j; S_1)$, $\mathscr{C}_2(j; S_1, S_2)$, and $\mathscr{C}_3(j; S_3)$, we have to continually check during the evolution of $G \in \bigl[G_j, \textnormal{min}\bigl(G_{j + 1}, G_{\textnormal{T}}^{(\textnormal{N} \rightarrow \textnormal{I})}(j)\bigr)\bigr)$ whether an injection $i : 1 \leq i \leq j$ belongs to $S_1$, $S_2$, or $S_3$. Therefore, we specify two different kinds of updates. The first kind occurs when a new injection enters the system, while the second kind is due to either NID-IID or IID-FID transitions. We start with the initial update at the time of the $j$th injection verifying
\begin{equation*} 
\begin{split}
& G_j - G_i < x_i/\xi_i^{(\textnormal{N} \rightarrow \textnormal{I})} \hspace{2.28cm} \textnormal{for the $i$th injection being in} \,\, S_1 \\
& x_i/\xi_i^{(\textnormal{N} \rightarrow \textnormal{I})} \leq G_j - G_i < \xi_i^{(\textnormal{I} \rightarrow \textnormal{F})} x_i \,\,\,\,\,\,\,\,\, \textnormal{for the $i$th injection being in} \,\, S_2 \\
& \xi_i^{(\textnormal{I} \rightarrow \textnormal{F})} x_i \leq G_j - G_i \hspace{2.49cm} \textnormal{for the $i$th injection being in} \,\, S_3 \, .
\end{split}
\end{equation*}
Next, since all transition values $G_{\textnormal{T}}^{(\textnormal{N} \rightarrow \textnormal{I})}(i)$ and $G_{\textnormal{T}}^{(\textnormal{I} \rightarrow \textnormal{F})}(i)$ are known, they can be arranged as an ordered $2 j$-tuple representing an increasing sequence. Assuming that a certain number of these values is contained in the time interval under consideration, whenever $G$ reaches one of them, all sets $S_1$, $S_2$, and $S_3$ have to be updated accordingly, i.e., the corresponding injection is moved from either $S_1$ to $S_2$ or from $S_2$ to $S_3$ and the constants involved change. For more details on the updating see Section \ref{C} and Appendix \ref{app:-Constants:Gluing-Updating}.

\subsubsection{Intermediate-injection Domain Solution} \label{IIASS}

\noindent In the IID $G_{\textnormal{T}}^{(\textnormal{N} \rightarrow \textnormal{I})}(j) \leq G < \textnormal{min}\bigl(G_{j + 1}, G_{\textnormal{T}}^{(\textnormal{I} \rightarrow \textnormal{F})}(j)\bigr)$, where at least the $j$th injection is in $S_2$, the previous $j - 1$ injections can reside in their respective NIDs, IIDs, or FIDs. Thus, we have to discuss the four cases  
\begin{align} 
\frac{\textnormal{d}G}{\textnormal{d}t} & = \mathscr{C}_1(j; S_2) & & \hspace{-3.6cm} \textnormal{for} \,\,\,\,\,\,\, S_1 = \emptyset = S_3 \label{IIDODE1} \\ \nonumber \\
\frac{\textnormal{d}G}{\textnormal{d}t} & = \mathscr{C}_1(j; S_1, S_2) & & \hspace{-3.6cm} \textnormal{for} \,\,\,\,\,\,\, S_1 \not= \emptyset, S_3 = \emptyset \label{IIDODE2} \\ \nonumber \\
\frac{\textnormal{d}G}{\textnormal{d}t} & = \mathscr{C}_1(j; S_2) + \frac{\mathscr{C}_3(j; S_3)}{G^2} & & \hspace{-3.6cm} \textnormal{for} \,\,\,\,\,\,\, S_1 = \emptyset, S_3 \not= \emptyset \label{IIDODE3} \\ \nonumber \\
\frac{\textnormal{d}G}{\textnormal{d}t} & = \mathscr{C}_1(j; S_1, S_2) + \frac{\mathscr{C}_3(j; S_3)}{G^2} & & \hspace{-3.6cm} \textnormal{for} \,\,\,\,\,\,\, S_1 \not= \emptyset \not= S_3 \, . \label{IIDODE4}
\end{align}
As these coincide structurally with the ODEs (\ref{NIDODE1NEW}) and (\ref{NIDODE2NEW}) as well as (\ref{NIDODE3}) and (\ref{NIDODE4}) of the NID, we can directly write down their solutions. For Eqs.\ (\ref{IIDODE1}) and (\ref{IIDODE2}), we get
\begin{equation} \label{iidsol1}
G(t) = \mathscr{C}_1(j) \, t + d_1(j) \,\,\,\,\,\,\, \textnormal{for} \,\,\,\,\,\,\, t_{\textnormal{T}}^{(\textnormal{N} \rightarrow \textnormal{I})}(j) \leq t < \textnormal{min}\bigl(t_{j + 1}, t_{\textnormal{T}}^{(\textnormal{I} \rightarrow \textnormal{F})}(j)\bigr) \, , \,\,\,\,\, d_1 = \textnormal{const.} \in \mathbb{R} \, ,
\end{equation}
whereas for Eqs.\ (\ref{IIDODE3}) and (\ref{IIDODE4}), we obtain in case $G_{\textnormal{T}}^{(\textnormal{N} \rightarrow \textnormal{I})}(j) < G_{\textnormal{T}}^{(\textnormal{I})}(j) < G_{\textnormal{T}}^{(\textnormal{I} \rightarrow \textnormal{F})}(j)$
\begin{equation} \label{iidsol2}
G(t) = 
\begin{cases} 
\bigl(3 \, \mathscr{C}_3(j) \, t + d_2(j)\bigr)^{1/3} & \, \textnormal{for} \,\,\,\,\, t_{\textnormal{T}}^{(\textnormal{N} \rightarrow \textnormal{I})}(j)  \leq t < \textnormal{min}\bigl(t_{j + 1}, t_{\textnormal{T}}^{(\textnormal{I})}(j)\bigr) \\  
\mathscr{C}_1(j) \, t + d_3(j) & \, \textnormal{for} \,\,\,\,\, \textnormal{min}\bigl(t_{j + 1}, t_{\textnormal{T}}^{(\textnormal{I})}(j)\bigr) \leq t < \textnormal{min}\bigl(t_{j + 1}, t_{\textnormal{T}}^{(\textnormal{I} \rightarrow \textnormal{F})}(j)\bigr) \, , \,\,\,\,\, d_2, d_3 = \textnormal{const.} \in \mathbb{R} \, ,
\end{cases}
\end{equation}
if $G_{\textnormal{T}}^{(\textnormal{N} \rightarrow \textnormal{I})}(j) < G_{\textnormal{T}}^{(\textnormal{I} \rightarrow \textnormal{F})}(j) \leq G_{\textnormal{T}}^{(\textnormal{I})}(j)$ 
\begin{equation} \label{iidsol3}
G(t) = \bigl(3 \, \mathscr{C}_3(j) \, t + d_4(j)\bigr)^{1/3} \,\,\,\,\,\, \textnormal{for} \,\,\,\,\, t_{\textnormal{T}}^{(\textnormal{N} \rightarrow \textnormal{I})}(j) \leq t < \textnormal{min}\bigl(t_{j + 1}, t_{\textnormal{T}}^{(\textnormal{I} \rightarrow \textnormal{F})}(j)\bigr) \, , \,\,\,\,\, d_5 = \textnormal{const.} \in \mathbb{R} \, ,
\end{equation}
and if $G_{\textnormal{T}}^{(\textnormal{I})}(j) \leq G_{\textnormal{T}}^{(\textnormal{N} \rightarrow \textnormal{I})}(j)$ 
\begin{equation} \label{iidsol4}
G(t) = \mathscr{C}_1(j) \, t + d_5(j) \,\,\,\,\,\, \textnormal{for} \,\,\,\,\, t_{\textnormal{T}}^{(\textnormal{N} \rightarrow \textnormal{I})}(j) \leq t < \textnormal{min}\bigl(t_{j + 1}, t_{\textnormal{T}}^{(\textnormal{I} \rightarrow \textnormal{F})}(j)\bigr) \, , \,\,\,\,\, d_5 = \textnormal{const.} \in \mathbb{R} \, ,
\end{equation}
with the transition time
\begin{equation*} 
t_{\textnormal{T}}^{(\textnormal{I})}(j) = t_{\textnormal{T}}^{(\textnormal{N} \rightarrow \textnormal{I})}(j) + \frac{1}{\mathscr{C}_1(j)} \, \Biggl(\sqrt{\frac{\mathscr{C}_3(j)}{\mathscr{C}_1(j)}} \, \Biggl[1 - \frac{\pi}{4} + \arctan{\Biggl(\sqrt{\frac{\mathscr{C}_1(j)}{\mathscr{C}_3(j)}} \, G_{\textnormal{T}}^{(\textnormal{N} \rightarrow \textnormal{I})}(j)\Biggr)}\Biggr] - G_{\textnormal{T}}^{(\textnormal{N} \rightarrow \textnormal{I})}(j)\Biggr) 
\end{equation*}
associated to the transition value
\begin{equation*} 
G_{\textnormal{T}}^{(\textnormal{I})}(j) = \sqrt{\frac{\mathscr{C}_3(j)}{\mathscr{C}_1(j)}} \, .
\end{equation*}
The derivation of the integration constants $d_1, ..., d_5$ can be found in Appendix \ref{app:-Constants:Gluing-Updating}.

\subsubsection{Far-injection Domain Solution} \label{FIASS}

\noindent In the FID $G_{\textnormal{T}}^{(\textnormal{I} \rightarrow \textnormal{F})}(j) \leq G < G_{j + 1}$, the $j$th injection is, per definition, an element of $S_3$. As before, the previous $j - 1$ injections can be in their respective NIDs, IIDs, or FIDs depending on the initial injection parameters and, therefore, we have to evaluate the four ODEs
\begin{align} 
\frac{\textnormal{d}G}{\textnormal{d}t} & = D_0 + \frac{\mathscr{C}_3(j; S_3)}{G^2} + \frac{\mathscr{C}_4(j; S_3)}{G^3} & & \hspace{-3.5cm} \textnormal{for} \,\,\,\,\,\,\, S_1 = \emptyset = S_2 \label{FIDODE1} \\ \nonumber \\
\frac{\textnormal{d}G}{\textnormal{d}t} & = \mathscr{C}_1(j; S_1) + \frac{\mathscr{C}_3(j; S_3)}{G^2} & & \hspace{-3.5cm} \textnormal{for} \,\,\,\,\,\,\, S_1 \not= \emptyset, S_2 = \emptyset \label{FIDODE2} \\ \nonumber \\
\frac{\textnormal{d}G}{\textnormal{d}t} & = \mathscr{C}_1(j; S_2) + \frac{\mathscr{C}_3(j; S_3)}{G^2} & & \hspace{-3.5cm} \textnormal{for} \,\,\,\,\,\,\, S_1 = \emptyset, S_2 \not= \emptyset \label{FIDODE3} \\ \nonumber \\
\frac{\textnormal{d}G}{\textnormal{d}t} & = \mathscr{C}_1(j; S_1, S_2) + \frac{\mathscr{C}_3(j; S_3)}{G^2} & & \hspace{-3.5cm} \textnormal{for} \,\,\,\,\,\,\, S_1 \not= \emptyset \not= S_2 \, . \label{FIDODE4}
\end{align}
Since Eqs.\ (\ref{FIDODE2}), (\ref{FIDODE3}), and (\ref{FIDODE4}) are also structurally identical to the ODEs (\ref{NIDODE3}) and (\ref{NIDODE4}), we can once again directly write down their solutions, yielding for $G_{\textnormal{T}}^{(\textnormal{I} \rightarrow \textnormal{F})}(j) < G_{\textnormal{T}, 1}^{(\textnormal{F})}(j) < G_{j + 1}$
\begin{equation} \label{fidsol2}
G(t) = 
\begin{cases} 
\bigl(3 \, \mathscr{C}_3(j) \, t + e_1(j)\bigr)^{1/3} & \, \textnormal{for} \,\,\,\,\, t_{\textnormal{T}}^{(\textnormal{I} \rightarrow \textnormal{F})}(j) \leq t < t_{\textnormal{T}, 1}^{(\textnormal{F})}(j) \\  
\mathscr{C}_1(j) \, t + e_2(j) & \, \textnormal{for} \,\,\,\,\, t_{\textnormal{T}, 1}^{(\textnormal{F})}(j) \leq t < t_{j + 1} \, , \,\,\,\,\,\,\,\,\,\,\,\,\,\,\,\,\, e_1, e_2 = \textnormal{const.} \in \mathbb{R} \, ,
\end{cases}
\end{equation}
for $G_{\textnormal{T}}^{(\textnormal{I} \rightarrow \textnormal{F})}(j) < G_{j + 1} \leq G_{\textnormal{T}, 1}^{(\textnormal{F})}(j)$ 
\begin{equation} \label{fidsol3}
G(t) = \bigl(3 \, \mathscr{C}_3(j) \, t + e_3(j)\bigr)^{1/3} \,\,\,\,\,\, \textnormal{for} \,\,\,\,\, t_{\textnormal{T}}^{(\textnormal{I} \rightarrow \textnormal{F})}(j) \leq t < t_{j + 1} \, , \,\,\,\,\, e_3 = \textnormal{const.} \in \mathbb{R} \, ,
\end{equation}
and for $G_{\textnormal{T}, 1}^{(\textnormal{F})}(j) \leq G_{\textnormal{T}}^{(\textnormal{I} \rightarrow \textnormal{F})}(j) < G_{j + 1}$ 
\begin{equation} \label{fidsol4}
G(t) = \mathscr{C}_1(j) \, t + e_4(j) \,\,\,\,\,\, \textnormal{for} \,\,\,\,\, t_{\textnormal{T}}^{(\textnormal{I} \rightarrow \textnormal{F})}(j) \leq t < t_{j + 1} \, , \,\,\,\,\, e_4 = \textnormal{const.} \in \mathbb{R} \, ,
\end{equation}
where 
\begin{equation*} 
t_{\textnormal{T}, 1}^{(\textnormal{F})}(j) = t_{\textnormal{T}}^{(\textnormal{I} \rightarrow \textnormal{F})}(j) + \frac{1}{\mathscr{C}_1(j)} \, \Biggl(\sqrt{\frac{\mathscr{C}_3(j)}{\mathscr{C}_1(j)}} \, \Biggl[1 - \frac{\pi}{4} + \arctan{\Biggl(\sqrt{\frac{\mathscr{C}_1(j)}{\mathscr{C}_3(j)}} \, G_{\textnormal{T}}^{(\textnormal{I} \rightarrow \textnormal{F})}(j)\Biggr)}\Biggr] - G_{\textnormal{T}}^{(\textnormal{I} \rightarrow \textnormal{F})}(j)\Biggr) 
\end{equation*}
is the transition time corresponding to the transition value
\begin{equation*} 
G_{\textnormal{T}, 1}^{(\textnormal{F})}(j) = \sqrt{\frac{\mathscr{C}_3(j)}{\mathscr{C}_1(j)}} \, .
\end{equation*}
An approximate analytical solution of Eq.\ (\ref{FIDODE1}) can be obtained in a similar way as for the NID ODEs (\ref{NIDODE3}) and (\ref{NIDODE4}) by first applying separation of variables, resulting in an integral equation of the form
\begin{equation} \label{INT}
\int \frac{G^3 \, \textnormal{d}G}{D_0 \, G^3 + \mathscr{C}_3 \, G + \mathscr{C}_4} = t + e_5 \, , \,\,\,\,\, e_5 = \textnormal{const.} \in \mathbb{R} \, .
\end{equation}
This integral equation is shown to be approximately equivalent to a specific transcendental equation for which we subsequently derive asymptotic solutions for $G$ that are continued up to an intermediate transition point (if both asymptotic ends exist, else we only consider a continuation of the solution of the present asymptotic end over the entire domain), where they are glued together continuously. In more detail, since $\mathscr{C}_4/(\mathscr{C}_3 \, G) \ll 1$, we employ a first-order geometric series approximation in the integrand of (\ref{INT})
\begin{equation} \label{CALC}
\begin{split}
\int \frac{G^3 \, \textnormal{d}G}{D_0 \, G^3 + \mathscr{C}_3 \, G + \mathscr{C}_4} & = \frac{G}{D_0} - \frac{1}{D_0} \, \int \frac{\mathscr{C}_3 \, G + \mathscr{C}_4}{D_0 \, G^3 + \mathscr{C}_3 \, G + \mathscr{C}_4} \, \textnormal{d}G = \frac{G}{D_0} - \frac{1}{D_0} \, \int \frac{\textnormal{d}G}{1 + \displaystyle \frac{D_0 \, G^3}{\mathscr{C}_3 \, G + \mathscr{C}_4}} \\ \\
& = \frac{G}{D_0} - \frac{1}{D_0} \, \int \frac{\textnormal{d}G}{1 + \displaystyle \frac{D_0 \, G^2}{\mathscr{C}_3} \biggl(1 + \frac{\mathscr{C}_4}{\mathscr{C}_3 \, G}\biggr)^{- 1}} = \frac{G}{D_0} - \frac{1}{D_0} \, \int \frac{\textnormal{d}G}{1 + \displaystyle \frac{D_0 \, G^2}{\mathscr{C}_3}\sum_{n = 0}^{\infty} \, \biggl(- \frac{\mathscr{C}_4}{\mathscr{C}_3 \, G}\biggr)^n} \\ \\
& \approx \frac{G}{D_0} - \frac{1}{D_0} \,\int \frac{\textnormal{d}G}{1 + \displaystyle \frac{D_0}{\mathscr{C}_3} \, \biggl(G - \frac{\mathscr{C}_4}{2 \, \mathscr{C}_3}\biggr)^2 - \frac{D_0 \, \mathscr{C}_4^2}{4 \, \mathscr{C}_3^3}} \approx \frac{G}{D_0} - \frac{1}{D_0} \, \int \frac{\textnormal{d}G}{\displaystyle 1 + \frac{D_0}{\mathscr{C}_3} \, \biggl(G - \frac{\mathscr{C}_4}{2 \, \mathscr{C}_3}\biggr)^2} \, .
\end{split}
\end{equation}
Note that in the first step, we included the multiplicative identity $D_0/D_0$ and the neutral element $(\mathscr{C}_3 \, G + \mathscr{C}_4) - (\mathscr{C}_3 \, G + \mathscr{C}_4)$ in the numerator of the integrand, giving the splitting into two terms. Then, by means of simple algebraic manipulations, we expressed the integrand in a form suitable for the substitution of the geometric series. For reasons of computational simplicity, we dropped the small second-order contribution $- D_0 \, \mathscr{C}_4^2/(4 \, \mathscr{C}_3^3)$ in the last step. The final integral in (\ref{CALC}) is solved by an arctan function. Thus, introducing the parameter $\beta = \beta(j) := D_0^{1/2} \, \mathscr{C}_4(j)/\bigl(2 \, \mathscr{C}_3^{3/2}(j)\bigr)$, Eq.\ (\ref{INT}) results in
\begin{equation} \label{TE2}
\frac{G}{D_0} - \frac{\mathscr{C}_3^{1/2}}{D_0^{3/2}} \, \arctan{\biggl(\beta \, \biggl[\frac{2 \, \mathscr{C}_3 \, G}{\mathscr{C}_4} - 1\biggr]\biggr)} = t + e_5 \, .
\end{equation}
In order to find an approximate analytical solution of this transcendental equation, we again determine $G$ asymptotically for both small and large arguments of the arctan function. Therefore, using the third-order approximation of the arctan function for small arguments $\beta \, (2 \, \mathscr{C}_3 \, G/\mathscr{C}_4 - 1) \ll 1$, the associated asymptotic solution becomes 
\begin{equation*} 
G(t) \simeq (3 \, \mathscr{C}_3 \, t + \mathfrak{e}_0)^{1/3} + \frac{\mathscr{C}_4}{2 \, \mathscr{C}_3} \, , \,\,\,\,\, \mathfrak{e}_0 = \textnormal{const.} \in \mathbb{R} \, .
\end{equation*}
Further, because $\arctan{(x)}_{| x \gg 1} \approx \pi/2$, the asymptotic solution for large arguments $\beta \, (2 \, \mathscr{C}_3 \, G/\mathscr{C}_4 - 1) \gg 1$ reads
\begin{equation*} 
G(t) \simeq D_0 \, t + \mathfrak{e}'_0 \, , \,\,\,\,\, \mathfrak{e}'_0 = \textnormal{const.} \in \mathbb{R} \, .
\end{equation*}
Extending the domains of these solutions up to -- and connecting them continuously at -- the transition point
\begin{equation} \label{TP2}
G_{\textnormal{T}, 2}^{(\textnormal{F})}(j) = \sqrt{\frac{\mathscr{C}_3(j)}{D_0}} + \frac{\mathscr{C}_4(j)}{2 \, \mathscr{C}_3(j)} \, ,
\end{equation}
which is derived from the transition condition $\beta \, (2 \, \mathscr{C}_3 \, G/\mathscr{C}_4 - 1) = 1$ and corresponds to the transition time 
\begin{equation}\label{tTf}
t_{\textnormal{T}, 2}^{(\textnormal{F})}(j) = t_{\textnormal{T}}^{(\textnormal{I} \rightarrow \textnormal{F})}(j) + \frac{1}{D_0} \, \Biggl(\sqrt{\frac{\mathscr{C}_3(j)}{D_0}} \, \Biggl[1 - \frac{\pi}{4} + \arctan{\biggl(\beta(j) \, \biggl[\frac{2 \, \mathscr{C}_3(j) \, G_{\textnormal{T}}^{(\textnormal{I} \rightarrow \textnormal{F})}(j)}{\mathscr{C}_4(j)} - 1\biggr]\biggr)}\Biggr] - G_{\textnormal{T}}^{(\textnormal{I} \rightarrow \textnormal{F})}(j) + \frac{\mathscr{C}_4(j)}{2 \, \mathscr{C}_3(j)}\Biggr) \, ,
\end{equation}
we obtain for $G_{\textnormal{T}}^{(\textnormal{I} \rightarrow \textnormal{F})}(j) < G_{\textnormal{T}, 2}^{(\textnormal{F})}(j) < G_{j + 1}$ the piecewise-defined solution 
\begin{equation} \label{S2a}
G(t) = \begin{cases} \displaystyle \bigl(3 \, \mathscr{C}_3(j) \, t + e_6(j)\bigr)^{1/3} + \frac{\mathscr{C}_4(j)}{2 \, \mathscr{C}_3(j)} & \, \textnormal{for} \,\,\,\,\, t_{\textnormal{T}}^{(\textnormal{I} \rightarrow \textnormal{F})}(j) \leq t < t_{\textnormal{T}, 2}^{(\textnormal{F})}(j) \\  
D_0 \, t + e_7(j) & \, \textnormal{for} \,\,\,\,\, t_{\textnormal{T}, 2}^{(\textnormal{F})}(j) \leq t < t_{j + 1} \, , \,\,\,\,\,\,\,\,\,\,\,\,\,\,\,\,\, e_6, e_7 = \textnormal{const.} \in \mathbb{R} \, . \end{cases}
\end{equation}
In case only one asymptotic end exists, that is, for $G_{\textnormal{T}}^{(\textnormal{I} \rightarrow \textnormal{F})}(j) < G_{j + 1} \leq G_{\textnormal{T}, 2}^{(\textnormal{F})}(j)$ or $G_{\textnormal{T}, 2}^{(\textnormal{F})}(j) \leq G_{\textnormal{T}}^{(\textnormal{I} \rightarrow \textnormal{F})}(j) < G_{j + 1}$, we get
\begin{equation}\label{AddSol}
G(t) = \bigl(3 \, \mathscr{C}_3(j) \, t + e_8(j)\bigr)^{1/3} + \frac{\mathscr{C}_4(j)}{2 \, \mathscr{C}_3(j)} \,\,\,\,\,\, \textnormal{for} \,\,\,\,\, t_{\textnormal{T}}^{(\textnormal{I} \rightarrow \textnormal{F})}(j) \leq t < t_{j + 1} \, , \,\,\,\,\, e_8 = \textnormal{const.} \in \mathbb{R} \, ,
\end{equation}
or 
\begin{equation} \label{S2b}
G(t) = D_0 \, t + e_9(j) \,\,\,\,\,\, \textnormal{for} \,\,\,\,\, t_{\textnormal{T}}^{(\textnormal{I} \rightarrow \textnormal{F})}(j) \leq t < t_{j + 1} \, , \,\,\,\,\, e_9 = \textnormal{const.} \in \mathbb{R} \, ,
\end{equation}
respectively. We remark that the transition time (\ref{tTf}) was computed by fixing the integration constant 
\begin{equation*}
e_5 = \frac{1}{D_0} \, \Biggl(G_{\textnormal{T}}^{(\textnormal{I} \rightarrow \textnormal{F})} - \sqrt{\frac{\mathscr{C}_3}{D_0}} \, \arctan{\biggl(\beta \, \biggl[\frac{2 \, \mathscr{C}_3 \, G_{\textnormal{T}}^{(\textnormal{I} \rightarrow \textnormal{F})}}{\mathscr{C}_4} - 1\biggr]\biggr)}\Biggr) - t_{\textnormal{T}}^{(\textnormal{I} \rightarrow \textnormal{F})} 
\end{equation*}
in Eq.\ (\ref{TE2}) imposing the initial condition $G(t = t_{\textnormal{T}}^{(\textnormal{I} \rightarrow \textnormal{F})}) = G_{\textnormal{T}}^{(\textnormal{I} \rightarrow \textnormal{F})}$ and substituting the transition value (\ref{TP2}). The determination of the integration constants $e_1, ..., e_9$ is given in Appendix \ref{app:-Constants:Gluing-Updating}.

\subsection{Solution of the Relativistic Kinetic Equation for $\{t \in \mathbb{R}_{\geq 0}\}$} \label{C}

\noindent The complete solution of Eq.\ (\ref{GE}) is derived as follows. Beginning with the single-injection domain (SID), the solution branch $G(t \, | \, 0 \leq t < t_2)$ is given for $0 \leq t < \textnormal{min}\bigl(t_2, t_{\textnormal{T}}^{(\textnormal{N} \rightarrow \textnormal{I})}(1)\bigr)$ by Sol.\ (\ref{NIRSOL1NEW}) (in which $j = 1$ as for the other SID solutions below) with $\mathscr{C}_1 = \mathscr{C}_1(1; S_1 = \{1\})$, for 
$t_{\textnormal{T}}^{(\textnormal{N} \rightarrow \textnormal{I})}(1) \leq t < \textnormal{min}\bigl(t_2, t_{\textnormal{T}}^{(\textnormal{I} \rightarrow \textnormal{F})}(1)\bigr)$ by Sol.\ (\ref{iidsol1}) with $\mathscr{C}_1 = \mathscr{C}_1(1; S_2 = \{1\})$, and for $t_{\textnormal{T}}^{(\textnormal{I} \rightarrow \textnormal{F})}(1) \leq t < t_2$ by
\begin{center}
\begin{varwidth}{6.0in}
\begin{itemize}
\item Sol.\ (\ref{S2a}) \,\,\, for \,\, $t_{\textnormal{T}}^{(\textnormal{I} \rightarrow \textnormal{F})}(1) < t_{\textnormal{T}, 2}^{(\textnormal{F})}(1) < t_2$ ,
\item Sol.\ (\ref{AddSol}) \,\,\, for \,\, $t_{\textnormal{T}}^{(\textnormal{I} \rightarrow \textnormal{F})}(1) < t_2 \leq t_{\textnormal{T}, 2}^{(\textnormal{F})}(1)$ ,
\item Sol.\ (\ref{S2b}) \,\,\, for \,\, $t_{\textnormal{T}, 2}^{(\textnormal{F})}(1) \leq t_{\textnormal{T}}^{(\textnormal{I} \rightarrow \textnormal{F})}(1) < t_2$ 
\end{itemize}
\end{varwidth}
\end{center}
with $\mathscr{C}_1 = D_0$ and $\mathscr{C}_3 = \mathscr{C}_3(1; S_3 = \{1\})$. We point out that in the SID, Eq.\ (\ref{GE}) can also be solved by directly applying separation of variables, resulting in \cite{SBM10}  
\begin{equation*} 
\int \frac{\mathscr{G}^2}{D_0 \, \mathscr{G}^2 + A_0 \, q_1} \, \textnormal{d}\mathscr{G}  = \frac{\mathscr{G}}{D_0} - \frac{1}{D_0} \, \int \frac{\textnormal{d}\mathscr{G}}{1 + \displaystyle \frac{D_0 \, \mathscr{G}^2}{A_0 \, q_1}} = \frac{\mathscr{G}}{D_0} - \frac{(A_0 \, q_1)^{1/2}}{D_0^{3/2}} \, \arctan{\Biggl(\sqrt{\frac{D_0}{A_0 \, q_1}} \, \mathscr{G}\Biggr)} = t + f_1 \, ,
\end{equation*}
where $\mathscr{G} := G + x_1$ and $f_1 = \textnormal{const.} \in \mathbb{R}$. Using once again the method of matched asymptotic expansions with the transition time 
\begin{equation*}
t_{\textnormal{T}}^{(\textnormal{S})} = \frac{1}{D_0} \, \left(\sqrt{\frac{A_0 \, q_1}{D_0}} \, \Biggl[1 - \frac{\pi}{4} + \arctan{\Biggl(\sqrt{\frac{D_0}{A_0 \, q_1}} \, x_1\Biggr)}\Biggr] - x_1\right) 
\end{equation*}
leads for $0 < t_{\textnormal{T}}^{(\textnormal{S})} < t_2$ to the approximate solution 
\begin{equation} \label{sAS1}
G(t) = \begin{cases} (3 \, A_0 \, q_1 \, t + f_2)^{1/3} - x_1 & \, \textnormal{for} \,\,\,\,\, 0 \leq t < t_{\textnormal{T}}^{(\textnormal{S})} \\  
D_0 \, t + f_3 & \, \textnormal{for} \,\,\,\,\, t_{\textnormal{T}}^{(\textnormal{S})} \leq t < t_2 \, , \,\,\,\,\, f_2, f_3 = \textnormal{const.} \in \mathbb{R} \, , \end{cases}
\end{equation}
for $0 < t_2 \leq t_{\textnormal{T}}^{(\textnormal{S})}$ to 
\begin{equation} \label{sAS2}
G(t) = (3 \, A_0 \, q_1 \, t + f_4)^{1/3} - x_1 \,\,\,\,\,\,\, \textnormal{for} \,\,\,\,\, 0 \leq t < t_2 \, , \,\,\,\,\, f_4 = \textnormal{const.} \in \mathbb{R} \, , 
\end{equation}
and otherwise for $t_{\textnormal{T}}^{(\textnormal{S})} \leq 0 < t_2$ to 
\begin{equation} \label{sAS3}
G(t) = D_0 \, t + f_5 \,\,\,\,\,\,\, \textnormal{for} \,\,\,\,\, 0 \leq t < t_2 \, , \,\,\,\,\, f_5 = \textnormal{const.} \in \mathbb{R} \, .
\end{equation}
The integration constants $f_1$, $f_2$, $f_4$, and $f_5$ are fixed by the initial condition $G(t = t_1 = 0) = G_1 = 0$
\begin{equation*}
\begin{split}
& f_1 = \frac{x_1}{D_0} - \frac{(A_0 \, q_1)^{1/2}}{D_0^{3/2}} \, \arctan{\Biggl(\sqrt{\frac{D_0}{A_0 \, q_1}} \, x_1\Biggr)} \\ 
& f_2 = f_4 = x_1^3 \\ 
& f_5 = 0 \, ,
\end{split}
\end{equation*}
whereas the integration constant $f_3$ is determined by the transition condition $G\bigl(t = t_{\textnormal{T}}^{(\textnormal{S})} \, | \, 0 \leq t < t_{\textnormal{T}}^{(\textnormal{S})}\bigr) = G\bigl(t = t_{\textnormal{T}}^{(\textnormal{S})} \, | \, t_{\textnormal{T}}^{(\textnormal{S})} \leq t < t_2\bigr)$ 
\begin{equation*}
f_3 = \bigl(3 \, A_0 \, q_1 \, t_{\textnormal{T}}^{(\textnormal{S})} + x_1^3\bigr)^{1/3} - x_1 - D_0 \, t_{\textnormal{T}}^{(\textnormal{S})} \, .
\end{equation*}
In the following, we use the more exact SID approximation (\ref{sAS1})-(\ref{sAS3}). The double-injection domain (DID) solution branch $G(t \, | \, t_2 \leq t < t_3)$ is given for $t_2 \leq t < \textnormal{min}\bigl(t_3, t_{\textnormal{T}}^{(\textnormal{N} \rightarrow \textnormal{I})}(2)\bigr)$ by Sol.\ (\ref{NIRSOL1NEW}) (with $j = 2$ as for the other DID solutions below) if $S_2 \, \substack{= \\ \not=} \, \emptyset, S_3 = \emptyset$ and by
\begin{center}
\begin{varwidth}{6.0in}
\begin{itemize}
\item Sol.\ (\ref{S1a}) \,\,\, for \,\, $t_2 < t_{\textnormal{T}}^{(\textnormal{N})}(2) < \textnormal{min}\bigl(t_3, t_{\textnormal{T}}^{(\textnormal{N} \rightarrow \textnormal{I})}(2)\bigr)$ ,
\item Sol.\ (\ref{S1c}) \,\,\, for \,\, $t_2 < \textnormal{min}\bigl(t_3, t_{\textnormal{T}}^{(\textnormal{N} \rightarrow \textnormal{I})}(2)\bigr) \leq t_{\textnormal{T}}^{(\textnormal{N})}(2)$ ,
\item Sol.\ (\ref{S1b}) \,\,\, for \,\, $t_{\textnormal{T}}^{(\textnormal{N})}(2) \leq t_2 < \textnormal{min}\bigl(t_3, t_{\textnormal{T}}^{(\textnormal{N} \rightarrow \textnormal{I})}(2)\bigr)$ 
\end{itemize}
\end{varwidth}
\end{center}
if $S_2 = \emptyset, S_3 \not= \emptyset$ (with the specific arguments of the constants $\mathscr{C}_1$ and $\mathscr{C}_2$ associated to the respective cases as for the other DID solutions below). For $t_{\textnormal{T}}^{(\textnormal{N} \rightarrow \textnormal{I})}(2) \leq t < \textnormal{min}\bigl(t_3, t_{\textnormal{T}}^{(\textnormal{I} \rightarrow \textnormal{F})}(2)\bigr)$, we employ Sol.\ (\ref{iidsol1}) if 
$S_1 \, \substack{= \\ \not=} \, \emptyset, S_3 = \emptyset$ and 
\begin{center}
\begin{varwidth}{6.0in}
\begin{itemize}
\item Sol.\ (\ref{iidsol2}) \,\,\, for \,\, $t_{\textnormal{T}}^{(\textnormal{N} \rightarrow \textnormal{I})}(2) < t_{\textnormal{T}}^{(\textnormal{I})}(2) < \textnormal{min}\bigl(t_3, t_{\textnormal{T}}^{(\textnormal{I} \rightarrow \textnormal{F})}(2)\bigr)$ ,
\item Sol.\ (\ref{iidsol3}) \,\,\, for \,\, $t_{\textnormal{T}}^{(\textnormal{N} \rightarrow \textnormal{I})}(2) < \textnormal{min}\bigl(t_3, t_{\textnormal{T}}^{(\textnormal{I} \rightarrow \textnormal{F})}(2)\bigr) \leq t_{\textnormal{T}}^{(\textnormal{I})}(2)$ ,
\item Sol.\ (\ref{iidsol4}) \,\,\, for \,\, $t_{\textnormal{T}}^{(\textnormal{I})}(2) \leq t_{\textnormal{T}}^{(\textnormal{N} \rightarrow \textnormal{I})}(2)$ 
\end{itemize}
\end{varwidth}
\end{center}
if $S_1 =\emptyset, S_3 \not= \emptyset$. Finally, for $t_{\textnormal{T}}^{(\textnormal{I} \rightarrow \textnormal{F})}(2) \leq t < t_3$, we apply
\begin{center}
\begin{varwidth}{6.0in}
\begin{itemize}
\item Sol.(\ref{fidsol2}) \,\,\, for \, $t_{\textnormal{T}}^{(\textnormal{I} \rightarrow \textnormal{F})}(2) < t_{\textnormal{T}, 1}^{(\textnormal{F})}(2) < t_3$ ,
\item Sol.(\ref{fidsol3}) \,\,\, for \, $t_{\textnormal{T}}^{(\textnormal{I} \rightarrow \textnormal{F})}(2) < t_3 \leq t_{\textnormal{T}, 1}^{(\textnormal{F})}(2)$ ,
\item Sol.(\ref{fidsol4}) \,\,\, for \, $t_{\textnormal{T}, 1}^{(\textnormal{F})}(2) \leq t_{\textnormal{T}}^{(\textnormal{I} \rightarrow \textnormal{F})}(2) < t_3$ 
\end{itemize}
\end{varwidth}
\end{center}
if either $S_1 \not= \emptyset, S_2 = \emptyset$ or $S_1 = \emptyset, S_2 \not= \emptyset$ and 
\begin{center}
\begin{varwidth}{6.0in}
\begin{itemize}
\item Sol.(\ref{S2a}) \,\,\, for \, $t_{\textnormal{T}}^{(\textnormal{I} \rightarrow \textnormal{F})}(2) < t_{\textnormal{T}, 2}^{(\textnormal{F})}(2) < t_3$ ,
\item Sol.(\ref{AddSol}) \,\,\, for \, $t_{\textnormal{T}}^{(\textnormal{I} \rightarrow \textnormal{F})}(2) < t_3 \leq t_{\textnormal{T}, 2}^{(\textnormal{F})}(2)$ ,
\item Sol.(\ref{S2b}) \,\,\, for \, $t_{\textnormal{T}, 2}^{(\textnormal{F})}(2) \leq t_{\textnormal{T}}^{(\textnormal{I} \rightarrow \textnormal{F})}(2) < t_3$ 
\end{itemize}
\end{varwidth}
\end{center}
if $S_1 = \emptyset = S_2$. The initial value $G_2$ is fixed by requiring continuity of the SID and DID solution branches at the time of the second injection, yielding
\begin{equation*}
G_2 = \begin{cases} 
G\bigl(t = t_2 \, | \, t_{\textnormal{T}}^{(\textnormal{S})} \leq t < t_2 \, ; \, \textnormal{Sol.}(\ref{sAS1})\bigr) & \, \textnormal{for} \,\,\,\,\, 0 < t_{\textnormal{T}}^{(\textnormal{S})} < t_2 \\ \\
G\bigl(t = t_2 \, | \, 0 \leq t < t_2 \, ; \, \textnormal{Sol.}(\ref{sAS2})\bigr) & \, \textnormal{for} \,\,\,\,\, 0 < t_2 \leq t_{\textnormal{T}}^{(\textnormal{S})} \\ \\ 
G\bigl(t = t_2 \, | \, 0 \leq t < t_2 \, ; \, \textnormal{Sol.}(\ref{sAS3})\bigr) & \, \textnormal{for} \,\,\,\,\, t_{\textnormal{T}}^{(\textnormal{S})} \leq 0 < t_2 \, . 
\end{cases}
\end{equation*}
In addition to the initial DID updating of constants at $t = t_2$, we have to perform NID-IID updates at $t = t_{\textnormal{T}}^{(\textnormal{N} \rightarrow \textnormal{I})}(1)$ and/or $t = t_{\textnormal{T}}^{(\textnormal{N} \rightarrow \textnormal{I})}(2)$ if $\bigl\{t_{\textnormal{T}}^{(\textnormal{N} \rightarrow \textnormal{I})}(1), t_{\textnormal{T}}^{(\textnormal{N} \rightarrow \textnormal{I})}(2)\bigr\} \cap [t_2, t_3) \not= \emptyset$ as well as  IID-FID updates at $t = t_{\textnormal{T}}^{(\textnormal{I} \rightarrow \textnormal{F})}(1)$ and/or $t = t_{\textnormal{T}}^{(\textnormal{I} \rightarrow \textnormal{F})}(2)$ if $\bigl\{t_{\textnormal{T}}^{(\textnormal{I} \rightarrow \textnormal{F})}(1), t_{\textnormal{T}}^{(\textnormal{I} \rightarrow \textnormal{F})}(2)\bigr\} \cap [t_2, t_3) \not= \emptyset$. Assuming, for example, that the transition times $t_{\textnormal{T}}^{(\textnormal{N} \rightarrow \textnormal{I})}(1)$, $t_{\textnormal{T}}^{(\textnormal{I} \rightarrow \textnormal{F})}(1)$, and $t_{\textnormal{T}}^{(\textnormal{N} \rightarrow \textnormal{I})}(2)$ are contained in this interval with the order $t_2 < t_{\textnormal{T}}^{(\textnormal{N} \rightarrow \textnormal{I})}(1) < t_{\textnormal{T}}^{(\textnormal{N} \rightarrow \textnormal{I})}(2) < t_{\textnormal{T}}^{(\textnormal{I} \rightarrow \textnormal{F})}(1) < t_3$, we have to update the initial DID sets $S_1$, $S_2$, and $S_3$ thrice. More precisely, at the time of the second injection $t = t_2$, both the first and the second injection are contained in $S_1$ while $S_2$ and $S_3$ are empty. During the temporal progression toward the upper bound $t_3$, the first injection switches from $S_1$ to $S_2$ at $t = t_{\textnormal{T}}^{(\textnormal{N} \rightarrow \textnormal{I})}(1)$, whereas the second injection continues to be in $S_1$. At $t = t_{\textnormal{T}}^{(\textnormal{N} \rightarrow \textnormal{I})}(2)$, also the second injection switches over to $S_2$, leaving $S_1$ empty. Last, at $t = t_{\textnormal{T}}^{(\textnormal{I} \rightarrow \textnormal{F})}(1)$, the first injection switches from $S_2$ to $S_3$. This amounts to the following updating sequence:
\begin{itemize}
\item $t_2 \leq t < t_{\textnormal{T}}^{(\textnormal{N} \rightarrow \textnormal{I})}(1): \hspace{1.49cm} \mathscr{C}_1(2; S_1 = \{1, 2\}, S_2 = \emptyset) = \displaystyle D_0 + A_0 \, \Biggl(\frac{q_1}{x_1^2} +  \frac{q_2}{x_2^2} \, \biggl[1 + \frac{2 \, G_2}{x_2}\biggr]\Biggr) \, ,$
\item[] \hspace{4.41cm} $\mathscr{C}_2(2; S_1 = \{1, 2\}, S_2 = \emptyset) = \displaystyle - 2 \, A_0 \, \biggl(\frac{q_1}{x_1^3} + \frac{q_2}{x_2^3}\biggr) \, ,$ 
\item[] \hspace{4.41cm} $\mathscr{C}_3(2; S_3 = \emptyset) = \mathscr{C}_4(2; S_3 = \emptyset) = 0 \, ,$
\item $\displaystyle t_{\textnormal{T}}^{(\textnormal{N} \rightarrow \textnormal{I})}(1) \leq t < t_{\textnormal{T}}^{(\textnormal{N} \rightarrow \textnormal{I})}(2): \hspace{0.38cm} \mathscr{C}_1(2; S_1 = \{2\}, S_2 = \{1\}) = D_0 + A_0 \, \Biggl(\frac{q_1}{2 \, x_1^2} + \frac{q_2}{x_2^2} \, \biggl[1 + \frac{2 \, G_2}{x_2}\biggr]\Biggr) \, ,$ 
\item[] \hspace{4.41cm} $\mathscr{C}_2(2; S_1 = \{2\}, S_2 = \{1\}) = \displaystyle - A_0 \, \biggl(\frac{q_1}{4 \, x_1^3} + \frac{2 \, q_2}{x_2^3}\biggr) \, ,$ 
\item[] \hspace{4.41cm} $\mathscr{C}_3(2; S_3 = \emptyset) = \mathscr{C}_4(2; S_3 = \emptyset) = 0 \, ,$
\item $\displaystyle t_{\textnormal{T}}^{(\textnormal{N} \rightarrow \textnormal{I})}(2) \leq t < t_{\textnormal{T}}^{(\textnormal{I} \rightarrow \textnormal{F})}(1): \hspace{0.40cm} \mathscr{C}_1(2; S_1 = \emptyset, S_2 = \{1, 2\}) = D_0 + \frac{A_0}{2} \, \Biggl(\frac{q_1}{x_1^2} + \frac{q_2}{x_2^2} \, \biggl[1 + \frac{G_2}{2 \, x_2}\biggr]\Biggr) \, ,$ 
\item[] \hspace{4.41cm} $\mathscr{C}_2(2; S_1 = \emptyset, S_2 = \{1, 2\}) = \displaystyle - \frac{A_0}{4} \, \biggl(\frac{q_1}{x_1^3} + \frac{q_2}{x_2^3} \biggr) \, ,$
\item[] \hspace{4.41cm} $\mathscr{C}_3(2; S_3 = \emptyset) = \mathscr{C}_4(2; S_3 = \emptyset) = 0 \, ,$
\item $ \displaystyle t_{\textnormal{T}}^{(\textnormal{I} \rightarrow \textnormal{F})}(1) \leq t < t_3: \hspace{1.51cm} \mathscr{C}_1(2; S_1 = \emptyset, S_2 = \{2\}) = D_0 + \frac{A_0 \, q_2}{2 \, x_2^2} \, \biggl(1 + \frac{G_2}{2 \, x_2}\biggr) \, ,$ 
\item[] \hspace{4.41cm} $\mathscr{C}_2(2; S_1 = \emptyset, S_2 = \{2\}) = \displaystyle - \frac{A_0 \, q_2}{4 \, x_2^3} \, ,$
\item[] \hspace{4.41cm} $\mathscr{C}_3(2; S_3 = \{1\}) = A_0 \, q_1 \, , \,\,\,\,\, \textnormal{and} \,\,\,\,\, \mathscr{C}_4(2; S_3 = \{1\}) = - 2 \, A_0 \, q_1 \, x_1  \, .$
\end{itemize}
Repeating this procedure for the remaining $m - 2$ injections results in the formal representation of $G$ for $t : 0 \leq t < \infty$ given by 
\begin{equation}\label{GFULL}
\begin{split}
& G(t \, | \, 0 \leq t < \infty) = H(t) \, H(t_2 - t) \, G_{\textnormal{SID}}(t) + \sum_{i = 2}^{m} \, \Bigl[H(t - t_i) \, H(t_{i + 1} - t) \, H\bigl(t_{\textnormal{T}}^{(\textnormal{N} \rightarrow \textnormal{I})}(i) - t\bigr) \, G_{\textnormal{NID}}(t; i) \\ \\ 
& + H\bigl(t - t_{\textnormal{T}}^{(\textnormal{N} \rightarrow \textnormal{I})}(i)\bigr) \, H(t_{i + 1} - t) \, H\bigl(t_{\textnormal{T}}^{(\textnormal{I} \rightarrow \textnormal{F})}(i) - t\bigr) \, G_{\textnormal{IID}}(t; i) + H\bigl(t - t_{\textnormal{T}}^{(\textnormal{I} \rightarrow \textnormal{F})}(i)\bigr) \, H(t_{i + 1} - t) \, G_{\textnormal{FID}}(t; i)\Bigr] \, .
\end{split}
\end{equation}
The SID contribution $G_{\textnormal{SID}}$, the NID, IID, and FID contributions $G_{\textnormal{NID}}$, $G_{\textnormal{IID}}$ and $G_{\textnormal{FID}}$, and the initial constants $G_i$ are stated explicitly in Appendix E. Note that here the updating of constants is suppressed for readability. It could, however, be written down explicitly similar to the updating of the integration constants presented in Appendix D.

\section{Synchrotron and Synchrotron Self-Compton Intensities} \label{SEC3}

\noindent In this section, we calculate the optically thin synchrotron intensity and the SSC intensity in the Thomson limit. The optically thick component of the synchrotron intensity is not considered because it was shown in \cite{SchlRö} that, for all frequencies and times, it provides only a small contribution to the high-energy SSC component.

\subsection{Synchrotron Intensity} \label{syncint}

\noindent The optically thin synchrotron intensity $I_{\textnormal{syn.}}(\epsilon, t)$ (with $[I_{\textnormal{syn.}}] = \textnormal{eV} \, \textnormal{s}^{- 1} \, \textnormal{cm}^{- 2} \, \textnormal{sr}^{- 1}$ and similar for the SSC intensity) for an isotropically distributed electron number density is given by
\begin{equation}\label{int0}
I_{\textnormal{syn.}}(\epsilon, t) = \frac{\mathcal{R}_0}{4 \, \pi} \, \int_0^{\infty} n(\gamma, t) \, P(\epsilon, \gamma) \, \textnormal{d}\gamma \, , 
\end{equation}
where the function  
\begin{equation}\label{BCSA}
P(\epsilon, \gamma) = P_0 \, \frac{\epsilon}{\gamma^2} \, CS\left(\frac{2 \, \epsilon}{3 \, \epsilon_0 \, \gamma^2}\right)
\end{equation}
is the pitch-angle-averaged spectral synchrotron power of a single electron in a magnetic field of strength $b$, $\epsilon := E_{\textnormal{syn.}}/(m_{\textnormal{e}} \, c^2)$ is the normalized synchrotron photon energy, $P_0 := 8.5 \times 10^{23} \, \textnormal{eV} \, \textnormal{s}^{- 1}$, and $\epsilon_0 := 2.3 \times 10^{- 14} \, b$ \cite{BG}. The $CS$ function is discussed in detail in Appendix \ref{app:CS}. Here, we employ the approximation 
\begin{equation}\label{CSI}
CS(z) \approx \frac{a_0}{z^{2/3} \, \bigl(1 + z^{1/3} \, \exp{(z)}\bigr)} \, , \,\,\,\, z \in \mathbb{R}_{\geq 0} \, ,
\end{equation}
where $a_0 := 1.15$. Substituting the electron number density (\ref{edd}) and the synchrotron power (\ref{BCSA}) with the $CS$ function (\ref{CSI}) into formula (\ref{int0}), we obtain for the optically thin synchrotron intensity
\begin{equation} \label{synint}
I_{\textnormal{syn.}}(\epsilon, t) = I_{0, \textnormal{syn.}} \, \epsilon^{1/3} \, \sum_{i = 1}^m  \, \frac{q_i \, H\left(t - t_i\right) \, \mathcal{Y}_i^{2/3}(t)}{1 + \displaystyle \biggl(\frac{2 \, \epsilon}{3 \, \epsilon_0}\biggr)^{1/3} \, \mathcal{Y}_i^{2/3}(t) \, \exp{\displaystyle \left(\frac{2 \, \epsilon}{3 \, \epsilon_0} \, \mathcal{Y}_i^2(t)\right)}} 
\end{equation}
with $I_{0, \textnormal{syn.}} := 3^{2/3} \, \mathcal{R}_0 \, P_0 \, a_0 \, \epsilon_0^{2/3}/(2^{8/3} \, \pi)$ and the abbreviation $\mathcal{Y}_i(t) := G(t) - G_i + x_i$. For comparisons with observational data or for generic case studies, that is, for fitting or plotting lightcurves, we have to compute the energy-integrated synchrotron intensity 
\begin{equation*}
\bar{I}_{\textnormal{syn.}}(t; \epsilon_{\textnormal{min.}}, \epsilon_{\textnormal{max.}}) = \int^{\epsilon_{\textnormal{max.}}}_{\epsilon_{\textnormal{min.}}} I_{\textnormal{syn.}}(\epsilon, t) \, \textnormal{d}\epsilon 
\end{equation*}
with a lower integration limit $\epsilon_{\textnormal{min.}}$ corresponding to the energy of the first data point in the fluence SED and an upper integration limit $\epsilon_{\textnormal{max.}}$ defined by the last. Using (\ref{synint}), this quantity yields 
\begin{equation*}
\bar{I}_{\textnormal{syn.}}(t; \epsilon_{\textnormal{min.}}, \epsilon_{\textnormal{max.}}) = \left(\frac{3 \, \epsilon_0}{2}\right)^{4/3} \, I_{0, \textnormal{syn.}} \, \sum_{i = 1}^m  \, \frac{q_i \, H\left(t - t_i\right)}{\mathcal{Y}_i^2(t)} \, \int^{\tau_{i, \textnormal{max.}}}_{\tau_{i, \textnormal{min.}}} \, \frac{\widetilde{\tau}^{1/3}}{1 + \widetilde{\tau}^{1/3} \, \exp{(\widetilde{\tau})}} \, \textnormal{d}\widetilde{\tau} \, ,
\end{equation*}
where $\tau_{i, \textnormal{min.}} := 2 \, \epsilon_{\textnormal{min.}} \, \mathcal{Y}_i^2/(3 \, \epsilon_0)$ and $\tau_{i, \textnormal{max.}} := 2 \, \epsilon_{\textnormal{max.}} \, \mathcal{Y}_i^2/(3 \, \epsilon_0)$. In order to solve the integral, we approximate the integrand by $\widetilde{\tau}^{1/3}$ for $\widetilde{\tau} \leq 1$ and $\exp{(- \widetilde{\tau})}$ for $\widetilde{\tau} > 1$. This is justified because the approximated $CS$ function (\ref{CSI}) is only adapted to the asymptotics $\widetilde{\tau} \ll 1$ and $\widetilde{\tau} \gg 1$ of the exact $CS$ function and extrapolated in between. Moreover, since $\tau_{i, \textnormal{min.}}$ and $\tau_{i, \textnormal{max.}}$ depend on the time $t$, we have to consider the three cases where $1 \leq \tau_{i, \textnormal{min.}}$, $\tau_{i, \textnormal{min.}} < 1 \leq \tau_{i, \textnormal{max.}}$, and $\tau_{i, \textnormal{max.}} < 1$. Accordingly, the integral results in
\begin{equation}\label{tauint}
\begin{split}
\int^{\tau_{i, \textnormal{max.}}}_{\tau_{i, \textnormal{min.}}} \, \frac{\widetilde{\tau}^{1/3}}{1 + \widetilde{\tau}^{1/3} \, \exp{(\widetilde{\tau})}} \, \textnormal{d}\widetilde{\tau} & \approx H(\tau_{i, \textnormal{min.}} - 1) \, \bigl[\exp{(- \tau_{i, \textnormal{min.}})} - \exp{(- \tau_{i, \textnormal{max.}})}\bigr] \\ 
& \hspace{0.4cm} + H(\tau_{i, \textnormal{max.}} - 1) \, H(1 - \tau_{i, \textnormal{min.}}) \, \biggl[\frac{3}{4} \, \Bigl(1 - \tau_{i, \textnormal{min.}}^{4/3}\Bigr) - \exp{(- \tau_{i, \textnormal{max.}})} + \exp{(- 1)}\biggr] \\ 
& \hspace{0.4cm} + \frac{3}{4} \, H(1 - \tau_{i, \textnormal{max.}}) \, \Bigl[\tau_{i, \textnormal{max.}}^{4/3} - \tau_{i, \textnormal{min.}}^{4/3}\Bigr] \, .
\end{split}
\end{equation}

\subsection{Synchrotron Self-Compton Intensity} \label{SSCint}

\noindent In the computation of the SSC intensity 
\begin{equation}\label{otssci}
I_{\textnormal{SSC}}(\epsilon_{\textnormal{s}}, t) = \frac{\mathcal{R}_0}{4 \, \pi} \, \int_0^{\infty} n(\gamma, t) \, P_{\textnormal{SSC}}(\epsilon_{\textnormal{s}}, \gamma, t) \, \textnormal{d}\gamma \, ,
\end{equation}
where $P_{\textnormal{SSC}}(\epsilon_{\textnormal{s}}, \gamma, t)$ is the SSC power of a single electron and $\epsilon_{\textnormal{s}} := E_{\textnormal{s}}/(m_{\textnormal{e}} \, c^2)$ is the normalized scattered photon energy, we have to employ the Thomson limit because the SSC radiative losses in (\ref{Cool}) are already restricted to the Thomson regime. In this limit, the SSC power reads \cite{FM, BG} 
\begin{equation*}
P_{\textnormal{SSC}}(\epsilon_{\textnormal{s}}, \gamma, t) = \frac{4}{3} \, \sigma_{\textnormal{T}} \, c \, \gamma^2 \, \mathcal{E}(\epsilon_{\textnormal{s}}, t) 
\end{equation*}
with the Thomson cross section $\sigma_{\textnormal{T}} = 6.65 \times 10^{-25} \, \textnormal{cm}^2$ and the total SSC photon energy density $\mathcal{E}(\epsilon_{\textnormal{s}}, t)$ (having the dimension $[\mathcal{E}] = \textnormal{eV} \, \textnormal{cm}^{- 3}$). For ultrarelativistic electrons with $\gamma \gg 1$ and synchrotron photon energies in the Thomson regime for which $\gamma \, \epsilon \ll 1$, the characteristic energy of the SSC-scattered photons is $\epsilon_{\textnormal{s}} \approx 4 \, \gamma^2 \, \epsilon$ \cite{FM}, corresponding to head-on collisions of the synchrotron photons with the electrons \cite{Jones, BG, Rey}. Thus, it is justified to apply a monochromatic approximation in the total SSC photon energy density in form of a Dirac distribution that spikes at this characteristic energy 
\begin{equation*}
\mathcal{E}(\epsilon_{\textnormal{s}}, t) = \frac{1}{4 \, \pi} \, \int_0^{\infty} \epsilon \, N(\epsilon, t) \, \delta\bigl(\epsilon_{\textnormal{s}} - 4 \,  \gamma^2 \, \epsilon\bigr) \, \textnormal{d}\epsilon \, ,
\end{equation*}
where 
\begin{equation*}
N(\epsilon, t) = \frac{4 \, \pi \, I_{\textnormal{syn.}}(\epsilon, t)}{c \, \epsilon} 
\end{equation*}
is the synchrotron photon number density. We remark that by assuming an isotropic, ultrarelativistic electron distribution, the synchrotron photon number density becomes inevitably isotropically distributed, too. Substituting the latter formulas into (\ref{otssci}) and using Fubini's theorem, we obtain
\begin{equation} \label{SSCIntensity}
I_{\textnormal{SSC}}(\epsilon_{\textnormal{s}}, t) = \frac{\mathcal{R}_0 \, \sigma_{\textnormal{T}} \, \epsilon_{\textnormal{s}}}{12 \, \pi} \, \int_0^{\infty} \frac{I_{\textnormal{syn.}}(\epsilon, t)}{\epsilon} \, \int_0^{1/\epsilon} n(\gamma, t) \, \delta\bigl(\epsilon_{\textnormal{s}} - 4 \, \gamma^2 \, \epsilon\bigr) \, \textnormal{d}\gamma \, \textnormal{d}\epsilon \, ,
\end{equation}
in which the upper $\gamma$-integration limit arises from the restriction to the Thomson regime. With the electron number density (\ref{edd}) and the synchrotron intensity (\ref{synint}), the SSC intensity (\ref{SSCIntensity}) yields, after having performed both integrations,
\begin{equation} \label{SSCTINT}
I_{\textnormal{SSC}}(\epsilon_{\textnormal{s}}, t) = I_{0, \textnormal{SSC}} \, \epsilon_{\textnormal{s}}^{1/3} \, \sum_{i, j = 1}^m \, \frac{q_i \, q_j \, H\left(t - t_i\right) \, H\left(t - t_j\right) \, H\left(1 - \displaystyle \frac{\epsilon_{\textnormal{s}}}{4} \, \mathcal{Y}_j(t)\right) \, \bigl[\mathcal{Y}_i(t) \, \mathcal{Y}_j(t)\bigr]^{2/3}}{1 + \displaystyle \biggl(\frac{\epsilon_{\textnormal{s}}}{6 \, \epsilon_0}\biggr)^{1/3} \, \bigl[\mathcal{Y}_i(t) \, \mathcal{Y}_j(t)\bigr]^{2/3} \, \exp{\left(\frac{\epsilon_{\textnormal{s}}}{6 \, \epsilon_0} \, \bigl[\mathcal{Y}_i(t) \, \mathcal{Y}_j(t)\bigr]^2\right)}} \, ,
\end{equation}
where $I_{0, \textnormal{SSC}} := \mathcal{R}_0 \, \sigma_{\textnormal{T}} \, I_{0, \textnormal{syn.}}/(2^{2/3} \, 12 \, \pi)$. The double sum, with the index $i$ referring to the $i$th synchrotron photon population and the index $j$ to the $j$th electron population, accounts for all combinations of SSC scattering between the various electron and synchrotron photon populations. For the corresponding energy-integrated SSC intensity, we find
\begin{equation*}
\begin{split}
\bar{I}_{\textnormal{SSC}}(t; \epsilon_{\textnormal{s}, \textnormal{min.}}, \epsilon_{\textnormal{s}, \textnormal{max.}}) & = (6 \, \epsilon_0)^{4/3} \, I_{0, \textnormal{SSC}} \, \sum_{i, j = 1}^m  \, \frac{q_i \, q_j \, H\left(t - t_i\right) \, H\left(t - t_j\right)}{\bigl[\mathcal{Y}_i(t) \, \mathcal{Y}_j(t)\bigr]^2} \\ \\
& \hspace{0.4cm} \times \int^{\textnormal{min}(\tau_{i j, \textnormal{max.}}, \, 2 \, \mathcal{Y}_i^2 \, \mathcal{Y}_j/(3 \, \epsilon_0))}_{\tau_{i j, \textnormal{min.}}} \, \frac{\widetilde{\tau}^{1/3}}{1 + \widetilde{\tau}^{1/3} \, \exp{(\widetilde{\tau})}} \, \textnormal{d}\widetilde{\tau} \, ,
\end{split}
\end{equation*}
where $\tau_{i j, \textnormal{min.}} := \epsilon_{\textnormal{s}, \textnormal{min.}} \, (\mathcal{Y}_i \, \mathcal{Y}_j)^2/(6 \, \epsilon_0)$ and $\tau_{i j, \textnormal{max.}} := \epsilon_{\textnormal{s}, \textnormal{max.}} \, (\mathcal{Y}_i \, \mathcal{Y}_j)^2/(6 \, \epsilon_0)$. The integral is given by (\ref{tauint}), however, with adapted integration limits.

\section{Synchrotron and Synchrotron Self-Compton Fluences} \label{SEC4}

\noindent We compute the total fluences associated with the synchrotron intensity (\ref{synint}) and the SSC intensity (\ref{SSCTINT}). For this purpose, we derive a general expression for the total fluence that, on the one hand, employs the function $G$ and, on the other hand, explicitly displays the various approximate cases of the Jacobian determinant of the integration measure. For simplicity, the updating of constants is once more suppressed. With $(\varepsilon, I, F) \in \bigl\{(\epsilon, I_{\textnormal{syn.}}, F_{\textnormal{syn.}}), (\epsilon_{\textnormal{s}}, I_{\textnormal{SSC}}, F_{\textnormal{SSC}})\bigr\}$, the total fluence $F$ (for which $[F] = \textnormal{eV} \, \textnormal{cm}^{- 2} \, \textnormal{sr}^{- 1}$) is given by 
\begin{equation} \label{genfluence}
F(\varepsilon) = \int_0^\infty I(\varepsilon, t) \, \textnormal{d}t = \sum_{i = 1}^{m} \, \int_{G_i}^{G_{i + 1}} I(\varepsilon, G) \, \frac{\textnormal{d}t}{\textnormal{d}G} \, \textnormal{d}G \, .
\end{equation}
The Jacobian determinant yields 
\begin{equation} \label{derfl1}
\frac{\textnormal{d}t}{\textnormal{d}G} = \begin{cases} 
\displaystyle \frac{(G + x_1)^2}{D_0 \, (G + x_1)^2 + A_0 \, q_1} & \,\, \textnormal{for} \,\,\,\,\, 0 \leq G < G_2 \\ \\
\displaystyle \frac{1}{\mathscr{C}_1(i)} & \,\, \textnormal{for the domains of validity of Eqs.} \, (\ref{NIDODE1NEW}), (\ref{NIDODE2NEW}), (\ref{IIDODE1}), \, \textnormal{and} \, (\ref{IIDODE2}) \\ \\ 
\displaystyle \frac{G^2}{\mathscr{C}_1(i) \, G^2 + \mathscr{C}_3(i)} & \,\, \textnormal{for the domains of validity of Eqs.} \, (\ref{NIDODE3}), (\ref{NIDODE4}), (\ref{IIDODE3}), (\ref{IIDODE4}), (\ref{FIDODE2}) \textnormal{-} (\ref{FIDODE4}) \\ \\
\displaystyle \frac{G^3}{D_0 \, G^3 + \mathscr{C}_3(i) \, G + \mathscr{C}_4(i)} & \,\, \textnormal{for} \,\,\,\,\, G_{\textnormal{T}}^{(\textnormal{I} \rightarrow \textnormal{F})}(i) \leq G < G_{i + 1} \,\,\,\,\, \textnormal{and} \,\,\,\,\, S_1 = \emptyset = S_2 \, ,
\end{cases}
\end{equation}
where $i: \, 2 \leq i \leq m$. Substituting (\ref{derfl1}) into (\ref{genfluence}), we obtain the expression
\begin{equation*} 
\begin{split}
F(\varepsilon) & = \int_0^{G_2} \frac{(G + x_1)^2 \, I(\varepsilon, G)}{D_0 \, (G + x_1)^2 + A_0 \, q_1} \, \textnormal{d}G + \sum_{i = 2}^m \, \Biggl[\int_{G_i}^{\textnormal{min}\bigl(G_{i + 1}, G_{\textnormal{T}}^{(\textnormal{N} \rightarrow \textnormal{I})}(i)\bigr)} \mathscr{B}_{\textnormal{NID}}(G; i) \, I(\varepsilon, G) \, \textnormal{d}G \\ \\
& \hspace{0.4cm} + \int^{\textnormal{min}\bigl(G_{i + 1}, G_{\textnormal{T}}^{(\textnormal{I} \rightarrow \textnormal{F})}(i)\bigr)}_{\textnormal{min}\bigl(G_{i + 1}, G_{\textnormal{T}}^{(\textnormal{N} \rightarrow \textnormal{I})}(i)\bigr)} \mathscr{B}_{\textnormal{IID}}(G; i) \, I(\varepsilon, G) \, \textnormal{d}G + \int_{\textnormal{min}\bigl(G_{i + 1}, G_{\textnormal{T}}^{(\textnormal{I} \rightarrow \textnormal{F})}(i)\bigr)}^{G_{i + 1}} \mathscr{B}_{\textnormal{FID}}(G; i) \, I(\varepsilon, G) \, \textnormal{d}G \Biggr] \, ,
\end{split}
\end{equation*}
in which 
\begin{equation*} 
\begin{split}
\mathscr{B}_{\textnormal{NID}}(G; i) & := \frac{\bigl[1 - \chi(S_2)\bigr] \, \bigl[1 - \chi(S_3)\bigr]}{\mathscr{C}_1(i; S_1)} + \frac{\chi(S_2) \, \bigl[1 - \chi(S_3)\bigr]}{\mathscr{C}_1(i; S_1, S_2)} \\ \\
& \hspace{0.4cm} + \frac{\bigl[1 - \chi(S_2)\bigr] \, \chi(S_3) \, G^2}{\mathscr{C}_1(i; S_1) \, G^2 + \mathscr{C}_3(i; S_3)} + \frac{\chi(S_2) \, \chi(S_3) \, G^2}{\mathscr{C}_1(i; S_1, S_2) \, G^2 + \mathscr{C}_3(i; S_3)} \, ,
\end{split} 
\end{equation*}
\begin{equation*}  
\begin{split}
\mathscr{B}_{\textnormal{IID}}(G; i) & := \frac{\bigl[1 - \chi(S_1)\bigr] \, \bigl[1 - \chi(S_3)\bigr]}{\mathscr{C}_1(i; S_2)} + \frac{\chi(S_1) \, \bigl[1 - \chi(S_3)\bigr]}{\mathscr{C}_1(i; S_1, S_2)} \\ \\
& \hspace{0.4cm} + \frac{\bigl[1 - \chi(S_1)\bigr] \, \chi(S_3) \, G^2}{\mathscr{C}_1(i; S_2) \, G^2 + \mathscr{C}_3(i; S_3)} + \frac{\chi(S_1) \, \chi(S_3) \, G^2}{\mathscr{C}_1(i; S_1, S_2) \, G^2 + \mathscr{C}_3(i; S_3)} \, ,
\end{split} 
\end{equation*}
and
\begin{equation*}  
\begin{split}
\mathscr{B}_{\textnormal{FID}}(G; i) & := \frac{\bigl[1 - \chi(S_1)\bigr] \, \bigl[1 - \chi(S_2)\bigr] \, G^3}{D_0 \, G^3 + \mathscr{C}_3(i; S_3) \, G + \mathscr{C}_4(i; S_3)} + \frac{\chi(S_1) \, \bigl[1 - \chi(S_2)\bigr] \, G^2}{\mathscr{C}_1(i; S_1) \, G^2 + \mathscr{C}_3(i; S_3)} \\ \\
& \hspace{0.4cm} + \frac{\bigl[1 - \chi(S_1)\bigr] \, \chi(S_2) \, G^2}{\mathscr{C}_1(i; S_2) \, G^2 + \mathscr{C}_3(i; S_3)} + \frac{\chi(S_1) \, \chi(S_2) \, G^2}{\mathscr{C}_1(i; S_1, S_2) \, G^2 + \mathscr{C}_3(i; S_3)} 
\end{split}
\end{equation*}
with the characteristic function
\begin{equation*}
\chi(S_k) := \begin{cases} 1 & \,\,\, \textnormal{for} \,\,\, S_k \not= \emptyset \\  
0 & \,\,\, \textnormal{for} \,\,\, S_k = \emptyset \, , \,\,\,\,\,\,\, k \in \{1, 2, 3\} \, . 
\end{cases} 
\end{equation*}
In the following, for illustrative purposes, we calculate in detail both the synchrotron and the SSC NID fluence integrals for the cases $S_2 \, \substack{= \\ \not=} \, \emptyset, S_3 = \emptyset$. The remaining integrals can be solved in an analogous manner.

\subsection{Synchrotron Fluence}

\noindent With $I = I_{\textnormal{syn.}}(\epsilon, G)$ according to (\ref{synint}), the synchrotron NID fluence integral for $S_2 \, \substack{= \\ \not=} \, \emptyset, S_3 = \emptyset$ reads 
\begin{equation*}
\begin{split}
\mathscr{I}_{\textnormal{syn.}}^{\textnormal{NID}}(\epsilon; i) & := \frac{1}{\mathscr{C}_1(i)} \, \int_{G_i}^{\textnormal{min}\bigl(G_{i + 1}, G_{\textnormal{T}}^{(\textnormal{N} \rightarrow \textnormal{I})}(i)\bigr)} I_{\textnormal{syn.}}(\epsilon, G)\, \textnormal{d}G \\ \\ 
& \hspace{0.085cm} = \frac{I_{0, \textnormal{syn.}} \, \epsilon^{1/3}}{\mathscr{C}_1(i)} \, \sum_{l = 1}^i \, q_l \, \int_{G_i}^{\textnormal{min}\bigl(G_{i + 1}, G_{\textnormal{T}}^{(\textnormal{N} \rightarrow \textnormal{I})}(i)\bigr)} \frac{\mathcal{Y}_l^{2/3}(G)}{1 + \displaystyle \biggl(\frac{2 \, \epsilon}{3 \, \epsilon_0}\biggr)^{1/3} \, \mathcal{Y}_l^{2/3}(G) \, \exp{\displaystyle \left(\frac{2 \, \epsilon}{3 \, \epsilon_0} \, \mathcal{Y}_l^2(G)\right)}} \, \textnormal{d}G \, .
\end{split}
\end{equation*}
Using $\tau_l := 2 \, \epsilon \, \mathcal{Y}_l^2/(3 \, \epsilon_0)$ gives
\begin{equation}\label{I2ONE}
\mathscr{I}_{\textnormal{syn.}}^{\textnormal{NID}}(\epsilon; i) = \frac{(3 \, \epsilon_0)^{5/6} \, I_{0, \textnormal{syn.}}}{2^{11/6} \, \mathscr{C}_1(i)} \, \epsilon^{- 1/2} \, \sum_{l = 1}^i \, q_l \,  \int_{\tau_l(G_i)}^{\tau_l\bigl(\textnormal{min}\bigl(G_{i + 1}, G_{\textnormal{T}}^{(\textnormal{N} \rightarrow \textnormal{I})}(i)\bigr)\bigr)} \, \frac{ \textnormal{d}\widetilde{\tau}}{\widetilde{\tau}^{1/6} \, \bigl(1 + \widetilde{\tau}^{1/3} \, \exp{(\widetilde{\tau})}\bigr)} \, .
\end{equation}
Similar to the evaluation of the energy-integrated intensities (cf.\ the paragraph directly above formula (\ref{tauint})), we approximate the integrand for $\widetilde{\tau} \leq 1$ by $\widetilde{\tau}^{- 1/6}$ and for $\widetilde{\tau} > 1$ by
$\widetilde{\tau}^{- 1/2} \, \exp{(- \widetilde{\tau})}$. Thus, solving the integral, (\ref{I2ONE}) yields
\begin{equation*}
\begin{split}
& \mathscr{I}_{\textnormal{syn.}}^{\textnormal{NID}}(\epsilon; i) \approx \frac{(3 \, \epsilon_0)^{5/6} \, I_{0, \textnormal{syn.}}}{2^{11/6} \, \mathscr{C}_1(i)} \, \epsilon^{- 1/2} \, \sum_{l = 1}^i \, q_l \, \Biggl[H\bigl(\tau_l(G_i) - 1\bigr) \, \Gamma\biggl(\frac{1}{2}, \tau_l(G_i), \tau_l\bigl(\textnormal{min}\bigl(G_{i + 1}, G_{\textnormal{T}}^{(\textnormal{N} \rightarrow \textnormal{I})}(i)\bigr)\bigr)\biggr) \\ \\
& + H\Bigl(\tau_l\bigl(\textnormal{min}\bigl(G_{i + 1}, G_{\textnormal{T}}^{(\textnormal{N} \rightarrow \textnormal{I})}(i)\bigr)\bigr) - 1\Bigr) \, H\bigl(1- \tau_l(G_i)\bigr) \, \biggl[\Gamma\biggl(\frac{1}{2}, 1, \tau_l\bigl(\textnormal{min}\bigl(G_{i + 1}, G_{\textnormal{T}}^{(\textnormal{N} \rightarrow \textnormal{I})}(i)\bigr)\bigr)\biggr) + \frac{6}{5} \, \bigl(1 - \tau_l^{5/6}(G_i)\bigr)\biggr] \\ \\
& + \frac{6}{5} \, H\Bigl(1 - \tau_l\bigl(\textnormal{min}\bigl(G_{i + 1}, G_{\textnormal{T}}^{(\textnormal{N} \rightarrow \textnormal{I})}(i)\bigr)\bigr)\Bigr) \, \Bigl[\tau_l^{5/6}\bigl(\textnormal{min}\bigl(G_{i + 1}, G_{\textnormal{T}}^{(\textnormal{N} \rightarrow \textnormal{I})}(i)\bigr)\bigr) - \tau_l^{5/6}(G_i)\Bigr] \Biggr]
\end{split}
\end{equation*}
with the generalized incomplete gamma function defined for $a \in \mathbb{C}$ and $\textnormal{Re}{(a)} > 0$ by \cite{AbrSt} 
\begin{equation*} 
\Gamma(a, y, z) := \int_y^z t^{a - 1} \, \exp{(- t)} \, \textnormal{d}t \, .
\end{equation*}
In the special case $a = 1/2$, the generalized incomplete gamma function can be expressed in terms of the error function, namely $\Gamma(1/2, y, z) = \sqrt{\pi} \, \bigl[\textnormal{erf}(z) - \textnormal{erf}(y)\bigr]$.

\subsection{Synchrotron Self-Compton Fluence}

\noindent With $I = I_{\textnormal{SSC}}(\epsilon_{\textnormal{s}}, G)$ given in (\ref{SSCTINT}), the SSC NID fluence integral for $S_2 \, \substack{= \\ \not=} \, \emptyset, S_3 = \emptyset$ becomes
\begin{equation*}  
\begin{split}
\mathscr{I}_{\textnormal{SSC}}^{\textnormal{NID}}(\epsilon_{\textnormal{s}}; i) & := \frac{1}{\mathscr{C}_1(i)} \, \int_{G_i}^{\textnormal{min}\bigl(G_{i + 1}, G_{\textnormal{T}}^{(\textnormal{N} \rightarrow \textnormal{I})}(i)\bigr)} I_{\textnormal{SSC}}(\epsilon_{\textnormal{s}}, G) \, \textnormal{d}G \\ \\
& \hspace{0.085cm} = \frac{I_{0, \textnormal{SSC}} \, \epsilon_{\textnormal{s}}^{1/3}}{\mathscr{C}_1(i)} \, \sum_{k, l = 1}^i \, q_k \, q_l \, H\biggl(\frac{4}{\epsilon_{\textnormal{s}}} - G_i + G_l - x_l\biggr) \\ \\  
& \hspace{0.3cm} \times \int_{G_i}^{\textnormal{min}\bigl(G_{i + 1}, G_{\textnormal{T}}^{(\textnormal{N} \rightarrow \textnormal{I})}(i), 4/\epsilon_{\textnormal{s}} + G_l - x_l\bigr)} \frac{\bigl[\mathcal{Y}_k(G) \, \mathcal{Y}_l(G)\bigr]^{2/3}}{1 + \displaystyle \biggl(\frac{\epsilon_{\textnormal{s}}}{6 \, \epsilon_0}\biggr)^{1/3} \, \bigl[\mathcal{Y}_k(G) \, \mathcal{Y}_l(G)\bigr]^{2/3} \, \exp{\left(\frac{\epsilon_{\textnormal{s}}}{6 \, \epsilon_0} \, \bigl[\mathcal{Y}_k(G) \, \mathcal{Y}_l(G)\bigr]^2\right)}} \, \textnormal{d}G \, .
\end{split}
\end{equation*} 
An approximate analytical solution of the integral may be derived with the same method as in the case of the synchrotron fluence. However, one could also employ the methods used for the computations of the NID-IID and IID-FID transition times, which are explained in more detail in Appendix \ref{NIRFIRTT}, as follows. Applying the mean-value-theorem method, we first define the function
\begin{equation*} 
C^{\infty}\bigl(\mathbb{R}_{\geq 0}^2\bigr) \ni \mathscr{H}_{k l}(\epsilon_{\textnormal{s}}, G) := \frac{\bigl[\mathcal{Y}_k(G) \, \mathcal{Y}_l(G)\bigr]^{2/3}}{1 + \displaystyle \biggl(\frac{\epsilon_{\textnormal{s}}}{6 \, \epsilon_0}\biggr)^{1/3} \, \bigl[\mathcal{Y}_k(G) \, \mathcal{Y}_l(G)\bigr]^{2/3} \, \exp{\left(\frac{\epsilon_{\textnormal{s}}}{6 \, \epsilon_0} \, \bigl[\mathcal{Y}_k(G) \, \mathcal{Y}_l(G)\bigr]^2\right)}} \, .
\end{equation*}
Next, we assert that there exists a mean value $\xi_{i l} \in \bigl[G_i, \textnormal{min}\bigl(G_{i + 1}, G_{\textnormal{T}}^{(\textnormal{N} \rightarrow \textnormal{I})}(i), 4/\epsilon_{\textnormal{s}} + G_l - x_l\bigr)\bigr]$ such that
\begin{equation*} 
\begin{split}
\mathscr{I}_{\textnormal{SSC}}^{\textnormal{NID}}(\epsilon_{\textnormal{s}}; i) = \frac{I_{0, \textnormal{SSC}} \, \epsilon_{\textnormal{s}}^{1/3}}{\mathscr{C}_1(i)} \, \sum_{k, l = 1}^i \, & q_k \, q_l \, H\biggl(\frac{4}{\epsilon_{\textnormal{s}}} - G_i + G_l - x_l\biggr) \, \mathscr{H}_{k l}(\epsilon_{\textnormal{s}}, \xi_{i l}) \\ \\
& \times \Bigl[\textnormal{min}\bigl(G_{i + 1}, G_{\textnormal{T}}^{(\textnormal{N} \rightarrow \textnormal{I})}(i), 4/\epsilon_{\textnormal{s}} + G_l - x_l\bigr) - G_i\Bigr] \, .
\end{split}
\end{equation*}
Lastly, we choose the midpoint of the integration interval $\bigl[G_i + \textnormal{min}\bigl(G_{i + 1}, G_{\textnormal{T}}^{(\textnormal{N} \rightarrow \textnormal{I})}(i), 4/\epsilon_{\textnormal{s}} + G_l - x_l\bigr)\bigr]/2$ as an approximate value for $\xi_{i l}$. The trapezoid-approximation method on the other hand results in the expression
\begin{equation*}
\begin{split}
\mathscr{I}_{\textnormal{SSC}}^{\textnormal{NID}}(\epsilon_{\textnormal{s}}; i) & \approx \frac{I_{0, \textnormal{SSC}} \, \epsilon_{\textnormal{s}}^{1/3}}{\mathscr{C}_1(i)} \, \sum_{k, l = 1}^i \, q_k \, q_l \, H\biggl(\frac{4}{\epsilon_{\textnormal{s}}} - G_i + G_l - x_l\biggr) \, \frac{\textnormal{min}\bigl(G_{i + 1}, G_{\textnormal{T}}^{(\textnormal{N} \rightarrow \textnormal{I})}(i), 4/\epsilon_{\textnormal{s}} + G_l - x_l\bigr) - G_i}{2} \\ \\
& \hspace{2.9cm} \times \Bigl[\mathscr{H}_{k l}(\epsilon_{\textnormal{s}}, G_i) + \mathscr{H}_{k l}\bigl(\epsilon_{\textnormal{s}}, \textnormal{min}\bigl(G_{i + 1}, G_{\textnormal{T}}^{(\textnormal{N} \rightarrow \textnormal{I})}(i), 4/\epsilon_{\textnormal{s}} + G_l - x_l\bigr)\bigr)\Bigr] \, .
\end{split}
\end{equation*}

\section{Lightcurves and Fluence Spectral Energy Distributions} \label{MSTVBL}

\noindent To obtain realistic lightcurves, we have to consider injections of finite duration time and radiative transport inside the emission region. But since, for reasons of computational feasibility, we have only employed Dirac distributions for the time profile of our source function and did not concern ourselves with any details of radiative transport, we have to mimic both by modeling each flare via a sequence of suitably distributed, instantaneous injections. In more detail, we partition the $i$th flare into $n_i$ separate injections labeled by a subscript $p \in \{1, ..., n_i\}$, which are induced equidistantly over the entire emission region given by $2 \,  \mathcal{R}_0$, i.e., the $p$th injection of the $i$th flare occurs at the time $t_i + \kappa_{i p}$ with
\begin{equation*}
\kappa_{i p} := \frac{p - 1}{n_i - 1} \, \frac{2 \, \mathcal{R}_0}{c} \, ,
\end{equation*}
and for each of which we use the quantities derived in Section \ref{SEC3}. Thus, the energy-integrated synchrotron intensity reads
\begin{equation} \label{smearedint}
\overline{\mathcal{I}}_{\textnormal{syn.}}(t; \epsilon_{\textnormal{min.}}, \epsilon_{\textnormal{max.}}) = \int^{\epsilon_{\textnormal{max.}}}_{\epsilon_{\textnormal{min.}}} \mathcal{I}_{\textnormal{syn.}}(\epsilon, t) \, \textnormal{d}\epsilon = \int^{\epsilon_{\textnormal{max.}}}_{\epsilon_{\textnormal{min.}}} \sum_{i = 1}^m \sum_{p = 1}^{n_i} q_i(p) \, H(t - t_i - \kappa_{i p}) \, I_{i p}(\epsilon, t) \, \textnormal{d}\epsilon \, ,
\end{equation}
where $\bigl(q_i(p)\bigr)_{p \in \{1, ..., n_i\}}$ for all $i: \, 1 \leq i \leq m$ is a 
specific distribution satisfying the constraint 
\begin{equation} \label{stconstr}
q_i = \sum_{p = 1}^{n_i} q_i(p) 
\end{equation}
and the function $I_{i p}(\epsilon, t)$ is, according to the original synchrotron intensity (\ref{synint}), defined by
\begin{equation} \label{int2ind}
I_{i p}(\epsilon, t) := \frac{I_{0, \textnormal{syn.}} \, \epsilon^{1/3} \, \mathcal{Y}_{i p}^{2/3}(t)}{1 + \displaystyle \biggl(\frac{2 \, \epsilon}{3 \, \epsilon_0}\biggr)^{1/3} \, \mathcal{Y}_{i p}^{2/3}(t) \, \exp{\displaystyle \left(\frac{2 \, \epsilon}{3 \, \epsilon_0} \, \mathcal{Y}_{i p}^2(t)\right)}} 
\end{equation}
with $\mathcal{Y}_{i p}(t) := G(t) - G(t_i + \kappa_{i p}) + x_i$. Similarly, one may construct a proper formula for the energy-integrated SSC intensity. We refrain from showing a parameter study for lightcurves because an adequate value for $n_i$, which is at least of the order $\mathcal{O}(10^4)$, results in numerical computations that would exceed our available CPU time by far. 
\begin{figure}[p]
\centering
\subfigure[][]{\includegraphics[width=0.48\columnwidth]{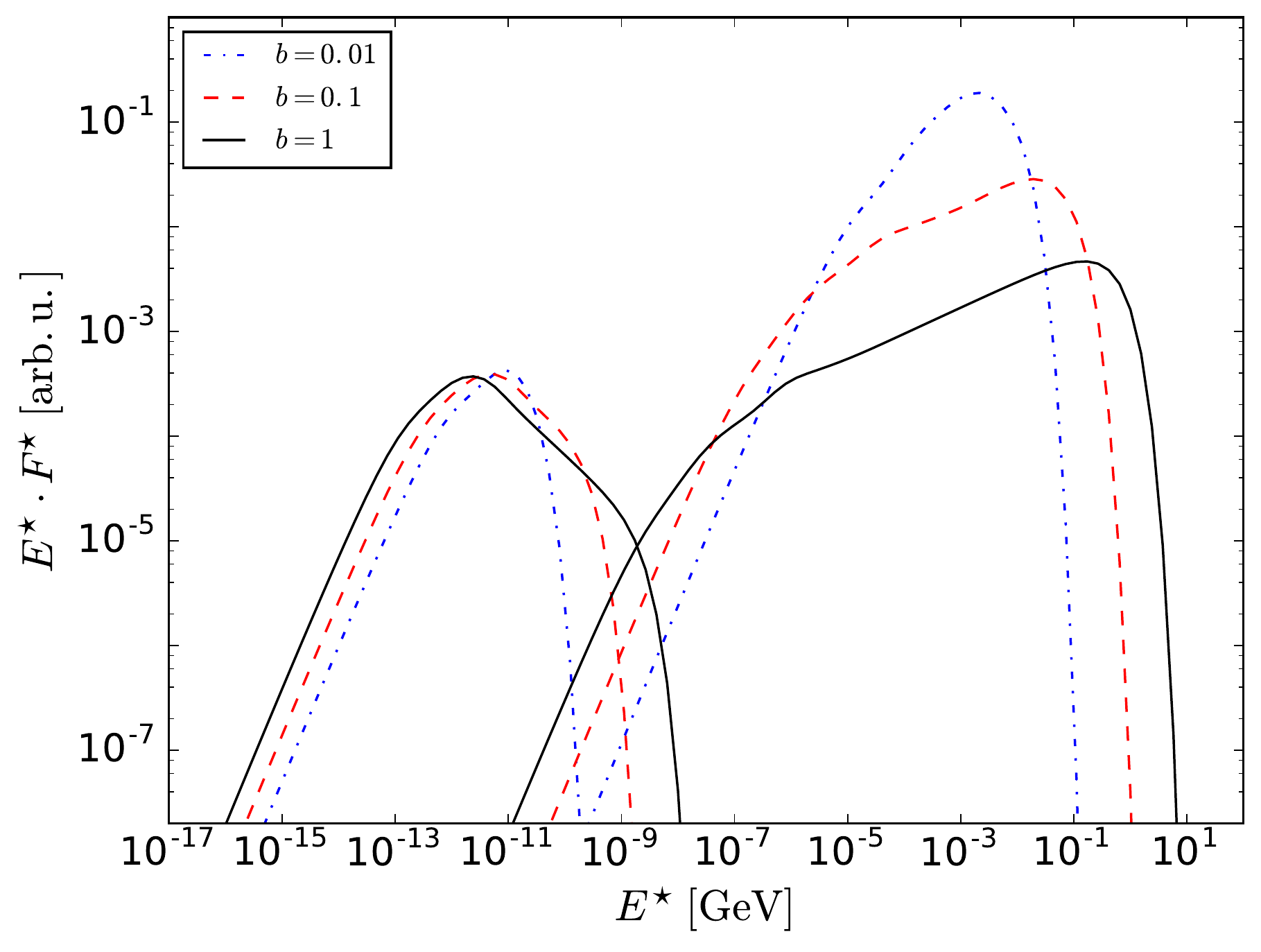}} \label{Var1} \\
\subfigure[][]{\includegraphics[width=0.48\columnwidth]{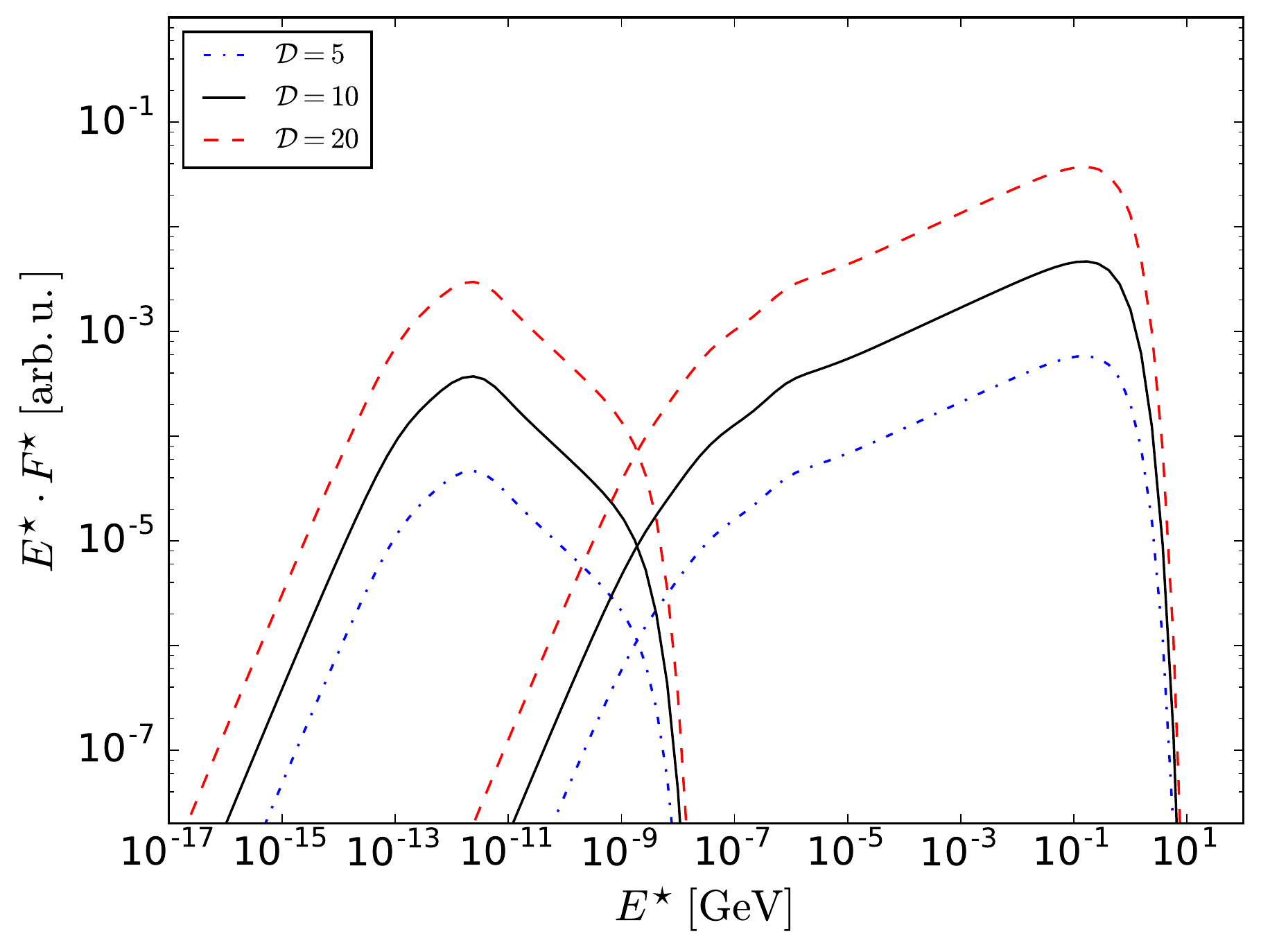}} \label{Var2} \\
\subfigure[][]{\includegraphics[width=0.48\columnwidth]{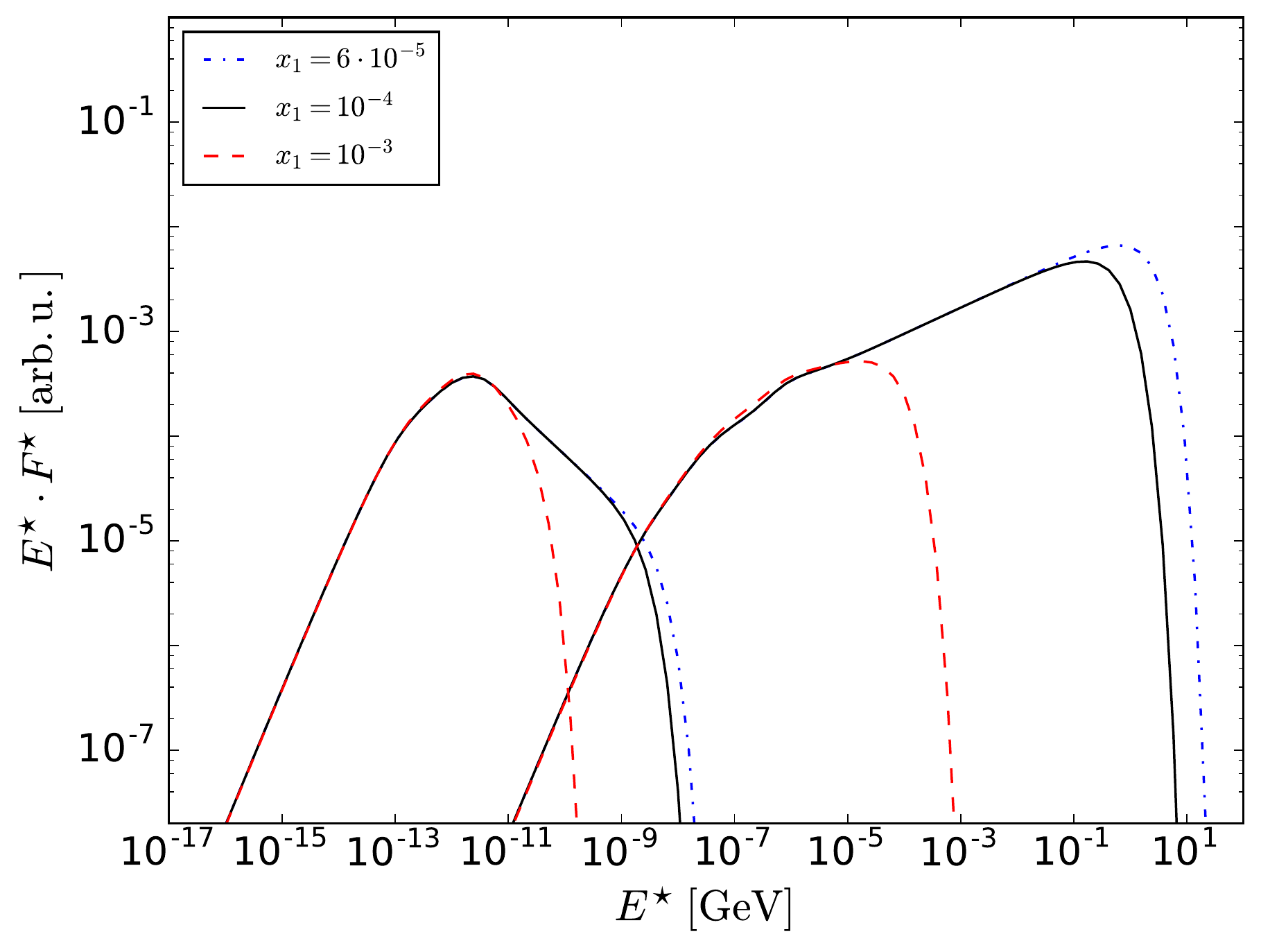}} \label{Var4} 
\caption[...]{Total synchrotron (left curves) and SSC (right curves) fluence SEDs for generic three-injection scenarios with magnetic field strengths, Doppler boost factors, and reciprocal initial electron energies $b \in \{0.01, 0.1, 1\}$, $\mathcal{D} = 10$, and $(x_i)_{i = 1, 2, 3} = (10^{- 4}, 10^{- 4}, 10^{- 4})$ for (a), $b = 1$,  $\mathcal{D} \in \{5, 10, 20\}$, and $(x_i)_{i = 1, 2, 3} = (10^{- 4}, 10^{- 4}, 10^{- 4})$ for (b), as well as $b = 1$, $\mathcal{D} = 10$, $x_1 \in \{0.6 \times 10^{- 4}, 10^{- 4}, 10^{- 3}\}$, and $(x_i)_{i = 2, 3} = (10^{- 4}, 10^{- 4})$ for (c). The injection times are always given by $(t_i)_{i = 1, 2, 3} = (0, 1.5, 3) \, 2 \, \mathcal{R}_0/c$ with the characteristic radius of the plasmoid $\mathcal{R}_0 = 10^{15} \, \textnormal{cm}$, the injection strengths by $(q_i)_{i = 1, 2, 3} = (1.5 \times 10^5 \, \textnormal{cm}^{- 3}, 2 \times 10^5 \, \textnormal{cm}^{- 3}, 5 \times 10^4 \, \textnormal{cm}^{- 3})$, and the cooling rate prefactors $D_0 = 1.3 \times 10^{- 9} \, b^2 \, \textnormal{s}^{- 1}$, $A_0 = 1.2 \times 10^{- 18} \, b^2 \, \textnormal{cm}^3 \, \textnormal{s}^{- 1}$ depend on the specific value of the magnetic field strength. The number of grid points used in the plotting of each curve amounts to $4 \times 10^5$.}
	\label{Figure1}
\end{figure}  
For the associated total fluence SEDs, it is an entirely different matter as we can illustrate their functional shapes by the special case $n_i = 1$ for all $i : 1 \leq i \leq m$, in which each flare is modeled by a single injection (cf.\ Section \ref{SEC4}), given that it yields lower and upper bounds. In the following, we first show by direct calculation that the total fluence of the synchrotron intensity $\mathcal{I}_{\textnormal{syn.}}(\epsilon, t)$ defined in (\ref{smearedint}) is indeed bounded from below and above, up to specific constant factors, by the simple $n_i = 1$ case. To this end, we substitute $\mathcal{I}_{\textnormal{syn.}}(\epsilon, t)$ into the expression for the total fluence and employ the variable $G$
\begin{equation} \label{realtotfl}
\mathcal{F}_{\textnormal{syn}.}(\epsilon) = \int_0^{\infty} \mathcal{I}_{\textnormal{syn}.}(\epsilon, t) \, \textnormal{d}t = \sum_{i = 1}^m \sum_{p = 1}^{n_i} q_i(p) \int_{t_i + \kappa_{i p}}^{\infty} I_{i p}(\epsilon, t) \, \textnormal{d}t = \sum_{i = 1}^m \sum_{p = 1}^{n_i} q_i(p) \int_{G(t_i + \kappa_{i p})}^{\infty} I_{i p}(\epsilon, G) \, \frac{\textnormal{d}t}{\textnormal{d}G} \, \textnormal{d}G \, .
\end{equation}
From Eq.\ (\ref{GEm}), it can be deduced that the Jacobian determinant is positive and bounded from below and above by
\begin{equation} \label{measurebounds}
\biggl(D_0 + A_0 \, \sum_{i = 1}^{j} \, \frac{q_i}{x_i^2}\biggr)^{- 1} \leq \frac{\textnormal{d}t}{\textnormal{d}G} \leq \frac{1}{D_0} \, ,
\end{equation}
and since $I_{i p}$ is non-negative, we can, therefore, bound the fluence (\ref{realtotfl}) from below and above by
\begin{equation*}
\sum_{i = 1}^m \biggl(D_0 + A_0 \, \sum_{k = 1}^{i} \, \frac{q_k}{x_k^2}\biggr)^{- 1} \sum_{p = 1}^{n_i} q_i(p) \int_{G(t_i + \kappa_{i p})}^{\infty} I_{i p}(\epsilon, G) \, \textnormal{d}G \leq \mathcal{F}_{\textnormal{syn}.}(\epsilon) \leq \frac{1}{D_0} \, \sum_{i = 1}^m \sum_{p = 1}^{n_i} q_i(p) \int_{G(t_i + \kappa_{i p})}^{\infty} I_{i p}(\epsilon, G) \, \textnormal{d}G \, .
\end{equation*}
Then, applying the transformation $\widetilde{G}_{i p} := G - G(t_i + \kappa_{i p}) + G_i$ and defining $I_i := I_{i 1}$ (cf.\ formula (\ref{int2ind})), we obtain 
\begin{equation*}
\sum_{i = 1}^m \biggl(D_0 + A_0 \, \sum_{k = 1}^{i} \, \frac{q_k}{x_k^2}\biggr)^{- 1} \sum_{p = 1}^{n_i} q_i(p) \int_{G_i}^{\infty} I_{i}(\epsilon, \widetilde{G}_{i p}) \, \textnormal{d}\widetilde{G}_{i p} \leq \mathcal{F}_{\textnormal{syn}.}(\epsilon) \leq \frac{1}{D_0} \, \sum_{i = 1}^m \sum_{p = 1}^{n_i} q_i(p) \int_{G_i}^{\infty} I_{i}(\epsilon, \widetilde{G}_{i p}) \, \textnormal{d}\widetilde{G}_{i p} \, .
\end{equation*}
Because the integral has the same value for fixed $i$ and arbitrary $p$, we can always drop the subscript $p$ in the argument of the integrand and the integration measure. Hence, by means of the constraint (\ref{stconstr}), we get  
\begin{equation*} 
\sum_{i = 1}^m \biggl(D_0 + A_0 \, \sum_{k = 1}^{i} \, \frac{q_k}{x_k^2}\biggr)^{- 1} \, q_i \int_{G_i}^{\infty} I_i(\epsilon, G) \, \textnormal{d}G = \sum_{i = 1}^m \biggl(D_0 + A_0 \, \sum_{k = 1}^{i} \, \frac{q_k}{x_k^2}\biggr)^{- 1} \biggl(\sum_{p = 1}^{n_i} q_i(p)\biggr) \int_{G_i}^{\infty} I_i(\epsilon, \widetilde{G}_i) \, \textnormal{d}\widetilde{G}_i \leq \mathcal{F}_{\textnormal{syn}.}(\epsilon)
\end{equation*}
for the lower bound and
\begin{equation*} 
\mathcal{F}_{\textnormal{syn}.}(\epsilon) \leq \frac{1}{D_0} \, \sum_{i = 1}^m \biggl(\sum_{p = 1}^{n_i} q_i(p)\biggr) \int_{G_i}^{\infty} I_i(\epsilon, \widetilde{G}_i) \, \textnormal{d}\widetilde{G}_i = \frac{1}{D_0} \, \sum_{i = 1}^m q_i \int_{G_i}^{\infty} I_i(\epsilon, G) \, \textnormal{d}G 
\end{equation*}
for the upper bound. Transforming the integration measure back to the time variable $t$ and using the inverse of (\ref{measurebounds}) leads to
\begin{equation*}
D_0 \, \sum_{i = 1}^m \biggl(D_0 + A_0 \, \sum_{k = 1}^{i} \, \frac{q_k}{x_k^2}\biggr)^{- 1} \, q_i \int_{t_i}^{\infty} I_i(\epsilon, t) \, \textnormal{d}t \leq \mathcal{F}_{\textnormal{syn}.}(\epsilon) \leq \frac{1}{D_0} \, \sum_{i = 1}^m \biggl(D_0 + A_0 \, \sum_{k = 1}^{i} \frac{q_k}{x_k^2}\biggr) \, q_i \int_{t_i}^{\infty} I_i(\epsilon, t) \, \textnormal{d}t \, .
\end{equation*}
Since the second sum on both the left-hand side and the right-hand side can be bounded from above by $\sum_{k = 1}^{m} q_k/x_k^2$, we find
\begin{equation} \label{singmultinjflrel}
\biggl(1 + \frac{A_0}{D_0} \, \sum_{k = 1}^{m} \frac{q_k}{x_k^2}\biggr)^{- 1} \, F_{\textnormal{syn}.}(\epsilon) \leq \mathcal{F}_{\textnormal{syn}.}(\epsilon) \leq \biggl(1 + \frac{A_0}{D_0} \, \sum_{k = 1}^{m} \frac{q_k}{x_k^2}\biggr) \, F_{\textnormal{syn}.}(\epsilon) \, ,
\end{equation}
where $F_{\textnormal{syn}.}(\epsilon) = \sum_{i = 1}^m \int_{t_i}^{\infty} q_i \, I_i(\epsilon, t) \, \textnormal{d}t$, proving the claim. 

In Figure \ref{Figure1} (a)-(c), we show a parameter study for the total synchrotron and SSC fluence SEDs (given in arbitrary units), always considering three injections of ultrarelativistic electron populations into the emission region, each modeling one flare according to the estimate (\ref{singmultinjflrel}). We separately vary the nondimensional magnetic field strength $b$, the Doppler factor of the plasmoid $\mathcal{D}$, and the first reciprocal normalized initial electron energy $x_1$, leaving the remaining free parameters fixed. Details on the values of these parameters and the specific variations can be found in the figure caption. Note that in all three subfigures, the black, solid SEDs correspond to the same set of parameters, thus, serving as a reference curve in the parameter study. Varying only the magnetic field strength (Figure \ref{Figure1} (a)), toward higher values, we can see an increase of both the maximum synchrotron energy and emissivity as expected. This increase naturally leads to a reduction in the SSC emission, as the leftover SSC energy budget of the electrons becomes lower for larger and higher energetic synchrotron emission. However, the maximum SSC energy increases inasmuch as the maximum synchrotron energy can now reach larger values that add to the SSC scattering process. Further, shifting the Doppler factor toward higher values results in the typical increase in the apparent luminosity (Figure \ref{Figure1} (b)), which is well-known from studies of the physical effects of relativistic beaming, i.e., aberration, the Doppler effect, time dilation, and retardation. In Figure \ref{Figure1} (c), as we vary the reciprocal normalized initial electron energy of the first injection, towards larger values, both the maximum synchrotron energy as well as the maximum SSC energy increase, hence, broadening the SED curves on their right slopes according to the specific nonlinear cooling behavior and the strength of the parameter variation. Note that we could also create a plot in which we vary the strengths $q_i$, $i \in \{1, 2, 3\}$, of the different injections. However, we refrain from showing such a plot because increasing the injection strengths towards larger values obliges us to raise the number of grid points (i.e., the number of time steps, which is currently $4 \times 10^5$ amounting to a computation time of approximately four days) in direct proportion, as the injection strengths and the time variable are coupled by multiplication. This would result in computation times ranging from several months to years. We further remark in passing that in order to detect patterns due to variations in the number of injections or due to their nonlinear coupling, a larger parameter study than the one presented here is necessary. Finally, but most importantly, the resulting functional shapes of the various plots indicate that our model can reproduce the typical fluence SEDs of blazars visible in observational data.

\section{Summary and Outlook} \label{S&O}

\noindent We introduced a fully analytical, time-dependent leptonic one-zone model for the flaring of blazars that employs combined synchrotron and SSC radiative losses of multiple interacting, ultrarelativistic electron populations. Our model assumes several injections of electrons into the emission region as the cause of the flaring, which differs from common blazar models where only a single injection is considered. This is, from a physical point of view, more realistic since blazar jets may extend over distances of the order of tens of kiloparsecs and, thus, it is most likely that there is a pick up of more than just one particle population from interstellar and intergalactic clouds. At the same time, it further assumes both radiative cooling processes to occur simultaneously, as would be the case in any physical scenario. In more detail, applying Laplace transformations and the method of matched asymptotic expansions, we derived an approximate analytical solution of the relativistic kinetic equation of the volume-averaged differential electron number density for several successively and instantaneously injected, mono-energetic, spatially isotropically distributed, interacting electron populations, which are subjected to linear, time-independent synchrotron radiative losses and nonlinear, time-dependent SSC radiative losses in the Thomson limit. Using this solution, we computed the optically thin synchrotron intensity, the SSC intensity in the Thomson limit, as well as the corresponding total fluences. Moreover, we mimicked finite injection durations and radiative transport by modeling flares in terms of sequences of instantaneous injections. Ultimately, we presented a parameter study for the total synchrotron and SSC fluence SEDs for a generic three-injection scenario with variations of the magnetic field strength, the Doppler factor, and the initial electron energies, showing that our model can reproduce the characteristic broad-band SED shapes seen in observational data. We point out that the SSC radiative loss term considered here is strictly valid only in the Thomson regime, limiting the applicability of the model to at most GeV blazars. Nonetheless, it can be generalized to describe TeV blazars by using the full Klein-Nishina cross section in the SSC energy loss rate. This leads to a model for which similar yet technically more involved methods apply. Further, in order to make our simple analytical model more realistic, terms accounting for spatial diffusion and for electron escape could be added to the kinetic equation. Also, more elaborate source functions, e.g., with a power law energy dependence, a time dependence in form of rectangular functions for finite injection durations, and with a proper spatial dependence may be considered. However, judging from the complexity of our more elementary analysis, this is most likely only possible via direct numerical evaluation of the associated kinetic equation. 

\vspace{0.2cm}

\begin{acknowledgments}
\noindent The authors are grateful to Horst Fichtner, Reinhard Schlickeiser, and Michael Zacharias for useful discussions and comments. 
\end{acknowledgments}

\vspace{0.2cm}

\begin{appendix}

\section{Laplace Transformation Method} \label{app:LM}

\noindent We derive the solution $R(x, G)$ of the PDE (\ref{dglr}) by using a composition of two Laplace transformations. First, applying a Laplace transformation with respect to the reciprocal electron energy $x$
\begin{equation*}
\mathcal{L}_w(x) \, \left[\, \cdot \,\right] := \int_0^{\infty} \, \left[\, \cdot \,\right] \, \exp{(- w \, x)} \, \textnormal{d}x
\end{equation*}
to Eq.\ (\ref{dglr}) gives
\begin{equation*}
\int_0^{\infty} \frac{\partial R}{\partial G} \, \exp{(- w \, x)} \, \textnormal{d}x + \int_0^{\infty} \frac{\partial R}{\partial x} \, \exp{(- w \, x)} \, \textnormal{d}x = \sum_{i = 1}^{m} \, q_i \, \delta(G - G_i) \int_0^{\infty} \delta(x - x_i) \, \exp{(- w \, x)} \, \textnormal{d}x \, .
\end{equation*}
Evaluating the second integral on the left-hand side via integration by parts and employing the Laplace transform of $R$ 
\begin{equation}\label{LTR1} 
K(w, G) := \int_0^{\infty} R(x, G) \, \exp{(- w \, x)} \, \textnormal{d}x \, , 
\end{equation}
we obtain
\begin{equation*} 
\frac{\partial K}{\partial G} + R \, \exp{(- w \, x)}_{|x \rightarrow \infty} - R(0, G) + w \, K = \sum_{i = 1}^{m} \, q_i \, \delta(G - G_i) \, \exp{(- w \, x_i)} \, \bigl[H(x - x_i)_{|x \rightarrow \infty} - H(x - x_i)_{|x = 0}\bigr]  
\end{equation*}
and, hence, 
\begin{equation}\label{KEQWBT}
\frac{\partial K}{\partial G} - R(0, G) + w \, K = \sum_{i = 1}^{m} \, q_i \, \delta(G - G_i) \, \exp{(- w \, x_i)} \, . 
\end{equation}
As the normalized initial electron energies are finite and bounded from above by $\gamma_i < 1.9 \times 10^4 \, b^{-1/3}$ for all $i: \, 1 \leq i \leq m$ (due to the restriction to the Thomson regime) and only radiation loss processes are considered, we know that the electron number density has support
\begin{equation*}
\textnormal{supp} \, n(\gamma, t) = \bigl\{(\gamma, t) \in \mathbb{R}^2_{\geq 0} \, | \, \gamma \leq \gamma_{\textnormal{max.}} \,\,\, \textnormal{with} \,\,\, \gamma_{\textnormal{max.}} := \textnormal{max}\{\gamma_i \, | \, 1 \leq i \leq m\}\bigr\} \, .	
\end{equation*}
Thus, we can write $n(\gamma, t) = H(\gamma_{\textnormal{max.}} - \gamma) \, n(\gamma, t)$, yielding for the second term on the left-hand side of Eq.\ (\ref{KEQWBT})
\begin{equation*}
R(0, G) = \lim_{x \searrow 0} \, \left(H(x - x_{\textnormal{max.}}) \, \frac{n(x, G)}{x^2}\right) = 0 \, ,
\end{equation*}
where $x_{\textnormal{max.}} := 1/\gamma_{\textnormal{max.}}$. Accordingly, we find
\begin{equation}\label{EA1} 
\frac{\partial K}{\partial G} + w \, K = \sum_{i = 1}^{m} \, q_i \, \delta(G - G_i) \, \exp{(- w \, x_i)} \, .
\end{equation}
Secondly, applying a Laplace transformation with respect to the function $G$
\begin{equation*}
\mathcal{L}_s(G) \, \left[\, \cdot \,\right] := \int_0^{\infty} \, \left[\, \cdot \,\right] \, \exp{(- s \, G)} \, \textnormal{d}G
\end{equation*}
to Eq.\ (\ref{EA1}) results in
\begin{equation} \label{EA2}
\int_0^{\infty} \frac{\partial K}{\partial G} \, \exp{(- s \, G)} \, \textnormal{d}G + w \int_0^{\infty} K \, \exp{(- s \, G)} \, \textnormal{d}G = \sum_{i = 1}^{m} \, q_i \, \exp{(- w \, x_i)} \int_0^{\infty} \delta(G - G_i) \, \exp{(- s \, G)} \, \textnormal{d}G \, . 
\end{equation}
This Laplace transformation implies that $G$ is non-negative. With the Laplace transform of $K$
\begin{equation*}
M(w, s) := \int_0^{\infty} K(w, G) \, \exp{(- s \, G)} \, \textnormal{d}G 
\end{equation*}
and integration by parts as before, Eq.\ (\ref{EA2}) reads
\begin{equation*}
K \, \exp{(- s \, G)}_{|G \rightarrow \infty} - K(w, 0) + (w + s) \, M = \sum_{i = 1}^{m} \, q_i \, \exp{\big(- (w \, x_i + s \, G_i)\big)} \bigl[H(G - G_i)_{|G \rightarrow \infty} - H(G - G_i)_{|G = 0}\bigr] \, . 
\end{equation*}
Since the Heaviside functions are not well-defined at the jump discontinuities at $G = G_i$ for $i : 1 \leq i \leq m$, this equation reduces to
\begin{equation}\label{EA4}
- K(w, 0) + (w + s) \, M = \sum_{i = 1}^{m} \, q_i \, \exp{\big(- (w \, x_i + s \, G_i)\big)} - q_1 \, \exp{(- w \, x_1)} \, H(G)_{|G = 0} \, , 
\end{equation}
containing an unspecified contribution in the last term on the right-hand side, which is handled as follows. At $G = 0$, no energy losses have yet occurred. Therefore, the electron number density is of the form $n(x, 0) = n_1 \, \delta(x - x_1)$, where $n_1 = q_1 \, x_1^2$. Substituting this density into (\ref{LTR1}) by employing the relation $R(x, G) = n(x, G)/x^2$, we get
\begin{equation*}
K(w, 0) = n_1 \int_0^{\infty} \frac{\exp{(- w \, x)}}{x^2} \, \delta(x - x_1) \, \textnormal{d}x= q_1 \, \exp{(- w \, x_1)} \, .
\end{equation*}
Choosing $H(G)_{|G = 0} = 1$, we can use the last term on the right-hand side of Eq.\ (\ref{EA4}) in order to compensate this function, yielding
\begin{equation*}
M(w, s) = \frac{1}{w + s} \, \sum_{i = 1}^{m} \, q_i \, \exp{\big(- (w \, x_i + s \, G_i)\big)} \, .
\end{equation*}
We point out that both $M$ and the sum are positive. Hence, $s > - w$, which allows us to apply the inverse Laplace transformations with respect to the variables $s$ and $w$ to $M$. This gives the solution of Eq.\ (\ref{dglr})
\begin{equation*}
\begin{split}
R(x, G) = \mathcal{L}_w^{-1}(x) \, \mathcal{L}_s^{-1}(G) \, M(w, s) & = \mathcal{L}_w^{-1}(x) \, \sum_{i = 1}^{m} \, q_i \, H(G - G_i) \, \exp{\big(- w \, (G - G_i + x_i)\big)} \\
& = \sum_{i = 1}^{m} \, q_i \, H(G - G_i) \, \delta(x - x_i - G + G_i) \, .
\end{split}
\end{equation*}

\section{Approximation Errors and Optimal Domains} \label{AEaOD}

\noindent The errors of the NID, IID, and FID approximations (\ref{APPROX1}), (\ref{APPROX2}), and (\ref{APPROX4}) yield
\begin{align}
E_u^{\textnormal{NID}}(G) & = \frac{(G - G_u)^2}{x_u^2 \, (G - G_u + x_u)} \, \biggl[\frac{2}{x_u} + \frac{1}{G - G_u + x_u}\biggr] \nonumber \\ \nonumber \\
E_v^{\textnormal{IID}}(G) & = \frac{G - G_v - x_v}{4 \, x_v \, (G - G_v + x_v)} \, \biggl[\frac{G - G_v}{x_v^2} - \frac{2}{G - G_v + x_v}\biggr] \nonumber \\ \nonumber \\
E_w^{\textnormal{FID}}(G) & = \frac{(G_w - x_w)^2}{G^2 \, (G - G_w + x_w)} \, \biggl[\frac{2}{G} + \frac{1}{G - G_w + x_w}\biggr] \, . \nonumber
\end{align}
Evaluating the NID and IID errors at the NID-IID transition point $G = G_{\textnormal{T}}^{(\textnormal{N} \rightarrow \textnormal{I})}$ and the IID and FID errors at the IID-FID transition point $G = G_{\textnormal{T}}^{(\textnormal{I} \rightarrow \textnormal{F})}$, we obtain  
\begin{align}
E_u^{\textnormal{NID}}\bigl(G_{\textnormal{T}}^{(\textnormal{N} \rightarrow \textnormal{I})}(u)\bigr) & = \frac{2 + 3 \, \xi_u^{(\textnormal{N} \rightarrow \textnormal{I})}}{x_u^2 \, \xi_u^{(\textnormal{N} \rightarrow \textnormal{I})} \, (1 + \xi_u^{(\textnormal{N} \rightarrow \textnormal{I})})^2} \nonumber \\ \nonumber \\
E_v^{\textnormal{IID}}\bigl(G_{\textnormal{T}}^{(\textnormal{N} \rightarrow \textnormal{I})}(v)\bigr) & = \frac{\xi_v^{(\textnormal{N} \rightarrow \textnormal{I})} - 1}{4 \, x_v^2 \, (\xi_v^{(\textnormal{N} \rightarrow \textnormal{I})} + 1)} \, \Biggl[\frac{2 \, \xi_v^{(\textnormal{N} \rightarrow \textnormal{I})}}{\xi_v^{(\textnormal{N} \rightarrow \textnormal{I})} + 1} - \frac{1}{\xi_v^{(\textnormal{N} \rightarrow \textnormal{I})}}\Biggr] \nonumber
\end{align}
and
\begin{align}
E_v^{\textnormal{IID}}\bigl(G_{\textnormal{T}}^{(\textnormal{I} \rightarrow \textnormal{F})}(v)\bigr) & = \frac{\xi_v^{(\textnormal{I} \rightarrow \textnormal{F})} - 1}{4 \, x_v^2 \, (\xi_v^{(\textnormal{I} \rightarrow \textnormal{F})} + 1)} \, \Biggl[\xi_v^{(\textnormal{I} \rightarrow \textnormal{F})} - \frac{2}{\xi_v^{(\textnormal{I} \rightarrow \textnormal{F})} + 1}\Biggr] \nonumber \\ \nonumber \\
E_w^{\textnormal{FID}}\bigl(G_{\textnormal{T}}^{(\textnormal{I} \rightarrow \textnormal{F})}(w)\bigr) & =  \frac{(G_w - x_w)^2}{x_w \, (\xi_w^{(\textnormal{I} \rightarrow \textnormal{F})} + 1) \, (G_w + \xi_w^{(\textnormal{I} \rightarrow \textnormal{F})} \, x_w)^2} \, \Biggl[\frac{1}{x_w \, (\xi_w^{(\textnormal{I} \rightarrow \textnormal{F})} + 1)} + \frac{2}{G_w + \xi_w^{(\textnormal{I} \rightarrow \textnormal{F})} \, x_w}\Biggr] \, . \nonumber
\end{align}
We derive optimal values for the interval parameters $\xi_i^{(\textnormal{N} \rightarrow \textnormal{I})}$ and $\xi_i^{(\textnormal{I} \rightarrow \textnormal{F})}$ by requiring that the total error at each transition point becomes minimal. To this end, we first impose the sufficient conditions
\begin{equation} \label{MINCOND}
\begin{split}
\frac{\partial}{\partial \xi_i^{(\textnormal{N} \rightarrow \textnormal{I})}} \Bigl(\big\|E_i^{\textnormal{NID}}\bigl(G_{\textnormal{T}}^{(\textnormal{N} \rightarrow \textnormal{I})}(i)\bigr)\big\| + \big\|E_i^{\textnormal{IID}}\bigl(G_{\textnormal{T}}^{(\textnormal{N} \rightarrow \textnormal{I})}(i)\bigr)\big\|\Bigr) & = 0 \\ \\
\frac{\partial}{\partial \xi_i^{(\textnormal{I} \rightarrow \textnormal{F})}} \Bigl(\big\|E_i^{\textnormal{IID}}\bigl(G_{\textnormal{T}}^{(\textnormal{I} \rightarrow \textnormal{F})}(i)\bigr)\big\| + \big\|E_i^{\textnormal{FID}}\bigl(G_{\textnormal{T}}^{(\textnormal{I} \rightarrow \textnormal{F})}(i)\bigr)\big\|\Bigr) & = 0 \, , 
\end{split}
\end{equation}
which result in the solution $\xi_i^{(\textnormal{N} \rightarrow \textnormal{I})} \approx 4.73$ and an equation for the zeros of a seventh-order polynomial in $\xi_i^{(\textnormal{I} \rightarrow \textnormal{F})}$ given by
\begin{equation} \label{7thop}
(\xi_i^{(\textnormal{I} \rightarrow \textnormal{F})})^3 + 3 \, \xi_i^{(\textnormal{I} \rightarrow \textnormal{F})} \, (\xi_i^{(\textnormal{I} \rightarrow \textnormal{F})} + 1) - 7 - \frac{8 \, (G_i - x_i)^2}{(G_i + \xi_i^{(\textnormal{I} \rightarrow \textnormal{F})} \, x_i)^2} \, \Biggl(1 + \frac{x_i \, (\xi_i^{(\textnormal{I} \rightarrow \textnormal{F})} + 1) \, \bigl(2 \, G_i + x_i \, [3 + 5 \, \xi_i^{(\textnormal{I} \rightarrow \textnormal{F})}]\bigr)}{(G_i + \xi_i^{(\textnormal{I} \rightarrow \textnormal{F})} \, x_i)^2}\Biggr) = 0 \, ,
\end{equation}
respectively. We verify that the total error $\big\|E_i^{\textnormal{NID}}\bigl(G_{\textnormal{T}}^{(\textnormal{N} \rightarrow \textnormal{I})}(i)\bigr)\big\| + \big\|E_i^{\textnormal{IID}}\bigl(G_{\textnormal{T}}^{(\textnormal{N} \rightarrow \textnormal{I})}(i)\bigr)\big\|$ has a minimum at $\xi_i^{(\textnormal{N} \rightarrow \textnormal{I})} \approx 4.73$ by means of the second derivative test 
\begin{equation*}
\frac{\partial^2}{(\partial \xi_i^{(\textnormal{N} \rightarrow \textnormal{I})})^2} \Bigl(\big\|E_i^{\textnormal{NID}}\bigl(G_{\textnormal{T}}^{(\textnormal{N} \rightarrow \textnormal{I})}(i)\bigr)\big\| + \big\|E_i^{\textnormal{IID}}\bigl(G_{\textnormal{T}}^{(\textnormal{N} \rightarrow \textnormal{I})}(i)\bigr)\big\|\Bigr)_{\big| \xi_i^{(\textnormal{N} \rightarrow \textnormal{I})} = 4.73} > 0 \, .
\end{equation*}
From the Abel-Ruffini theorem, we know that there exists no general analytical solution to Eq.\ (\ref{7thop}). Hence, one has to employ a root-finding algorithm to compute numerical approximations. 

One may obtain a more accurate overall approximation of Eq.\ (\ref{GEmform}) by using the more general IID approximation 
\begin{equation*}
\begin{split}
\frac{1}{(G - G_v + x_v)^2} & = \frac{1}{x_v^2} \, \sum_{n = 0}^{\infty} \frac{1}{n!} \, \frac{\textnormal{d}^n}{\textnormal{d}G^n} \Biggl(\biggl[1 + \frac{G - G_v}{x_v}\biggr]^{- 2}\Biggr)_{\big| G - G_v = \, a_v \, x_v} \times (G - G_v - a_v \, x_v)^n \\ \\ 
& \approx \frac{1}{x_v^2 \, (1 + a_v)^2} \, \Biggl[1 - \frac{2}{1 + a_v} \, \biggl(\frac{G - G_v}{x_v} - a_v\biggr)\Biggr] 
\end{split}
\end{equation*}
with errors 
\begin{equation*}
\begin{split}
E_v^{\textnormal{IID}}\bigl(G_{\textnormal{T}}^{(\textnormal{N} \rightarrow \textnormal{I})}(v)\bigr) & = \frac{1}{x_v^2} \Biggl[\frac{(\xi_v^{(\textnormal{N} \rightarrow \textnormal{I})})^2}{(1 + \xi_v^{(\textnormal{N} \rightarrow \textnormal{I})})^2} - \frac{\xi_v^{(\textnormal{N} \rightarrow \textnormal{I})} \, [1 + 3 \, a_v] - 2}{\xi_v^{(\textnormal{N} \rightarrow \textnormal{I})} \, (1 + a_v)^3}\Biggr] \\ \\
E_v^{\textnormal{IID}}\bigl(G_{\textnormal{T}}^{(\textnormal{I} \rightarrow \textnormal{F})}(v)\bigr) & = \frac{1}{x_v^2} \Biggl[\frac{1}{(1 + \xi_v^{(\textnormal{I} \rightarrow \textnormal{F})})^2} - \frac{1 + 3 \, a_v - 2 \, \xi_v^{(\textnormal{I} \rightarrow \textnormal{F})}}{(1 + a_v)^3}\Biggr] \, ,
\end{split}
\end{equation*}
where the unspecified expansion point $a_v$ constitutes another degree of freedom. Note that our original IID approximation (\ref{APPROX2}) corresponds to expansion points which were set to $a_v = 1$ for all $v \in S_2$. Then, one can determine the optimal values for all $a_v$ such that the total errors  
\begin{equation*}
E_i^{\textnormal{tot}} := \big\|E_i^{\textnormal{NID}}\bigl(G_{\textnormal{T}}^{(\textnormal{N} \rightarrow \textnormal{I})}(i)\bigr)\big\| + \big\|E_i^{\textnormal{IID}}\bigl(G_{\textnormal{T}}^{(\textnormal{N} \rightarrow \textnormal{I})}(i)\bigr)\big\| + \big\|E_i^{\textnormal{IID}}\bigl(G_{\textnormal{T}}^{(\textnormal{I} \rightarrow \textnormal{F})}(i)\bigr)\big\| + \big\|E_i^{\textnormal{FID}}\bigl(G_{\textnormal{T}}^{(\textnormal{I} \rightarrow \textnormal{F})}(i)\bigr)\big\| \, , \,\,\,\,\, i \in \{1, ..., j\} \, ,
\end{equation*}
become minimal by making the first and second derivate tests 
\begin{equation*}
\frac{\partial E_i^{\textnormal{tot}}}{\partial a_i} = 0 \,\,\,\,\,\,\,\, \textnormal{and} \,\,\,\,\,\,\,\, \frac{\partial^2 E_i^{\textnormal{tot}}}{\partial a_i^2} > 0
\end{equation*}
under the constraints (\ref{MINCOND}) (and the associated second derivate tests). To guarantee that the minima do not coincide with expansion points at infinity, one has to further impose a constraint that limits the potential values for the expansion points to a suitable finite interval. We finally remark that, due to the complexity of the resulting equations, this procedure is feasible only numerically. Moreover, by numerical trial and error with standard blazar parameter values, we found out that the special case (\ref{APPROX2}) yields an adequate approximation close to the optimal values.

\section{NID-IID and IID-FID Transition Times} \label{NIRFIRTT}

\noindent In the following, we present two different methods for the determination of the transition times $t_{\textnormal{T}}^{(\textnormal{N} \rightarrow \textnormal{I})}(j)$ and $t_{\textnormal{T}}^{(\textnormal{I} \rightarrow \textnormal{F})}(j)$. To this end, it is advantageous to start from the integral equation representation of the ODE (\ref{GEm}) evaluated at the respective transition point $G_{\textnormal{T}}^{(\textnormal{N} \rightarrow \textnormal{I})}(j) = G_j + x_j/\xi_j^{(\textnormal{N} \rightarrow \textnormal{I})}$ or $G_{\textnormal{T}}^{(\textnormal{I} \rightarrow \textnormal{F})}(j) = G_j + \xi_j^{(\textnormal{I} \rightarrow \textnormal{F})} \, x_j$. In the former NID-IID case, we obtain
\begin{equation} \label{TT}
t_{\textnormal{T}}^{(\textnormal{N} \rightarrow \textnormal{I})}(j) = \int_{G_j}^{G_j + x_j/\xi_j^{(\textnormal{N} \rightarrow \textnormal{I})}} J^{-1}(\widetilde{G}) \, \textnormal{d}\widetilde{G} \,\,\,\,\,\,\, \textnormal{with} \,\,\,\,\,\,\, J^{-1}(\widetilde{G}) := \frac{1}{J(\widetilde{G})} = \Biggl(D_0 + A_0 \, \sum_{i = 1}^j \, \frac{q_i}{\bigl(\widetilde{G} - G_i + x_i\bigr)^2}\Biggr)^{- 1} \, .
\end{equation}
The first method applies the first mean value theorem for integration. Thus, as $J^{- 1} \in C^{\infty}(\mathbb{R}_{\geq 0})$, there exists a point $p_j \in \bigl[G_j, G_j + x_j/\xi_j^{(\textnormal{N} \rightarrow \textnormal{I})}\bigr]$ such that the transition time (\ref{TT}) can be written as 
\begin{equation*}
t_{\textnormal{T}}^{(\textnormal{N} \rightarrow \textnormal{I})}(j) = \frac{x_j}{\xi_j^{(\textnormal{N} \rightarrow \textnormal{I})}} \, J^{- 1}(p_j) \, .
\end{equation*}
Because the mean value theorem is merely an existence theorem, we have to approximate the value of $p_j$ by means of an additional input. Since $J^{- 1}$ is strictly increasing, a possible choice for $p_j$ is the midpoint $G_j + x_j/(2 \, \xi_j^{(\textnormal{N} \rightarrow \textnormal{I})})$ of the integration interval. The second method employs a trapezoid approximation. Again due to the strictly increasing functional shape of $J^{- 1}$, the integral in (\ref{TT}) can be approximated by the area $A(j)$ of a trapezoid, which is computed as the sum of the area $A_{\textnormal{r}}(j)$ of the rectangle defined by the distance between the endpoints $G_j$ and $G_j + x_j/\xi_j^{(\textnormal{N} \rightarrow \textnormal{I})}$ of the integration interval and the height $J^{- 1}(G_j)$ 
\begin{equation*}
A_{\textnormal{r}} = \frac{x_j}{\xi_j^{(\textnormal{N} \rightarrow \textnormal{I})}} \, J^{- 1}(G_j)
\end{equation*}
and the area $A_{\textnormal{t}}(j)$ of the right-angled triangle with one cathetus given by the distance between the endpoints and the other one by the height $J^{- 1}(G_j + x_j/\xi_j^{(\textnormal{N} \rightarrow \textnormal{I})}) - J^{- 1}(G_j)$ 
\begin{equation*}
A_{\textnormal{t}} = \frac{x_j}{2 \, \xi_j^{(\textnormal{N} \rightarrow \textnormal{I})}} \, \bigl(J^{- 1}(G_j + x_j/\xi_j^{(\textnormal{N} \rightarrow \textnormal{I})}) - J^{- 1}(G_j)\bigr) \, .
\end{equation*}
This leads to the formula
\begin{equation*} 
t_{\textnormal{T}}^{(\textnormal{N} \rightarrow \textnormal{I})}(j) = \frac{x_j}{2 \, \xi_j^{(\textnormal{N} \rightarrow \textnormal{I})}} \, \bigl(J^{- 1}(G_j) + J^{- 1}(G_j + x_j/\xi_j^{(\textnormal{N} \rightarrow \textnormal{I})})\bigr) 
\end{equation*}
for the transition time. To obtain a more accurate approximation, one could use additional supporting points in the interval $\bigl[G_j, G_j + x_j/\xi_j^{(\textnormal{N} \rightarrow \textnormal{I})}\bigr]$, giving rise to a finer trapezoid decomposition of the integral. Note that in the present study, the trapezoid method yields a better approximation of the transition time. The IID-FID transition time 
\begin{equation*}
t_{\textnormal{T}}^{(\textnormal{I} \rightarrow \textnormal{F})}(j) = \int_{G_j}^{G_j + \xi_j^{(\textnormal{I} \rightarrow \textnormal{F})} \, x_j} J^{-1}(\widetilde{G}) \, \textnormal{d}\widetilde{G} 
\end{equation*}
can be computed similarly, resulting, on the one hand, in the expression
\begin{equation*}
t_{\textnormal{T}}^{(\textnormal{I} \rightarrow \textnormal{F})}(j) = \xi_j^{(\textnormal{I} \rightarrow \textnormal{F})} \, x_j\, J^{- 1}\bigl(G_j + \xi_j^{(\textnormal{I} \rightarrow \textnormal{F})} \, x_j/2\bigr) 
\end{equation*}
for the mean value theorem method and, on the other hand, in the expression 
\begin{equation*} 
t_{\textnormal{T}}^{(\textnormal{I} \rightarrow \textnormal{F})}(j) = \frac{\xi_j^{(\textnormal{I} \rightarrow \textnormal{F})} \, x_j}{2} \, \bigl(J^{- 1}(G_j) + J^{- 1}(G_j + \xi_j^{(\textnormal{I} \rightarrow \textnormal{F})} \, x_j)\bigr) 
\end{equation*}
for the trapezoid method.

\section{Integration Constants -- Initial and Transition Conditions and Updating} \label{app:-Constants:Gluing-Updating}

\noindent The integration constants $c_2$ and $c_4, ..., c_7$ of the NID (see Sol.\ (\ref{NIRSOL1NEW}) and Sols.\ (\ref{S1a})-(\ref{S1b})), $d_1, ..., d_5$ of the IID (see Sols.\ (\ref{iidsol1})-(\ref{iidsol4})), as well as $e_1, ..., e_4$ and $e_6, ..., e_9$ of the FID (see Sols.\ (\ref{fidsol2})-(\ref{fidsol4}) and Sols.\ (\ref{S2a})-(\ref{S2b})) are determined. The NID integration constants $c_2$, $c_4$, $c_6$, and $c_7$ are fixed via the initial condition $G(t = t_j) = G_j$, whereas $c_5$ is fixed via the transition condition $G\bigl(t = t_{\textnormal{T}}^{(\textnormal{N})}(j) \, | \, t_j \leq t < t_{\textnormal{T}}^{(\textnormal{N})}(j)\bigr) = G\bigl(t = t_{\textnormal{T}}^{(\textnormal{N})}(j) \, | \, t_{\textnormal{T}}^{(\textnormal{N})}(j) \leq t < t_{\textnormal{T}}^{(\textnormal{N} \rightarrow \textnormal{I})}(j)\bigr)$. The initial condition provides continuity of $G$ at the transition from the $(j - 1)$th injection domain to the $j$th injection domain at $t = t_j$ and the transition condition between the nonlinear and linear NID solution branches at $t = t_{\textnormal{T}}^{(\textnormal{N})}(j)$. Similarly, the IID integration constants $d_1$, $d_2$, $d_4$, and $d_5$ are determined via the respective initial condition $G\bigl(t = t_{\textnormal{T}}^{(\textnormal{N} \rightarrow \textnormal{I})}(j)\bigr) = G_{\textnormal{T}}^{(\textnormal{N} \rightarrow \textnormal{I})}(j)$ and $d_3$ via the transition condition $G\bigl(t = t_{\textnormal{T}}^{(\textnormal{I})}(j) \, | \, t_{\textnormal{T}}^{(\textnormal{N} \rightarrow \textnormal{I})}(j) \leq t < t_{\textnormal{T}}^{(\textnormal{I})}(j)\bigr) = G\bigl(t = t_{\textnormal{T}}^{(\textnormal{I})}(j) \, | \, t_{\textnormal{T}}^{(\textnormal{I})}(j) \leq t < t_{\textnormal{T}}^{(\textnormal{I} \rightarrow \textnormal{F})}(j)\bigr)$, where the former condition yields continuity between the NID and IID solution branches at $t = t_{\textnormal{T}}^{(\textnormal{N} \rightarrow \textnormal{I})}(j)$ and the latter between the nonlinear and linear IID solution branches at $t = t_{\textnormal{T}}^{(\textnormal{I})}(j)$. Last, the FID integration constants $e_1$, $e_3$, $e_4$, $e_6$, $e_8$, and $e_9$ are specified by the initial condition $G\bigl(t = t_{\textnormal{T}}^{(\textnormal{I} \rightarrow \textnormal{F})}(j)\bigr) = G_{\textnormal{T}}^{(\textnormal{I} \rightarrow \textnormal{F})}(j)$ and the integration constants $e_2$ and $e_7$ by the transition conditions $G\bigl(t = t_{\textnormal{T}, 1}^{(\textnormal{F})}(j) \, | \, t_{\textnormal{T}}^{(\textnormal{I} \rightarrow \textnormal{F})}(j) \leq t < t_{\textnormal{T}, 1}^{(\textnormal{F})}(j)\bigr) = G\bigl(t = t_{\textnormal{T}, 1}^{(\textnormal{F})}(j) \, | \, t_{\textnormal{T}, 1}^{(\textnormal{F})}(j) \leq t < t_{j + 1}\bigr)$ and $G\bigl(t = t_{\textnormal{T}, 2}^{(\textnormal{F})}(j) \, | \, t_{\textnormal{T}}^{(\textnormal{I} \rightarrow \textnormal{F})}(j) \leq t < t_{\textnormal{T}, 2}^{(\textnormal{F})}(j)\bigr) = G\bigl(t = t_{\textnormal{T}, 2}^{(\textnormal{F})}(j) \, | \, t_{\textnormal{T}, 2}^{(\textnormal{F})}(j) \leq t < t_{j + 1}\bigr)$, respectively. These conditions guarantee continuity between the IID and FID solution branches at $t = t_{\textnormal{T}}^{(\textnormal{I} \rightarrow \textnormal{F})}(j)$ and between the nonlinear and linear FID solution branches at $t = t_{\textnormal{T}, 1}^{(\textnormal{F})}(j)$ and $t = t_{\textnormal{T}, 2}^{(\textnormal{F})}(j)$. Further, the integration constants have to be updated if $\bigl\{t_{\textnormal{T}}^{(\textnormal{N} \rightarrow \textnormal{I})}(i), t_{\textnormal{T}}^{(\textnormal{I} \rightarrow \textnormal{F})}(i)\bigr\} \cap [t_j, t_{j + 1}) \not= \emptyset$ for $i \in \{1, ..., j - 1\}$, as elements of $S_1$ switch over to $S_2$ and/or elements of $S_2$ switch over to $S_3$. Because these updates cause shifts in the NID, IID, and FID transition times $t_{\textnormal{T}}^{(\textnormal{N})}$, $t_{\textnormal{T}}^{(\textnormal{I})}$, $t_{\textnormal{T}, 1}^{(\textnormal{F})}$, and $t_{\textnormal{T}, 2}^{(\textnormal{F})}$ towards larger values, we have to account for updating conditions that cover transitions to other solution branches. Below, the determination of the NID integration constant $c_4$ for the case $S_2 \not= \emptyset \not= S_3$ is shown in detail. The remaining integration constants are computed accordingly. Applying the initial condition $G\bigl(t = t_j \, | \, t_j \leq t < \textnormal{min}\bigl(t_{j + 1}, t_{\textnormal{T}}^{(\textnormal{N})}(j)\bigr)\bigr) = G_j$ to the first branch of Sol.\ (\ref{S1a}), we obtain  
\begin{equation*} 
c_4(j; S_3) = G_j^3 - 3 \, \mathscr{C}_3(j; S_3) \, t_j \, .
\end{equation*}
As $c_4$ depends just on $S_3$, we have to consider only IID-FID updates. Hence, the conditions -- and the updated constants -- for all possible IID-FID transitions of the $i$th injection at $t_j < t = t_{\textnormal{T}}^{(\textnormal{I} \rightarrow \textnormal{F})}(i) < \textnormal{min}\bigl(t_{j + 1}, t_{\textnormal{T}}^{(\textnormal{N})}(j)\bigr)$ are:
\begin{itemize}
\item First branch Sol.\ (\ref{S1a}) $\rightarrow$ first branch Sol.\ (\ref{S1a}) \\
\begin{equation*} 
\begin{split}
& G\bigl(t = t_{\textnormal{T}}^{(\textnormal{I} \rightarrow \textnormal{F})}(i) \, | \, t_j \leq t < t_{\textnormal{T}}^{(\textnormal{N})}(j; S_3 \backslash \{i\})\bigr) =  G\bigl(t = t_{\textnormal{T}}^{(\textnormal{I} \rightarrow \textnormal{F})}(i) \, | \, t_j \leq t < t_{\textnormal{T}}^{(\textnormal{N})}(j; S_3 \cup \{i\})\bigr) \\ \\
& c_4\bigl(j; S_3 \cup \{i\}\bigr) = 3 \, \Bigl(\mathscr{C}_3\bigl(j; S_3 \backslash \{i\}\bigr) - \mathscr{C}_3\bigl(j; S_3 \cup \{i\}\bigr)\Bigr) \, t_{\textnormal{T}}^{(\textnormal{I} \rightarrow \textnormal{F})}(i) + c_4\bigl(j; S_3 \backslash \{i\}\bigr)
\end{split}
\end{equation*} 
\item First branch Sol.\ (\ref{S1a}) $\rightarrow$ Sol.\ (\ref{S1c}) \\
\begin{equation*} 
\begin{split}
& G\bigl(t = t_{\textnormal{T}}^{(\textnormal{I} \rightarrow \textnormal{F})}(i) \, | \, t_j \leq t < t_{\textnormal{T}}^{(\textnormal{N})}(j; S_3 \backslash \{i\})\bigr) =  G\bigl(t = t_{\textnormal{T}}^{(\textnormal{I} \rightarrow \textnormal{F})}(i) \, | \, t_j \leq t < t_{\textnormal{T}}^{(\textnormal{N} \rightarrow \textnormal{I})}(j)\bigr) \\ \\
& c_6\bigl(j; S_3 \cup \{i\}\bigr) = 3 \, \Bigl(\mathscr{C}_3\bigl(j; S_3 \backslash \{i\}\bigr) - \mathscr{C}_3\bigl(j; S_3 \cup \{i\}\bigr)\Bigr) \, t_{\textnormal{T}}^{(\textnormal{I} \rightarrow \textnormal{F})}(i) + c_4\bigl(j; S_3 \backslash \{i\}\bigr) \, .
\end{split}
\end{equation*}
\end{itemize}
Since $S_3 \cup \{i\} \not= \emptyset$ and $t_{\textnormal{T}}^{(\textnormal{N})}(j; S_3 \backslash \{i\}) < t_{\textnormal{T}}^{(\textnormal{N})}(j; S_3 \cup \{i\})$, it is obvious that transitions from the first branch of (\ref{S1a}) to (\ref{NIRSOL1NEW}), from the first to the second branch of (\ref{S1a}), and from the first branch of (\ref{S1a}) to (\ref{S1b}) are not possible.

\section{Components of $G(t \, | \, 0 \leq t < \infty)$ and Initial Values $G_i$} \label{app:D}

\noindent We specify the expressions for the SID, NID, IID, and FID components $G_{\textnormal{SID}}$, $G_{\textnormal{NID}}$, $G_{\textnormal{IID}}$, and $G_{\textnormal{FID}}$ of Sol.\ (\ref{GFULL}). First, the SID component simply yields 
\begin{equation*}
\begin{split}
G_{\textnormal{SID}}(t) & := H\bigl(t_{\textnormal{T}}^{(\textnormal{S})}\bigr) \, H\bigl(t_2 - t_{\textnormal{T}}^{(\textnormal{S})}\bigr) \\ \\
& \hspace{0.4cm} \times \Bigl[H\bigl(t_{\textnormal{T}}^{(\textnormal{S})} - t\bigr) \, G\bigl(t \, | \, 0 \leq t < t_{\textnormal{T}}^{(\textnormal{S})} \, ; \, \textnormal{Sol}.\ (\ref{sAS1})\bigr ) + H\bigl(t - t_{\textnormal{T}}^{(\textnormal{S})}\bigr) \, G\bigl(t \, | \, t_{\textnormal{T}}^{(\textnormal{S})} \leq t < t_2 \, ; \, \textnormal{Sol}.\ (\ref{sAS1})\bigr)\Bigr]  \\ \\
&  \hspace{0.4cm} + H\bigl(t_{\textnormal{T}}^{(\textnormal{S})} - t_2\bigr) \, G\bigl(t \, | \, 0 \leq t < t_2 \, ; \, \textnormal{Sol}.\ (\ref{sAS2})\bigr) + H\bigl(- t_{\textnormal{T}}^{(\textnormal{S})}\bigr) \, G\bigl(t \, | \, 0 \leq t < t_2 \, ; \, \textnormal{Sol}.\ (\ref{sAS3})\bigr) \, .
\end{split}
\end{equation*}
The more elaborate NID component is given by 
\begin{equation*}
\begin{split}
G_{\textnormal{NID}}(t; i) & := \bigl[1 - \chi(S_2)\bigr] \, \bigl[1 - \chi(S_3)\bigr] \, G\bigl(t \, | \, t_i \leq t < t_{\textnormal{T}}^{(\textnormal{N} \rightarrow \textnormal{I})}(i) \, ; \, \textnormal{Sol}.\ (\ref{NIRSOL1NEW}); S_2 = \emptyset = S_3\bigr) \\ \\
& \hspace{0.4cm} + \chi(S_2) \, \bigl[1 - \chi(S_3)\bigr] \, G\bigl(t \, | \, t_i \leq t < t_{\textnormal{T}}^{(\textnormal{N} \rightarrow \textnormal{I})}(i) \, ; \, \textnormal{Sol}.\ (\ref{NIRSOL1NEW}); S_2 \not= \emptyset, S_3 = \emptyset\bigr) \\ \\
& \hspace{0.4cm} + \bigl[1 - \chi(S_2)\bigr] \, \chi(S_3) \, G_{\textnormal{NID}}\bigl(t \, | \, t_i \leq t < t_{\textnormal{T}}^{(\textnormal{N} \rightarrow \textnormal{I})}(i) \, ; \, S_2 = \emptyset, S_3 \not= \emptyset\bigr) \\ \\
& \hspace{0.4cm} + \chi(S_2) \, \chi(S_3) \, G_{\textnormal{NID}}\bigl(t \, | \, t_i \leq t < t_{\textnormal{T}}^{(\textnormal{N} \rightarrow \textnormal{I})}(i) \, ; \, S_2 \not= \emptyset \not= S_3\bigr) \, ,
\end{split}
\end{equation*}
where 
\begin{equation*}
\begin{split}
& G_{\textnormal{NID}}\bigl(t \, | \, t_i \leq t < t_{\textnormal{T}}^{(\textnormal{N} \rightarrow \textnormal{I})}(i) \, ; \, S_2 \, \substack{= \\ \not=} \, \emptyset, S_3 \not= \emptyset\bigr) = H\bigl(t_{\textnormal{T}}^{(\textnormal{N})}(i) - t_i\bigr) \, H\bigl(t_{\textnormal{T}}^{(\textnormal{N} \rightarrow \textnormal{I})}(i) - t_{\textnormal{T}}^{(\textnormal{N})}(i)\bigr)\\ \\
& \hspace{2.0cm} \times \Bigl[H\bigl(t_{\textnormal{T}}^{(\textnormal{N})}(i) - t\bigr) \, G\bigl(t \, | \, t_i \leq t < t_{\textnormal{T}}^{(\textnormal{N})}(i) \, ; \, \textnormal{Sol.}(\ref{S1a}) \, ; \, S_2 \, \substack{= \\ \not=} \, \emptyset, S_3 \not= \emptyset\bigr) \\ \\
& \hspace{2.5cm} + H\bigl(t - t_{\textnormal{T}}^{(\textnormal{N})}(i)\bigr) \, G\bigl(t \, | \, t_{\textnormal{T}}^{(\textnormal{N})}(i) \leq t < t_{\textnormal{T}}^{(\textnormal{N} \rightarrow \textnormal{I})}(i) \, ; \, \textnormal{Sol.}(\ref{S1a}) \, ; \, S_2 \, \substack{= \\ \not=} \, \emptyset, S_3 \not= \emptyset\bigr)\Bigr] \\ \\
& \hspace{2.0cm} + H\bigl(t_{\textnormal{T}}^{(\textnormal{N})}(i) - t_{\textnormal{T}}^{(\textnormal{N} \rightarrow \textnormal{I})}(i)\bigr) \, G\bigl(t \, | \, t_i \leq t < t_{\textnormal{T}}^{(\textnormal{N} \rightarrow \textnormal{I})}(i) \, ; \, \textnormal{Sol.}(\ref{S1c}) \, ; \,  S_2 \, \substack{= \\ \not=} \, \emptyset, S_3 \not= \emptyset\bigr) \\ \\
& \hspace{2.0cm} + H\bigl(t_i - t_{\textnormal{T}}^{(\textnormal{N})}(i)\bigr) \, G\bigl(t \, | \, t_i \leq t < t_{\textnormal{T}}^{(\textnormal{N} \rightarrow \textnormal{I})}(i) \, ; \, \textnormal{Sol.}(\ref{S1b}) \, ; \,  S_2 \, \substack{= \\ \not=} \, \emptyset, S_3 \not= \emptyset\bigr)
\end{split}
\end{equation*}
and
\begin{equation*}
\chi(S_k) := \begin{cases} 1 & \,\,\, \textnormal{for} \,\,\, S_k \not= \emptyset \\  
0 & \,\,\, \textnormal{for} \,\,\, S_k = \emptyset \, ,
\end{cases} 
\end{equation*}
with $k \in \{1, 2, 3\}$, is the characteristic function. For the IID and FID components, we obtain similar expressions. On the one hand, we have
\begin{equation*}
\begin{split}
G_{\textnormal{IID}}(t; i) & := \bigl[1 - \chi(S_1)\bigr] \, \bigl[1 - \chi(S_3)\bigr] \, G\bigl(t \, | \, t_{\textnormal{T}}^{(\textnormal{N} \rightarrow \textnormal{I})}(i) \leq t < t_{\textnormal{T}}^{(\textnormal{I} \rightarrow \textnormal{F})}(i) \, ; \, \textnormal{Sol}.\ (\ref{iidsol1}) \, ; \, S_1 = \emptyset = S_3\bigr) \\ \\
& \hspace{0.4cm} + \chi(S_1) \, \bigl[1 - \chi(S_3)\bigr] \, G\bigl(t \, | \, t_{\textnormal{T}}^{(\textnormal{N} \rightarrow \textnormal{I})}(i) \leq t < t_{\textnormal{T}}^{(\textnormal{I} \rightarrow \textnormal{F})}(i) \, ; \, \textnormal{Sol}.\ (\ref{iidsol1}) \, ; \, S_1 \not= \emptyset, S_3 = \emptyset\bigr) \\ \\
& \hspace{0.4cm} + \bigl[1 - \chi(S_1)\bigr] \, \chi(S_3) \, G_{\textnormal{IID}}\bigl(t \, | \, t_{\textnormal{T}}^{(\textnormal{N} \rightarrow \textnormal{I})}(i) \leq t < t_{\textnormal{T}}^{(\textnormal{I} \rightarrow \textnormal{F})}(i) \, ; \, S_1 = \emptyset, S_3 \not= \emptyset\bigr) \\ \\
& \hspace{0.4cm} + \chi(S_1) \, \chi(S_3) \, G_{\textnormal{IID}}\bigl(t \, | \, t_{\textnormal{T}}^{(\textnormal{N} \rightarrow \textnormal{I})}(i) \leq t < t_{\textnormal{T}}^{(\textnormal{I} \rightarrow \textnormal{F})}(i) \, ; \, S_1 \not= \emptyset \not= S_3\bigr) \, ,
\end{split}
\end{equation*}
where 
\begin{equation*}
\begin{split}
& G_{\textnormal{IID}}\bigl(t \, | \, t_{\textnormal{T}}^{(\textnormal{N} \rightarrow \textnormal{I})}(i) \leq t < t_{\textnormal{T}}^{(\textnormal{I} \rightarrow \textnormal{F})}(i) \, ; \, S_1 \, \substack{= \\ \not=} \, \emptyset, S_3 \not= \emptyset\bigr) = H\bigl(t_{\textnormal{T}}^{(\textnormal{I})}(i) - t_{\textnormal{T}}^{(\textnormal{N} \rightarrow \textnormal{I})}(i)\bigr) \, H\bigl(t_{\textnormal{T}}^{(\textnormal{I} \rightarrow \textnormal{F})}(i) - t_{\textnormal{T}}^{(\textnormal{I})}(i)\bigr) \\ \\
& \hspace{2.0cm} \times \Bigl[H\bigl(t_{\textnormal{T}}^{(\textnormal{I})}(i) - t\bigr) \, G\bigl(t \, | \, t_{\textnormal{T}}^{(\textnormal{N} \rightarrow \textnormal{I})}(i) \leq t < t_{\textnormal{T}}^{(\textnormal{I})}(i) \, ; \, \textnormal{Sol.}(\ref{iidsol2}) \, ; \, S_1 \, \substack{= \\ \not=} \, \emptyset, S_3 \not= \emptyset\bigr) \\ \\
& \hspace{2.6cm} + H\bigl(t - t_{\textnormal{T}}^{(\textnormal{I})}(i)\bigr) \, G\bigl(t \, | \, t_{\textnormal{T}}^{(\textnormal{I})}(i) \leq t < t_{\textnormal{T}}^{(\textnormal{I} \rightarrow \textnormal{F})}(i) \, ; \, \textnormal{Sol.}(\ref{iidsol2}) \, ; \, S_1 \, \substack{= \\ \not=} \, \emptyset, S_3 \not= \emptyset\bigr)\Bigr] \\ \\
& \hspace{2.0cm} + H\bigl(t_{\textnormal{T}}^{(\textnormal{I})}(i) - t_{\textnormal{T}}^{(\textnormal{I} \rightarrow \textnormal{F})}(i)\bigr) \, G\bigl(t \, | \, t_{\textnormal{T}}^{(\textnormal{N} \rightarrow \textnormal{I})}(i) \leq t < t_{\textnormal{T}}^{(\textnormal{I} \rightarrow \textnormal{F})}(i) \, ; \, \textnormal{Sol.}(\ref{iidsol3}) \, ; \,  S_1 \, \substack{= \\ \not=} \, \emptyset, S_3 \not= \emptyset\bigr) \\ \\
& \hspace{2.0cm} + H\bigl(t_{\textnormal{T}}^{(\textnormal{N} \rightarrow \textnormal{I})}(i) - t_{\textnormal{T}}^{(\textnormal{I})}(i)\bigr) \, G\bigl(t \, | \, t_{\textnormal{T}}^{(\textnormal{N} \rightarrow \textnormal{I})}(i) \leq t < t_{\textnormal{T}}^{(\textnormal{I} \rightarrow \textnormal{F})}(i) \, ; \, \textnormal{Sol.}(\ref{iidsol4}) \, ; \,  S_1 \, \substack{= \\ \not=} \, \emptyset, S_3 \not= \emptyset\bigr) \, ,	
\end{split}
\end{equation*}
and on the other hand, we have
\begin{equation*}
\begin{split}
G_{\textnormal{FID}}(t; i) & := \bigl[1 - \chi(S_1)\bigr] \, \bigl[1 - \chi(S_2)\bigr] \, G_{\textnormal{FID}}\bigl(t \, | \, t_{\textnormal{T}}^{(\textnormal{I} \rightarrow \textnormal{F})}(i) \leq t < t_{i + 1} \, ; \, S_1 = \emptyset = S_2\bigr) \\ \\
& \hspace{0.4cm} + \chi(S_1) \, \bigl[1 - \chi(S_2)\bigr] \, G_{\textnormal{FID}}\bigl(t \, | \, t_{\textnormal{T}}^{(\textnormal{I} \rightarrow \textnormal{F})}(i) \leq t < t_{i + 1} \, ; \, S_1 \not= \emptyset, S_2 = \emptyset\bigr) \\ \\
& \hspace{0.4cm} + \bigl[1 - \chi(S_1)\bigr] \, \chi(S_2) \, G_{\textnormal{FID}}\bigl(t \, | \, t_{\textnormal{T}}^{(\textnormal{I} \rightarrow \textnormal{F})}(i) \leq t < t_{i + 1} \, ; \, S_1 = \emptyset, S_2 \not= \emptyset\bigr) \\ \\
& \hspace{0.4cm} + \chi(S_1) \, \chi(S_2) \, G_{\textnormal{FID}}\bigl(t \, | \, t_{\textnormal{T}}^{(\textnormal{I} \rightarrow \textnormal{F})}(i) \leq t < t_{i + 1} \, ; \, S_1 \not= \emptyset \not= S_2\bigr) \, ,
\end{split}
\end{equation*}
where
\begin{equation*}
\begin{split}
& G_{\textnormal{FID}}\bigl(t \, | \, t_{\textnormal{T}}^{(\textnormal{I} \rightarrow \textnormal{F})}(i) \leq t < t_{i + 1} \, ; \, S_1 = \emptyset = S_2\bigr) = H\bigl(t_{\textnormal{T}, 2}^{(\textnormal{F})}(i) - t_{\textnormal{T}}^{(\textnormal{I} \rightarrow \textnormal{F})}(i)\bigr) \, H\bigl(t_{i + 1} - t_{\textnormal{T}, 2}^{(\textnormal{F})}(i)\bigr) \\ \\
& \times \Bigl[H\bigl(t_{\textnormal{T}, 2}^{(\textnormal{F})}(i) - t\bigr) \, G\bigl(t \, | \, t_{\textnormal{T}}^{(\textnormal{I} \rightarrow \textnormal{F})}(i) \leq t < t_{\textnormal{T}, 2}^{(\textnormal{F})}(i) \, ; \, \textnormal{Sol.}(\ref{S2a}) \, ; \, S_1 = \emptyset = S_2\bigr) \\ \\
& \hspace{0.6cm} + H\bigl(t - t_{\textnormal{T}, 2}^{(\textnormal{F})}(i)\bigr) \, G\bigl(t \, | \, t_{\textnormal{T}, 2}^{(\textnormal{F})}(i) \leq t < t_{i + 1} \, ; \, \textnormal{Sol.}(\ref{S2a}) \, ; \, S_1 = \emptyset = S_2\bigr)\Bigr] \\ \\
& + H\bigl(t_{i + 1} - t_{\textnormal{T}}^{(\textnormal{I} \rightarrow \textnormal{F})}(i)\bigr) \Bigl[H\bigl(t_{\textnormal{T}, 2}^{(\textnormal{F})}(i) - t_{i + 1}\bigr) \, G\bigl(t \, | \, t_{\textnormal{T}}^{(\textnormal{I} \rightarrow \textnormal{F})}(i) \leq t < t_{i + 1} \, ; \, \textnormal{Sol.}(\ref{AddSol}) \, ; \,  S_1 = \emptyset = S_2\bigr) \\ \\
& \hspace{3.6cm} + H\bigl(t_{\textnormal{T}}^{(\textnormal{I} \rightarrow \textnormal{F})}(i) - t_{\textnormal{T}, 2}^{(\textnormal{F})}(i)\bigr) \, G\bigl(t \, | \, t_{\textnormal{T}}^{(\textnormal{I} \rightarrow \textnormal{F})}(i) \leq t < t_{i + 1} \, ; \, \textnormal{Sol.}(\ref{S2b}) \, ; \, S_1 = \emptyset = S_2\bigr)\Bigr]
\end{split}
\end{equation*}
as well as 
\begin{equation*}
\begin{split}
& G_{\textnormal{FID}}\bigl(t \, | \, t_{\textnormal{T}}^{(\textnormal{I} \rightarrow \textnormal{F})}(i) \leq t < t_{i + 1} \, ; \, S_1 \not= \emptyset, S_2 = \emptyset \,\,\, \textnormal{or} \,\,\, S_1 \, \substack{= \\ \not=} \, \emptyset, S_2 \not= \emptyset\bigr) = H\bigl(t_{\textnormal{T}, 1}^{(\textnormal{F})}(i) - t_{\textnormal{T}}^{(\textnormal{I} \rightarrow \textnormal{F})}(i)\bigr) \, H\bigl(t_{i + 1} - t_{\textnormal{T}, 1}^{(\textnormal{F})}(i)\bigr) \\ \\
& \times \Bigl[H\bigl(t_{\textnormal{T}, 1}^{(\textnormal{F})}(i) - t\bigr) \, G\bigl(t \, | \, t_{\textnormal{T}}^{(\textnormal{I} \rightarrow \textnormal{F})}(i) \leq t < t_{\textnormal{T}, 1}^{(\textnormal{F})}(i) \, ; \, \textnormal{Sol.}(\ref{fidsol2}) \, ; \, S_1 \not= \emptyset, S_2 = \emptyset \,\,\, \textnormal{or} \,\,\, S_1 \, \substack{= \\ \not=} \, \emptyset, S_2 \not= \emptyset\bigr) \\ \\
& \hspace{0.6cm} + H\bigl(t - t_{\textnormal{T}, 1}^{(\textnormal{F})}(i)\bigr) \, G\bigl(t \, | \, t_{\textnormal{T}, 1}^{(\textnormal{F})}(i) \leq t < t_{i + 1} \, ; \, \textnormal{Sol.}(\ref{fidsol2}) \, ; \, S_1 \not= \emptyset, S_2 = \emptyset \,\,\, \textnormal{or} \,\,\, S_1 \, \substack{= \\ \not=} \, \emptyset, S_2 \not= \emptyset\bigr)\Bigr] \\ \\
& + H\bigl(t_{i + 1} - t_{\textnormal{T}}^{(\textnormal{I} \rightarrow \textnormal{F})}(i)\bigr) \Bigl[H\bigl(t_{\textnormal{T}, 1}^{(\textnormal{F})}(i) - t_{i + 1}\bigr) \, G\bigl(t \, | \, t_{\textnormal{T}}^{(\textnormal{I} \rightarrow \textnormal{F})}(i) \leq t < t_{i + 1} \, ; \, \textnormal{Sol.}(\ref{fidsol3}) \, ; \, S_1 \not= \emptyset, S_2 = \emptyset \,\,\, \textnormal{or} \,\,\, S_1 \, \substack{= \\ \not=} \, \emptyset, S_2 \not= \emptyset\bigr) \\ \\
& \hspace{1.8cm} + H\bigl(t_{\textnormal{T}}^{(\textnormal{I} \rightarrow \textnormal{F})}(i) - t_{\textnormal{T}, 1}^{(\textnormal{F})}(i)\bigr) \, G\bigl(t \, | \, t_{\textnormal{T}}^{(\textnormal{I} \rightarrow \textnormal{F})}(i) \leq t < t_{i + 1} \, ; \, \textnormal{Sol.}(\ref{fidsol4}) \, ; \, S_1 \not= \emptyset, S_2 = \emptyset \,\,\, \textnormal{or} \,\,\, S_1 \, \substack{= \\ \not=} \, \emptyset, S_2 \not= \emptyset\bigr)\Bigr] \, .
\end{split}
\end{equation*}
Further, we determine the initial values $G_i$ for all $i : 3 \leq i \leq m$ by requiring continuity between the solution branches of the $(i -1)$th and $i$th injection domains. If the $i$th injection enters the system while the $(i - 1)$th injection is still in the NID, the initial values become in case $ S_2 \, \substack{= \\ \not=} \, \emptyset, S_3 = \emptyset$
\begin{equation*}
G_i = G\bigl(t = t_i \, | \, t_{i - 1} \leq t < t_{\textnormal{T}}^{(\textnormal{N} \rightarrow \textnormal{I})}(i - 1) \, ; \, \textnormal{Sol.}(\ref{NIRSOL1NEW})\bigr) \,\,\,\,\,\, \textnormal{for} \,\,\,\,\, t_i < t_{\textnormal{T}}^{(\textnormal{N} \rightarrow \textnormal{I})}(i - 1) 
\end{equation*}
and in case $ S_2 \, \substack{= \\ \not=} \, \emptyset, S_3 \not= \emptyset$
\begin{equation*}
G_i = \begin{cases} 
G\bigl(t = t_i \, | \, t_{i - 1} \leq t < t_{\textnormal{T}}^{(\textnormal{N})}(i - 1) \, ; \, \textnormal{Sol.}(\ref{S1a})\bigr) & \, \textnormal{for} \,\,\,\,\, t_i < t_{\textnormal{T}}^{(\textnormal{N})}(i - 1) < t_{\textnormal{T}}^{(\textnormal{N} \rightarrow \textnormal{I})}(i - 1) \\ \\
G\bigl(t = t_i \, | \, t_{\textnormal{T}}^{(\textnormal{N})}(i - 1) \leq t < t_{\textnormal{T}}^{(\textnormal{N} \rightarrow \textnormal{I})}(i - 1) \, ; \, \textnormal{Sol.}(\ref{S1a})\bigr) & \, \textnormal{for} \,\,\,\,\, t_{\textnormal{T}}^{(\textnormal{N})}(i - 1) \leq t_i < t_{\textnormal{T}}^{(\textnormal{N} \rightarrow \textnormal{I})}(i - 1) \\ \\ 
G\bigl(t = t_i \, | \, t_{i - 1} \leq t < t_{\textnormal{T}}^{(\textnormal{N} \rightarrow \textnormal{I})}(i - 1) \, ; \, \textnormal{Sol.}(\ref{S1c})\bigr) & \, \textnormal{for} \,\,\,\,\, t_i < t_{\textnormal{T}}^{(\textnormal{N} \rightarrow \textnormal{I})}(i - 1) \leq t_{\textnormal{T}}^{(\textnormal{N})}(i - 1) \\ \\
G\bigl(t = t_i \, | \, t_{i - 1} \leq t < t_{\textnormal{T}}^{(\textnormal{N} \rightarrow \textnormal{I})}(i - 1) \, ; \, \textnormal{Sol.}(\ref{S1b})\bigr) & \, \textnormal{for} \,\,\,\,\, t_{\textnormal{T}}^{(\textnormal{N})}(i - 1) \leq t_{i - 1} < t_i < t_{\textnormal{T}}^{(\textnormal{N} \rightarrow \textnormal{I})}(i - 1) \, .
\end{cases}
\end{equation*}
If, however, the $(i - 1)$th injection is in the IID, they result for $ S_1 \, \substack{= \\ \not=} \, \emptyset, S_3 = \emptyset$ in 
\begin{equation*}
G_i = G\bigl(t = t_i \, | \, t_{\textnormal{T}}^{(\textnormal{N} \rightarrow \textnormal{I})}(i - 1) \leq t < t_{\textnormal{T}}^{(\textnormal{I} \rightarrow \textnormal{F})}(i - 1) \, ; \, \textnormal{Sol.}(\ref{iidsol1})\bigr) \,\,\,\,\,\, \textnormal{for} \,\,\,\,\, t_{\textnormal{T}}^{(\textnormal{N} \rightarrow \textnormal{I})}(i - 1) \leq t_i < t_{\textnormal{T}}^{(\textnormal{I} \rightarrow \textnormal{F})}(i - 1) \, ,
\end{equation*}
whereas for  $S_1 \, \substack{= \\ \not=} \, \emptyset, S_3 \not= \emptyset$, they yield

\begin{equation*}
G_i = \begin{cases} 
G\bigl(t = t_i \, | \, t_{\textnormal{T}}^{(\textnormal{N} \rightarrow \textnormal{I})}(i - 1) \leq t < t_{\textnormal{T}}^{(\textnormal{I})}(i - 1) \, ; \, \textnormal{Sol.}(\ref{iidsol2})\bigr) & \, \textnormal{for} \,\,\,\,\, t_i < t_{\textnormal{T}}^{(\textnormal{I})}(i - 1) < t_{\textnormal{T}}^{(\textnormal{I} \rightarrow \textnormal{F})}(i - 1) \\ \\
G\bigl(t = t_i \, | \, t_{\textnormal{T}}^{(\textnormal{I})}(i - 1) \leq t < t_{\textnormal{T}}^{(\textnormal{I} \rightarrow \textnormal{F})}(i - 1) \, ; \, \textnormal{Sol.}(\ref{iidsol2})\bigr) & \, \textnormal{for} \,\,\,\,\, t_{\textnormal{T}}^{(\textnormal{I})}(i - 1) \leq t_i < t_{\textnormal{T}}^{(\textnormal{I} \rightarrow \textnormal{F})}(i - 1) \\ \\ 
G\bigl(t = t_i \, | \, t_{\textnormal{T}}^{(\textnormal{N} \rightarrow \textnormal{I})}(i - 1) \leq t < t_{\textnormal{T}}^{(\textnormal{I} \rightarrow \textnormal{F})}(i - 1) \, ; \, \textnormal{Sol.}(\ref{iidsol3})\bigr) & \, \textnormal{for} \,\,\,\,\, t_i < t_{\textnormal{T}}^{(\textnormal{I} \rightarrow \textnormal{F})}(i - 1) \leq t_{\textnormal{T}}^{(\textnormal{I})}(i - 1) \\ \\
G\bigl(t = t_i \, | \, t_{\textnormal{T}}^{(\textnormal{N} \rightarrow \textnormal{I})}(i - 1) \leq t < t_{\textnormal{T}}^{(\textnormal{I} \rightarrow \textnormal{F})}(i - 1) \, ; \, \textnormal{Sol.}(\ref{iidsol4})\bigr) & \, \textnormal{for} \,\,\,\,\, t_{\textnormal{T}}^{(\textnormal{I})}(i - 1) \leq t_{i - 1} < t_i < t_{\textnormal{T}}^{(\textnormal{I} \rightarrow \textnormal{F})}(i - 1) \, .
\end{cases}
\end{equation*}
Finally, when the $(i - 1)$th injection is already in the FID, we find in case $S_1 \not= \emptyset, S_2 = \emptyset$ or $S_1 \, \substack{= \\ \not=} \, \emptyset, S_2 \not= \emptyset$ 
\begin{equation*} 	
G_i = 	\begin{cases} 
G\bigl(t = t_i \, | \, t_{\textnormal{T}, 1}^{(\textnormal{F})}(i - 1) \leq t < t_i \, ; \, \textnormal{Sol.}(\ref{fidsol2})\bigr) & \, \textnormal{for} \,\,\,\,\, t_{\textnormal{T}}^{(\textnormal{I} \rightarrow \textnormal{F})}(i - 1) < t_{\textnormal{T}, 1}^{(\textnormal{F})}(i - 1) \leq t_i \\ \\
G\bigl(t = t_i \, | \, t_{\textnormal{T}}^{(\textnormal{I} \rightarrow \textnormal{F})}(i - 1) \leq t < t_i \, ; \, \textnormal{Sol.}(\ref{fidsol3})\bigr) & \, \textnormal{for} \,\,\,\,\, t_{\textnormal{T}}^{(\textnormal{I} \rightarrow \textnormal{F})}(i - 1) \leq t_i < t_{\textnormal{T}, 1}^{(\textnormal{F})}(i - 1) \\ \\
G\bigl(t = t_i \, | \, t_{\textnormal{T}}^{(\textnormal{I} \rightarrow \textnormal{F})}(i - 1) \leq t < t_i \, ; \, \textnormal{Sol.}(\ref{fidsol4})\bigr) & \, \textnormal{for} \,\,\,\,\, t_{\textnormal{T}, 1}^{(\textnormal{F})}(i - 1) \leq t_{\textnormal{T}}^{(\textnormal{I} \rightarrow \textnormal{F})}(i - 1) \leq t_i 
\end{cases}
\end{equation*}
and in case $S_1 = \emptyset = S_2$
\begin{equation*} 	
G_i = 	\begin{cases} 
G\bigl(t = t_i \, | \, t_{\textnormal{T}, 2}^{(\textnormal{F})}(i - 1) \leq t < t_i \, ; \, \textnormal{Sol.}(\ref{S2a})\bigr) & \, \textnormal{for} \,\,\,\,\, t_{\textnormal{T}}^{(\textnormal{I} \rightarrow \textnormal{F})}(i - 1) < t_{\textnormal{T}, 2}^{(\textnormal{F})}(i - 1) \leq t_i \\ \\
G\bigl(t = t_i \, | \, t_{\textnormal{T}}^{(\textnormal{I} \rightarrow \textnormal{F})}(i - 1) \leq t < t_i \, ; \, \textnormal{Sol.}(\ref{AddSol})\bigr) & \, \textnormal{for} \,\,\,\,\, t_{\textnormal{T}}^{(\textnormal{I} \rightarrow \textnormal{F})}(i - 1) \leq t_i < t_{\textnormal{T}, 2}^{(\textnormal{F})}(i - 1) \\ \\
G\bigl(t = t_i \, | \, t_{\textnormal{T}}^{(\textnormal{I} \rightarrow \textnormal{F})}(i - 1) \leq t < t_i \, ; \, \textnormal{Sol.}(\ref{S2b})\bigr) & \, \textnormal{for} \,\,\,\,\, t_{\textnormal{T}, 2}^{(\textnormal{F})}(i - 1) \leq t_{\textnormal{T}}^{(\textnormal{I} \rightarrow \textnormal{F})}(i - 1) \leq t_i \, .
\end{cases}
\end{equation*}

\section{The $CS$ Function} \label{app:CS}

\begin{figure}[t]
	\centering
	\includegraphics[width=0.49\columnwidth]{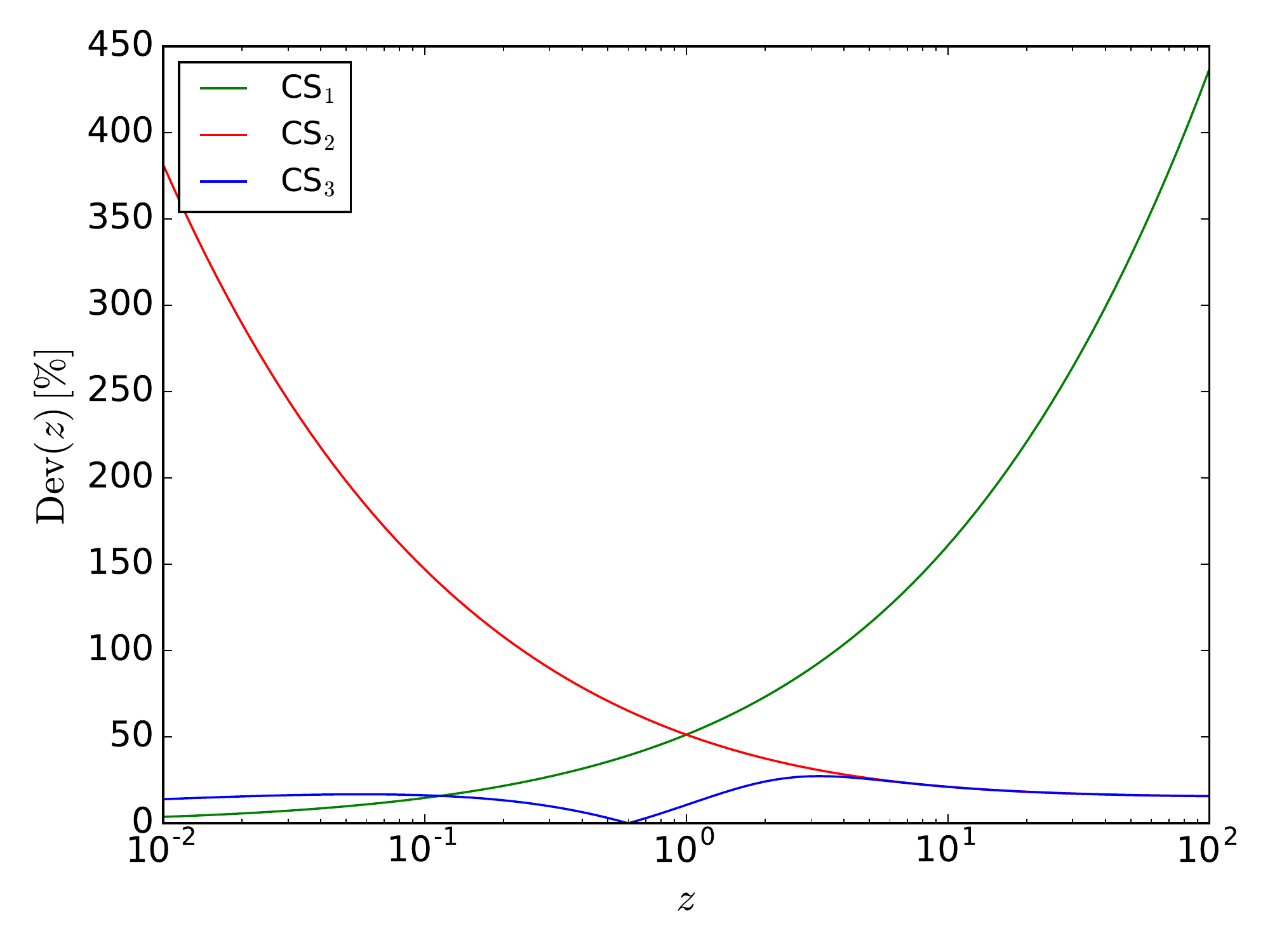}%
	\caption[...]{Absolute values of the relative deviations of the approximate functions $CS_n$ for all $n \in \{1, 2, 3\}$ with respect to the exact $CS$ function in percent.}
	\label{PICE}
\end{figure}

\noindent For $z \in \mathbb{R}_{\geq 0}$, the $CS$ function is defined by \cite{CS}
\begin{equation} \label{CSO}
\begin{split} 
CS(z) & := \frac{1}{\pi} \int_{0}^{\pi} \sin{(\theta)} \int_{z/\sin{(\theta)}}^{\infty} K_{5/3}(y) \, \textnormal{d}y \,\textnormal{d}\theta \\ \\
& \hspace{0.1cm} = W_{0, \, 4/3}(z) \, W_{0, \, 1/3}(z) - W_{1/2, \, 5/6}(z) \, W_{- 1/2, \, 5/6}(z) \, ,
\end{split}
\end{equation}
where $K_{a}$ is the modified Bessel function and $W_{a, \, b}$ denotes the Whittaker function \cite{AbrSt}. On account of the degree of complexity of (\ref{CSO}), one usually employs an approximate function that is adapted to its asymptotics 
\begin{equation*}
CS(z) \simeq \begin{cases} a_0 \, z^{-2/3} &  \,\,\, \textnormal{for} \,\,\,\, z \ll 1  \\ z^{-1} \, \exp{(- z)} & \,\,\, \textnormal{for} \,\,\,\, z \gg 1 \, , \end{cases}
\end{equation*}
where $a_0 := 1.15$. Standard approximations are, therefore, given by
\begin{equation*}
CS_1(z) := \frac{a_0 \, \exp{(- z)}}{z^{2/3}} \, , \,\,\,\,\, CS_2(z) := \frac{a_0 \, \exp{(- z)}}{z} \, , \,\,\,\,\,\, \textnormal{and} \,\,\,\,\,\, CS_3(z) := \frac{a_0}{z^{2/3} \, \bigl(1 + z^{1/3} \, \exp{(z)}\bigr)} \, .
\end{equation*}
Figure \ref{PICE} shows the absolute values of the relative deviations of these approximations with respect to (\ref{CSO}) in percent, that is, 
\begin{equation*}
\textnormal{Dev}(z) := 100 \, \left|\frac{CS(z) - CS_n(z)}{CS(z)}\right| \,\,\,\,\, \textnormal{for} \,\,\,\,\, n \in \{1, 2, 3\} \, .
\end{equation*}
Since the $CS$ function is primarily used for the computation of the synchrotron intensity, where the spectrum covers the energy range from radio waves to X-rays, i.e., with a lower energy limit of the order neV and an upper limit of the order keV, and $z = 2 \, \epsilon/(3 \, \epsilon_0 \, \gamma^2)$ with $\epsilon_0 \sim 10^{-14} \, b$, initial electron energies $\gamma_i \sim 10^4 \, b^{-1/3}$, and nondimensional magnetic field strength $b \sim 10^{- 3}$ up to $b \sim 10$, we have to consider values $z \ll 1$ up to $z \gg 1$. Thus, the only suitable choice is the approximate function $CS_3(z)$, which coincides with both asymptotic ends of (\ref{CSO}) and has the smallest overall deviation.

\section{Relativistic Beaming} \label{LTOF}

\noindent Accounting for relativistic beaming, the energy $\varepsilon \in \{\epsilon, \epsilon_{\textnormal{s}}\}$ and the time $t$, which are defined in the plasmoid rest frame, are related to the corresponding quantities in an observer frame (denoted with an asterisk) by
\begin{equation*}
\varepsilon^{\star} = \mathcal{D} \, \varepsilon \,\,\,\,\, \textnormal{and} \,\,\,\,\, t^{\star} = \frac{t}{\mathcal{D}} \, , 
\end{equation*}
where 
\begin{equation*}
\mathcal{D} := \frac{1}{\Gamma \bigl(1 - \beta \, \cos{(\theta^{\star})}\bigr)} 
\end{equation*}
is the boost factor, $\Gamma := 1/\sqrt{1 - \beta^2}$ the Lorentz factor of the plasmoid, $\beta := v/c$, and $\theta^{\star}$ the angle between the jet axis and the line of sight of the observer. For blazars, one can assume that $\theta^{\star} \searrow 0$ and, thus, 
\begin{equation*}
\mathcal{D} \simeq \sqrt{\frac{1 + \beta}{1 - \beta}} \, .
\end{equation*}
Furthermore, because the ratio $I/\varepsilon^3$ is Lorentz-invariant, i.e.,
\begin{equation*}
\frac{I(\varepsilon, t)}{\varepsilon^3} = \frac{I^{\star}(\varepsilon^{\star}, t^{\star})}{(\varepsilon^{\star})^3} \, ,
\end{equation*} 
one directly finds that the intensity and the fluence transform as
\begin{equation*}
I^{\star}(\varepsilon^{\star}, t^{\star}) = \mathcal{D}^3 \, I(\varepsilon, t) \,\,\,\,\, \textnormal{and} \,\,\,\,\, F^{\star}(\varepsilon^{\star}) = \mathcal{D}^2 \, F(\varepsilon) \, .
\end{equation*}

\section{Plotting Algorithm for the Fluence Spectral Energy Distributions} \label{numcode}

\noindent The numerical implementation of $G$, the synchrotron and SSC intensities, as well as the corresponding total fluence SEDs is carried out with Python. Here, we describe the functionality of the algorithm\footnote{The plotting algorithm is available on request via e-mail to christian.roeken@mathematik.uni-regensburg.de.} and the specific incorporation of the analytical formulas. The algorithm makes heavy use of the decimal package\footnote{For more detailed information on Python's decimal package, we refer to https://docs.python.org/2/library/decimal.html.}, which is designed for high floating point precision calculations. This is necessary because the range of possible values between  the synchrotron and SSC cooling rate prefactors $D_0$ and $A_0$, the injection strengths $q_i$, and the reciprocal initial electron energies $x_i$ spans several orders of magnitude, which may lead to a loss of accuracy in expressions where these parameters come up. This problem occurs in its most severe form in the evaluation of the formulas for the transition times $t_{\textnormal{T}}^{(\textnormal{N})}$, $t_{\textnormal{T}}^{(\textnormal{I})}$, $t_{\textnormal{T}, 1}^{(\textnormal{F})}$, and $t_{\textnormal{T}, 2}^{(\textnormal{F})}$, yielding deviations from the expected values by more than 50\% with standard floating point precision. But even with higher precision, it is preferable to avoid the evaluation of the transition times altogether. Thus, we compute $G$ on a grid, using fixed time steps. This makes it possible to read out the values of $G$ at the grid points and directly compare them to the values of $G_{\textnormal{T}}^{(\textnormal{N})}$, $G_{\textnormal{T}}^{(\textnormal{I})}$, $G_{\textnormal{T}, 1}^{(\textnormal{F})}$, and $G_{\textnormal{T}, 2}^{(\textnormal{F})}$ in order to determine the actual solution branch without referring to the transition times. 

The algorithm begins with the definitions of the free parameters, namely the injection times $t_i$, the injection strengths $q_i$, and the reciprocal initial electron energies $x_i$ for $i : 1 \leq i \leq m$, the nondimensional magnetic field strength $b$, the synchrotron and SSC cooling rate prefactors $D_0$ and $A_0$, the Lorentz boost $\mathcal{D}$ of the plasmoid, and the upper time boundary of the grid $t_{\textnormal{end}}$. The value of $t_{\textnormal{end}}$ is chosen as one and a half times the injection time of the final injection for a multiple-injection scenario or given by a sufficiently large value for a single-injection scenario. In a realistic setting, however, $t_{\textnormal{end}}$ corresponds to the end of the observation time. The time grid is set up homogeneously and linearly, and the number of grid points can be chosen arbitrarily. All computations are performed in the plasmoid rest frame. For the later evaluation of the fluence SEDs, the relevant quantities are transformed into the observer frame (see Appendix \ref{LTOF}). Ordered lists of the initial and transition values $G_i$, $G_{\textnormal{T}}^{(\textnormal{S})} = \sqrt{A_0 \, q_1/D_0}$, $G_{\textnormal{T}}^{(\textnormal{N})}$, $G_{\textnormal{T}}^{(\textnormal{I})}$, $G_{\textnormal{T}, 1}^{(\textnormal{F})}$, $G_{\textnormal{T}, 2}^{(\textnormal{F})}$, $G_{\textnormal{T}}^{(\textnormal{N} \rightarrow \textnormal{I})}$, and $G_{\textnormal{T}}^{(\textnormal{I} \rightarrow \textnormal{F})}$, as well as a list keeping track of the elements of the sets $S_1$, $S_2$, and $S_3$, and a list of the various solution branches are implemented. These lists are constantly updated during runtime. The algorithm constructs $G$ incrementally in two loops. The first loop covers $G$ from the time of the first injection $t = 0$ to the time of the second injection $t = t_2$ or -- in a scenario with only a single injection -- to the upper time boundary $t_{\textnormal{end}}$ using the solutions (\ref{sAS1})-(\ref{sAS3}). The second loop computes $G$ from the time of the second injection to the time $t_{\textnormal{end}}$ in case $t_2 < t_{\textnormal{end}}$ employing the solutions (\ref{NIRSOL1NEW}), (\ref{S1a})-(\ref{S1b}), (\ref{iidsol1})-(\ref{iidsol4}), (\ref{fidsol2})-(\ref{fidsol4}), and (\ref{S2a})-(\ref{S2b}). During each step, the current time $t_{\textnormal{cur.}}$ is incremented by a fixed value and $G_{\textnormal{cur.}} := G(t_{\textnormal{cur.}})$ is determined according to the proper solution branch, which is automatically selected via the above-mentioned lists. Moreover, at each grid point, the analytical expressions for $G$ are glued together continuously. In more detail, after initializing the values of $\mathscr{C}_1$, $\mathscr{C}_2$, $\mathscr{C}_3$, $\mathscr{C}_4$, as well as $G_{\textnormal{T}}^{(\textnormal{N} \rightarrow \textnormal{I})}$ and $G_{\textnormal{T}}^{(\textnormal{I} \rightarrow \textnormal{F})}$ at $t = 0$, the first loop starts its iteration, if $0 < G_{\textnormal{T}}^{(\textnormal{S})} < G_2$, in the first SID solution branch of (\ref{sAS1}). At each step, it evaluates $G$ at the current time grid point and checks whether $G_{\textnormal{cur.}}$ exceeds or equals $G_{\textnormal{T}}^{(\textnormal{S})}$, i.e., whether $G$ needs to be expressed by the second SID solution branch of (\ref{sAS1}). If necessary, $G$ is altered accordingly and connected to the previous solution branch continuously. (In case $0 < G_2 \leq G_{\textnormal{T}}^{(\textnormal{S})}$ or $G_{\textnormal{T}}^{(\textnormal{S})} \leq 0 < G_2$, the first loop makes use of the solutions (\ref{sAS2}) or (\ref{sAS3}), respectively.) Also, if $G_{\textnormal{cur.}}$ becomes larger than or equal to $G_{\textnormal{T}}^{(\textnormal{N} \rightarrow \textnormal{I})}(1)$ or $G_{\textnormal{T}}^{(\textnormal{I} \rightarrow \textnormal{F})}(1)$, the list for $S_1$, $S_2$, and $S_3$ is updated. The loop ends if $t_{\textnormal{cur.}}$ exceeds or equals either $t_2$ or $t_{\textnormal{end}}$. In the latter case, the numerical construction of $G$ is completed. The second loop constructs $G$ in a similar way as the first loop, but now more cases, which arise from the more elaborate structure of the analytical multiple injection solution, have to be taken into account. Once the second loop ends, the values of $G$ are known at every point of the grid. This allows us to evaluate the synchrotron and SSC intensities at each grid point by simply substituting these values into the corresponding analytical formulas (\ref{synint}) and (\ref{SSCTINT}). The associated total fluences are approximated by sums of the areas of rectangles, each of which is defined by the intensity at the left grid point and the size of the time step. Alternatively, one could implement the approximate analytical formulas derived in Section \ref{SEC4}. The total synchrotron and SSC fluence SEDs are then computed as the product of the respective energy and fluence. Since the number of grid points can be increased arbitrarily, the precision of these computations is limited only by the machine accuracy and the available CPU time.

\end{appendix}

\end{document}